\documentclass[a4paper,11pt]{article}
\pdfoutput=1 
\usepackage{jheppub} 
\usepackage{tikz}
\usetikzlibrary{snakes}
\usepackage{slashed}
\usepackage{ulem}
\usepackage[font=small,labelfont=bf,width=1\textwidth]{caption}
\usepackage{subfigure}
\usepackage{comment}
\usepackage[all]{xy}
\usepackage{multirow}
\usepackage{float}
\usepackage{mathtools}

\usepackage{color}

\newcommand \be{\begin{eqnarray}}
\newcommand \ee{\end{eqnarray}}

\numberwithin{equation}{section}

\DeclareMathOperator{\llangle}{\big\langle\hspace{-1.2mm}\big\langle\hspace{-.5mm}}
\DeclareMathOperator{\rrangle}{\hspace{-.5mm}\big\rangle\hspace{-1.2mm}\big\rangle}
\DeclareMathOperator{\Tr}{Tr}

\DeclareMathOperator{\sTr}{sTr}

\DeclareMathOperator{\diag}{diag}

\newcommand{\bea}{\begin{eqnarray}}
\newcommand{\eea}{\end{eqnarray}}
\newcommand{\beq}{\begin{equation}}
\newcommand{\eeq}{\end{equation}}
\newcommand{\bal}{\begin{equation}\begin{aligned}}
\newcommand{\eal}{\end{aligned} \end{equation}}

\newcommand{\cA}{{\mathcal A}}
\newcommand{\cB}{{\mathcal B}}
\newcommand{\cC}{{\mathcal C}}
\newcommand{\cD}{{\mathcal D}}

\newcommand{\cL}{{\mathcal L}}

\newcommand{\cN}{{\mathcal N}}
\newcommand{\cP}{{\mathcal P}}
\newcommand{\cQ}{{\mathcal Q}}
\newcommand{\cT}{{\mathcal T}}
\newcommand{\cR}{{\mathcal R}}
\newcommand{\cO}{{\mathcal O}}

\newcommand{\cW}{{\mathcal W}}

\newcommand{\cZ}{{\mathcal Z}}

\newcommand{\isEquivTo}[1]{\underset{#1}{\simeq}}

\title{Interpolating Wilson loops and enriched RG flows}

\author[a]{Luigi Castiglioni,}
\author[a]{Silvia Penati,}
\author[a]{Marcia Tenser,}
\author[b,c,d]{Diego Trancanelli}

\affiliation[a]{Dipartimento di Fisica, Universit\`a degli Studi di Milano--Bicocca and INFN, Sezione di Milano--Bicocca, Piazza della Scienza 3, 20126 Milano, Italy}
\affiliation[b]{Dipartimento di Scienze Fisiche, Informatiche e Matematiche, Universit\`a di Modena e Reggio Emilia, via G. Campi 213/A, 41125 Modena, Italy}
\affiliation[c]{INFN Sezione di Bologna, via Irnerio 46, 40126 Bologna, Italy}
\affiliation[d]{Institute of Physics, University of S\~ao Paulo, 05314-970 S\~ao Paulo, Brazil}

\emailAdd{l.castiglioni8@campus.unimib.it}
\emailAdd{silvia.penati@mib.infn.it}
\emailAdd{marciatenser@gmail.com}
\emailAdd{dtrancan@gmail.com} 

\abstract{
We study new $1/24$ BPS circular Wilson loops in ABJ(M) theory, which are defined in terms of several parameters that continuously interpolate between previously known $1/6$ BPS loops (both bosonic and fermionic) and $1/2$ BPS fermionic loops. We compute the expectation value of these operators up to second order in perturbation theory using a one-dimensional effective field theory approach. Within dimensional regularization, we find non-trivial $\beta$-functions for the parameters, which are marginally relevant deformations triggering RG flows from a UV fixed point represented by the $1/6$ BPS bosonic loop to an IR fixed point represented by a $1/2$ BPS fermionic loop. Generically, along all flows at least one supercharge of the theory is preserved, so that we refer to them as enriched RG flows. In particular, fixed points are connected through $1/6$ BPS fermionic operators. This holds at framing zero, which is a consequence of the regularization scheme employed. We also establish a g-theorem, relating the expectation values of the Wilson loops corresponding to the UV and IR fixed points of the flow, and discuss the one-dimensional defect SCFT living on the Wilson loop contour.
}

\arxivnumber{}

\begin{document}
\maketitle

\let\clearpage\relax
\section{Introduction}

Over the past few years, three-dimensional Chern-Simons-matter theories have been shown to display rich moduli spaces of supersymmetric line operators, starting with the discovery of $1/6$ BPS bosonic Wilson loops \cite{Drukker:2008zx,Chen:2008bp,Kluson:2008zrv,Rey:2008bh},  vortex loops \cite{Drukker:2008jm} and the $1/2$ BPS fermionic Wilson loop \cite{Drukker:2009hy} of the ABJ(M) theory \cite{Aharony:2008ug,Aharony:2008gk}. These studies have been subsequently extended to less supersymmetric settings like $\cN\ge 2$ quiver theories \cite{Gaiotto:2008sd,Imamura:2008dt,Hosomichi:2008jd,Hama:2010av} in, for example, \cite{Gaiotto:2007qi,Ouyang_2015,Cooke_2015,Ouyang:2015iza,Ouyang:2015bmy,Mauri:2017whf,Mauri_2018}, to continue with more recent attempts at a full classification of so-called hyperloop operators in \cite{drukker2020bps,Drukker:2020dvr,Drukker:2022ywj,Drukker:2022bff}. See \cite{Drukker:2019bev} for a review.
 
A characteristic feature in the construction of the BPS Wilson loops in these theories is the appearance of parametric families of operators interpolating between different amounts of preserved supersymmetries. One can in fact start from a given operator, be it bosonic as in \cite{Drukker:2020dvr} or fermionic as in \cite{Drukker:2022ywj}, choose a combination of supercharges it preserves and write down a deformation of that operator built out of the matter fields, which still preserves that supercharge. For special values of the parameters entering the definition of the deformation, supersymmetry enhancement is possible. This allows to interpolate continuously among different operators, preserving a varying number of supercharges of the theory.  

Given this plethora of BPS Wilson loops, it is natural to study the interpolations among them from the point of view of RG flows on defects, following the seminal work by Polchinski-Sully \cite{Polchinski_2011} and the subsequent literature, see for example \cite{Beccaria:2017rbe,Beccaria:2018ocq,Correa:2019rdk,Cuomo:2021rkm,Beccaria:2021rmj,Beccaria:2022bcr,Garay:2022szq,Aharony:2022ntz}.
In those cases, one has typically one parameter interpolating between BPS and non-supersymmetric operators, like the prototypical example of the $\zeta$-deformed operator introduced in \cite{Polchinski_2011} for $\cN=4$ super Yang-Mills in four dimensions, which interpolates between the ordinary, non-supersymmetric Wilson loop for $\zeta=0$ and the $1/2$ BPS Wilson-Maldacena loop \cite{Maldacena_1998} for $\zeta=1$. 

In this paper we initiate a study of such RG flows between Wilson loop operators in ABJ(M) theory. One main difference with respect to the cases mentioned above is that our flows are between operators that always preserve some supercharge, being therefore `enriched' flows: symmetries (in particular supersymmetries) are not completely broken along the flow, similarly to what has been considered in \cite{Cordova:2022lms}. Moreover, the flow spaces we consider are multi-dimensional,\footnote{For an example in four dimensions see \cite{Beccaria:2018ocq}, in which the deformation of the latitude Wilson loop in $\cN=4$ super Yang-Mills is considered.} as these Wilson loops are defined in terms of more than one parameter undergoing renormalization. 

\begin{figure}[]
    \centering
    \includegraphics[width=.7\textwidth]{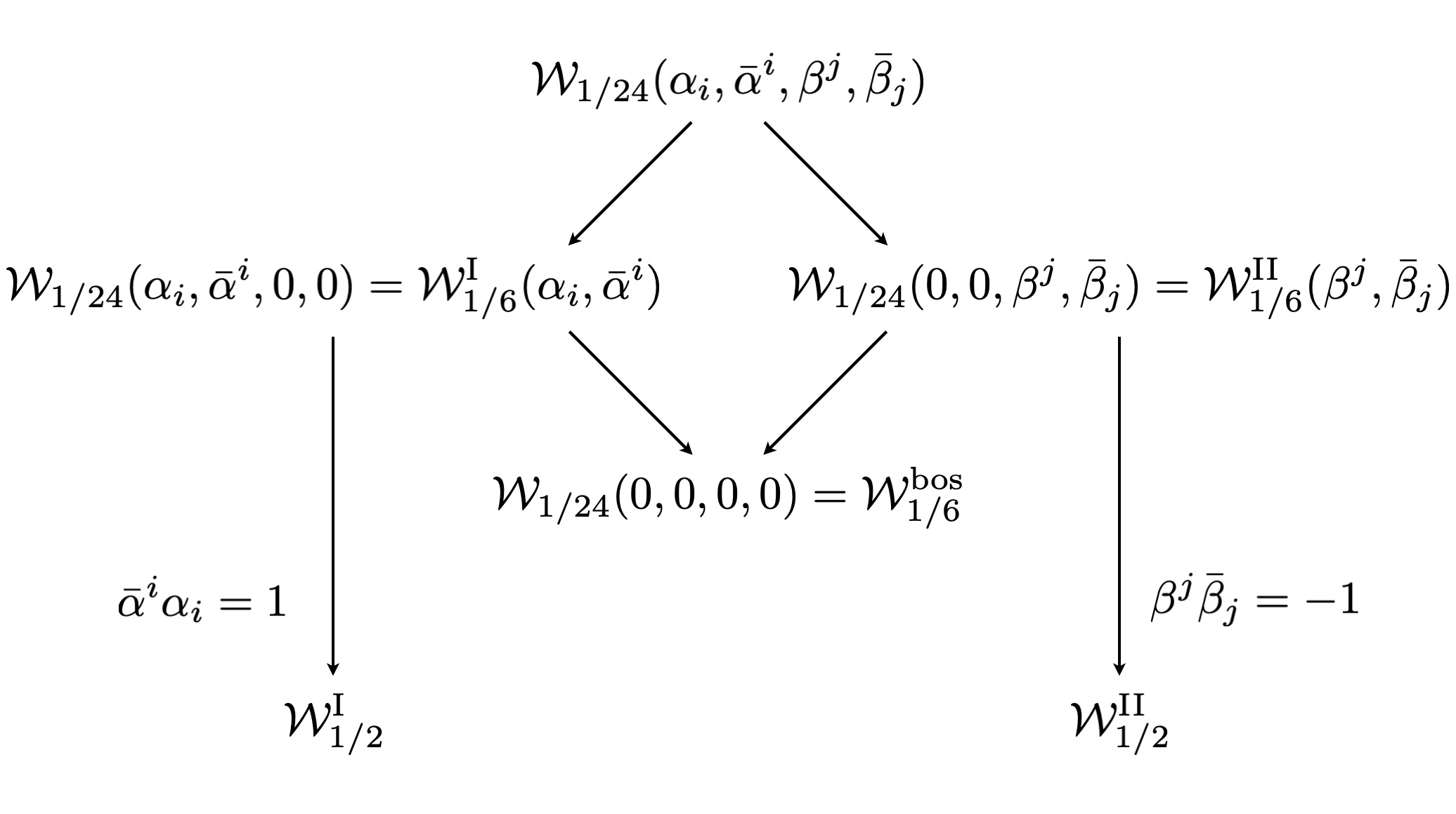}
    \caption{The interpolations among the various Wilson loops considered in this paper. }
    \label{fig:net}
\end{figure}

More specifically, we consider ABJ(M) theory on $\mathbb{R}^3$ and introduce a new BPS circular Wilson loop that preserves, generically, only one supercharge of the theory and is therefore $1/24$ BPS. We call it $\cW_{1/24}$. This operator has not been discussed before in ABJ(M), but its equivalent has appeared in the context of hyperloops in $\cN=4$ Chern-Simons-matter theories \cite{Drukker:2020dvr} and can be mapped to a corresponding operator in ABJ(M). This Wilson loop is defined in terms of a superconnection containing a coupling to the scalars and the fermions of the theory through 8 dimensionless parameters that we call $\alpha_i, \bar\alpha^i$ (with $i=1,2$) and $\beta^j,\bar\beta_j$ (with $j=3,4$). For generic values of these parameters the operator is $1/24$ BPS, as already mentioned. Selecting either $\alpha_i=\bar\alpha^i=0$ or $\beta^j=\bar\beta_j=0$, there is  supersymmetry enhancement and the loop from $1/24$ BPS becomes a $1/6$ BPS fermionic operator. In fact there are two different $1/6$ BPS operators that can be obtained in this way (one with vanishing alphas and one with vanishing betas), which we denote $\cW^\textrm{I}_{1/6}$ and $\cW^\textrm{II}_{1/6}$ \cite{Ouyang:2015iza,Ouyang:2015bmy}. If, moreover, the remaining parameters are set to a specific value, $\alpha_i\bar\alpha^i=1$ or $\bar\beta_j\beta^j=-1$, respectively, the $1/6$ BPS fermionic operators become the $1/2$ BPS fermionic Wilson loops $\cW_{1/2}^\textrm{I}$ and $\cW_{1/2}^\textrm{II}$ 
\cite{Drukker:2009hy}. These two loops differ by an overall sign in the scalar coupling, with $\cW_{1/2}^\textrm{I}$ being the loop with a mostly plus coupling originally introduced in \cite{Drukker:2009hy}. In this paper we shall be mainly interested in $\cW_{1/2}^\textrm{I}$. On the other hand, when all the parameters are turned off at the same time, $\alpha_i=\bar\alpha^i=\beta^j=\bar\beta_j=0$, one has the bosonic $1/6$ BPS Wilson loop $\cW^\textrm{bos}_{1/6}$ of \cite{Drukker:2008zx,Chen:2008bp,Kluson:2008zrv,Rey:2008bh}. This network of interpolations is summarized in figure~\ref{fig:net}.

At the classical level all Wilson loops in figure \ref{fig:net} are cohomologically equivalent, {\it i.e.} their expressions differ by a ${\cal Q}$-exact term, where ${\cal Q}$ is one of the preserved supercharges. In principle, this would imply that their vacuum expectation values (VEVs) should be all equal and independent of the parameters. However, this is true for instance for operators supported along straight lines, but it is no longer true on the circle, due to the well-known conformal anomaly \cite{gross} and framing effects. In fact, while supersymmetric localization requires framing 1, the regularization scheme we employ, namely dimensional regularization, is alternative to framing regularization and corresponds to framing 0.  This is the reason why the circular VEVs that we are going to compute carry a non-trivial dependence on the parameters, so providing interpolating BPS (enriched) flows.

We compute the vacuum expectation value of this $1/24$ BPS circular loop, for generic values of the parameters, up to two loops in perturbation theory at weak coupling. The way we do it is by mapping this problem to the computation of the two-point function of certain  auxiliary fields living on the one-dimensional theory along the Wilson loop contour. This is something that has been done in the past \cite{Samuel:1978iy,Gervais:1979fv} for ordinary Wilson loops, but we extend it to the case at hand, namely for Wilson loops defined in terms of superconnections. In particular, compared to the previous applications, in which the one-dimensional theory only contained a Fermi field, we have a theory with both commuting and anticommuting fields. The final planar-limit result for the Wilson loop VEV is given by 
\begin{equation}
\label{WVEVintro}
    \langle \cW_{1/24}
    \rangle = 1 -\frac{\pi^2}{6 k^2}\left[ N_1^2+N_2^2-4N_1N_2-3N_1N_2(\bar\alpha^i\alpha_i-\beta^j\bar\beta_j -1)^2 \right]+{\cal O}\left(\frac{1}{k^3}\right)\,.
\end{equation}
Here $N_1$ and $N_2$ are the ranks of the two gauge fields of the $U(N_1)_k \times U(N_2)_{-k}$ ABJ(M) theory, and $k$ is the Chern-Simons level. From this expression one obtains the VEVs of the $1/6$ BPS bosonic (all parameters equal to zero) and $1/6$ BPS fermionic operators (alphas or betas equal to zero), as well as of the $1/2$ BPS fermionic operator (last term in the square bracket equal to zero).

The coupling parameters undergo a non-trivial renormalization, leading to non-vanishing $\beta$-functions
\begin{equation}
\beta_{\alpha_k}= \frac{g^2}{4\pi}(N_1+N_2)\, (\bar\alpha^i\alpha_i + \beta^j\bar\beta_j-1)\alpha_k\,,
    \qquad
    \beta_{\beta^k}= \frac{g^2}{4\pi}(N_1+N_2) (\bar\alpha^i\alpha_i + \beta^j\bar\beta_j+1)\beta^k\,,  
\end{equation}
with similar expressions for the barred quantities. This shows that the Wilson loop parameters can be seen as marginally relevant deformations, triggering an RG flow from a UV fixed point represented by the $1/6$ BPS bosonic Wilson loop of ABJ(M) towards the $1/2$ BPS loop $\cW_{1/2}^{\text{I}}$. Such a flow is presented in figure \ref{fig:RGflow}.

This has a nice interpretation in terms of defects. In fact, it is well known that the bosonic $1/6$ BPS and fermionic $1/6$ and $1/2$ BPS operators describe one-dimensional superconformal theories (SCFTs) given by local operators inserted on the Wilson loop contour. Instead, the new $1/24$ BPS operator supports a defect which is no longer (super)conformal, as it does not preserve enough supersymmetries and the contour dependence of the scalar couplings breaks conformality. 

In this framework the RG flows depicted in figure \ref{fig:RGflow} can be interpreted as connecting different (super)conformal defects seated at the fixed points. Flowing along the green line of that figure we reach a non-trivial IR fixed point. In the defect theory at the UV fixed point we compute the anomalous dimension of the parametric perturbation and consistently find a small negative value, thus confirming that it corresponds to a marginally relevant deformation.

Finally, from \eqref{WVEVintro} and the $\beta$-functions we also establish a g-theorem, relating the VEVs of the Wilson loops corresponding to the UV and IR fixed points of the flows
\begin{equation}
    \log\langle \cW_{1/6}^\textrm{bos}\rangle > 
    \log\langle \cW_{1/2}^\textrm{I} \rangle\,,
\end{equation}
similarly to what has been done for Wilson loops in four dimensions in \cite{Cuomo:2021rkm,Beccaria:2021rmj}, with the main difference being that our flows are BPS, as stressed above.

As mentioned already above, the comparison of the Wilson loop VEV \eqref{WVEVintro} with the result coming from a matrix model computation \cite{Kapustin:2009kz,Marino:2009jd} requires taking into account framing issues, see chapter 6 of \cite{Drukker:2019bev} for a review. The regularization scheme employed in our perturbative computation amounts in fact to computing the VEV at framing $f=0$, whereas the matrix model computation yields a result valid for $f=1$. Moreover, at framing one the VEVs of all loops of figure \ref{fig:net} coincide, as these are all cohomologically equivalent operators. This is clearly not true for \eqref{WVEVintro}, which is obtained at framing zero.\footnote{Perturbative computations at framing 1 represent a hard open problem, especially regarding the evaluation of fermionic diagrams.} Note, in particular, how this VEV depends explicitly on the alpha and beta parameters of the deformation, which are not present in the definition of the matrix model insertion corresponding to these operators. The relation between the VEVs of $\cW_{1/24}$ at different framings is encoded in a phase, which we find empirically from our two-loop results to be given by
\begin{equation}
  \langle {\cal W}_{1/24} \rangle_{f=1} = \frac{N_1 e^{\frac{i\pi}{k}(N_1-(\bar\alpha^i\alpha_i-\beta^j\bar\beta_j)N_2)}+N_2 \, e^{\frac{i\pi}{k}((\bar\alpha^i\alpha_i-\beta^j\bar\beta_j)N_1-N_2)}}{N_1+N_2}\langle {\cal W}_{1/24}
     \rangle_{f= 0}\,.
\end{equation}
We expect this phase to receive corrections at higher order in perturbation theory, similarly to what happens for the $1/6$ BPS bosonic operator \cite{Bianchi:2016yzj}.

In this paper we also introduce a new operator: a $1/12$ BPS latitude Wilson loop defined in terms of an extra parameter, a latitude angle, along the lines of what has been done in $\cN=4$ super Yang-Mills in \cite{Drukker:2007dw,Drukker:2007yx,Drukker:2007qr} and in three-dimensional theories in \cite{Cardinali:2012ru,Bianchi:2014laa} and \cite{Drukker:2020dvr}. In a forthcoming publication \cite{CPTT}, we will generalize to this new setting the investigation of the present paper.

This paper is organized as follows.
In section \ref{sec:theory} we introduce the $1/24$ BPS circular Wilson loop, which is going to be the main character of our analysis, as well as the $1/12$ BPS latitude Wilson loop to be considered in the future. These operators are defined in terms of either traces or supertraces of superconnections. The former formulation simplifies the perturbative analysis while the latter is more natural for superconnections, so we discuss how to go from one to the other.
In section \ref{sec:renormalization} we consider an auxiliary problem in terms of one-dimensional fields which is suitable for studying the renormalization of the parameters of the $1/24$ BPS Wilson loop. This allows us to compute the $\beta$-functions of the Wilson loop parameters.
In section \ref{sec:WLVEV} we finally compute the vacuum expectation value of the $1/24$ BPS circular Wilson loop up to two loops in perturbation theory. This is the main result of this paper, together with the evaluation of the $\beta$-functions. 
In section \ref{sec:results} we collect and discuss our results. Specifically, we describe the RG flows among the different operators of figure \ref{fig:net} and plot an explicit example, we study the defect SCFT living on the Wilson loop, we establish the g-theorem mentioned above, and we compare the Wilson loop VEV with a matrix model computation. Finally, we conclude that section with some outlook. We collect some technical aspects in a series of appendices. In appendix \ref{app:abjm} we discuss our notation and conventions. In appendix \ref{sec:apxDorn} we derive the generalization of the one-dimensional auxiliary field method to Wilson loops defined in terms of superconnections. In appendix \ref{app:renormalization} we detail the computation of the various Feynman diagrams considered in the main text.

The reader who is not interested in technical details may skip section \ref{sec:renormalization} and go directly to section \ref{sec:Wresults}.
\section{Theory and supersymmetric loops}
\label{sec:theory}

\begin{figure}[H]
    \centering
    \begin{tikzpicture}
\draw[line width=.5mm] (6,2) circle (6mm);
\draw[line width=.5mm] (10,2) circle (6mm);
\draw[line width=.25mm,->] (6.57,2.1)--(9.41,2.1);
\draw[line width=.25mm,<-] (6.59,1.9)--(9.43,1.9);
\draw (6,2) node  []  {$A$};
\draw (10,2) node  []  {$\hat{A}$};
\draw (8,2.5) node  []  {${C_I\,\, \bar\psi^I}$};
\draw (8,1.5) node  []  {$\bar{C}^I\,\, \psi_I$};
\draw (6,1) node  []  {$k$};
\draw (10,1) node  []  {$-k$};
\end{tikzpicture}
    \caption{Quiver diagram of ABJ(M) theory. Below each node we include the level of the respective copy of the Chern-Simons action.}
    \label{fig:ABJMquiver}
\end{figure}
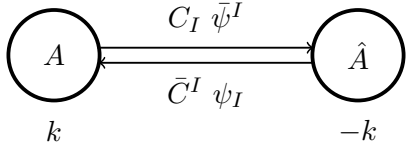

The field content of ABJ(M) can be depicted in terms of a quiver diagram as the one shown in figure \ref{fig:ABJMquiver}. It includes two gauge fields, $A$ and $\hat{A}$, with respective gauge groups $U(N_1)$ and $U(N_2)$. The matter sector has $SU(4)$ R-symmetry and is composed of scalars $C_I$ and fermions $\bar\psi^I$, $I=\{1,2,3,4\}$, in the $(\Box,\bar{\Box})$ representation of $U(N_1)\times U(N_2)$. By conjugation there are also $\bar{C}^I$ and $\psi_I$ in $(\bar{\Box},\Box)$.
Wilson loops preserving some amount of the 24 supercharges of the theory can be constructed by allowing for couplings to scalar bilinears. In this case the usual gauge connection $A$ is promoted to a bosonic connection $\cA$. Besides that, there is also the possibility of adding fermi fields, in which case the bosonic connection is further promoted to a superconnection $\cL$ \cite{Drukker:2009hy}.
Recently, the structure of these operators started being unravelled \cite{Drukker:2019bev} through the understanding that they are related via
\beq
\label{eqn:deformation}
\cL = \cL_0 + i\cQ G + G^2\,.
\eeq
The quantity $\cL_0$ is a composite bosonic connection complemented by a constant shift in one of the entries\footnote{In what follows $c$ will be either $\frac{1}{2}$ or $\frac{\cos\theta}{2}$, but this is not necessarily always the case \cite{Drukker:2022ywj}.}
\beq
\label{eqn:superconnection_bos}
\cL_0 = \begin{pmatrix}
\cA + c && 0 \\
0 && \hat\cA
\end{pmatrix}\,.
\eeq
The supercharge $\cQ$ is a suitable linear combination of supercharges preserved by $\cL_0$, and $G$ is an off-diagonal matrix comprised of scalars. These appear through a set of constant complex parameters that we denote as $\alpha_{i},\bar\alpha^{i},\beta^{j},\bar\beta_{j}$ (with $i=1,2$ and $j=3,4$), though they are not complex conjugates of each other. The construction \eqref{eqn:deformation} is such that $\cQ$ is always preserved by $\cL$, but one may find extra preserved supercharges depending on particular values of the parameters in $G$.

Below we consider two possible operators built from different choices of $\cL_0$. The first one is the $1/24$ BPS circular loop, the protagonist of the present analysis, and the second one is the $1/12$ BPS latitude loop of ABJ(M), which is going to be studied in detail in a future publication \cite{CPTT}. The construction of the latter is to a great extent parallel to the $\theta$-deformation considered in \cite{Drukker:2020dvr}, with the difference that what we mean by `latitude' here is an actual geometric latitude of the contour $x^\mu$ of the loop, instead of simply an internal $\theta$-deformation in the space of the couplings.


\subsection{1/24 BPS circular Wilson loop}
\label{sec:parametricWL}

The first operator that we are going to consider is supported along the circle
\beq
x^\mu = (0,\cos\tau,\sin\tau)\,.
\eeq
Its bosonic components can be separately charged under each node of the quiver
\bal
\label{eqn:bosonicconnections}
{\mathcal W}^\textrm{bos}  &= 
 \Tr \cP \exp\left(-i \oint \cA \,d\tau \right)\,,\qquad \cA = A_\mu \dot{x}^\mu - \frac{2 \pi i}{k}\vert\dot{x}\vert M_J^{\ I} C_I \bar{C}^J\,, \\
\hat{\mathcal W}^\textrm{bos} &= 
 \Tr \cP \exp\left(-i \oint \hat\cA \,d\tau \right)\,,\qquad \hat\cA = \hat{A}_\mu \dot{x}^\mu - \frac{2 \pi i}{k}\vert\dot{x}\vert M_J^{\ I} \bar{C}^J C_I\,.
\eal
When $M_J^{\ I}=\diag(-1,-1,1,1)$, they preserve the set of supercharges
\beq
\label{eqn:supercharges1/4}
Q_{12}^+ - i S_{12}^+ \,,\quad Q_{12}^- + i S_{12}^- \,,\quad Q_{34-} + i S_{34-} \,,\quad Q_{34+} - i S_{34+}\, ,
\eeq
and are therefore $1/6$ BPS operators \cite{Drukker:2008zx,Chen:2008bp,Kluson:2008zrv,Rey:2008bh}.

\begin{figure}[H]
\centering
\begin{tikzpicture}
\draw[decoration={snake,amplitude = .5mm,segment length=3.46mm},decorate,line width=.5mm] (6,2) circle (7mm);
\draw[line width=.5mm] (10,1.9) circle (6.5mm);
\draw[line width=.25mm,<-] (6.6,1.65) to [out=330,in=210] (9.4,1.65);
\draw[line width=.25mm,<-] (6.65,1.75) to [out=330,in=210] (9.35,1.75);
\draw[line width=.25mm,dashed,->] (6.5,1.5) to [out=330,in=210] (9.5,1.5);
\draw[line width=.25mm,dashed,->] (6.4,1.4) to [out=330,in=210] (9.58,1.4);
\draw[line width=.25mm,->] (6.68,2.2) to [out=30,in=150] (9.4,2.2);
\draw[line width=.25mm,->] (6.69,2.33) to [out=30,in=150] (9.45,2.3);
\draw[line width=.25mm,dashed,<-] (6.55,2.45) to [out=30,in=150] (9.55,2.42);
\draw[line width=.25mm,dashed,<-] (6.57,2.6) to [out=30,in=150] (9.6,2.53);
\draw (6,2) node  []  {};
\draw (10,2) node  []  {};
\draw (8,1.7) node  []  {$\alpha_{1}\,\alpha_2$};
\draw (8,0.5) node  []  {${\bar\alpha}^{1}\,\bar\alpha^2$};
\draw (8,2.2) node  []  {$\beta^3\,\beta^4$};
\draw (8,3.5) node  []  {${\bar\beta}_{3}\,\bar\beta_4$};
\end{tikzpicture}
\caption{The quiver diagram of the $1/24$ BPS Wilson loop in ABJ(M). Following the notation of \cite{drukker2020bps}, a squiggly circle, like the one on the left here, represents a node whose bosonic connection is shifted by the constant $c$. The couplings to (anti-)chiral fields are denoted by solid (dashed) arrows.}
\label{fig:1/24-quiver}
\end{figure}
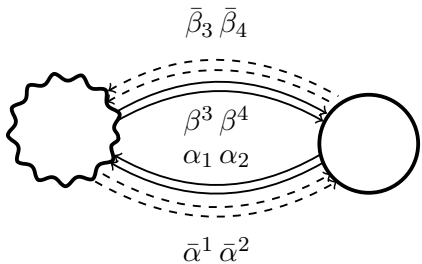

Their fermionic counterpart can be derived using the prescription outlined above. In this case, we take the supercharge $\cQ$ to be given by the linear combination
\beq
\cQ \equiv (Q_{12}^+ - i S_{12}^+) + (Q_{34+} - i S_{34+})\,.
\eeq
The constant shift is implemented in the composite bosonic connection $\cL_0$ such that $c=\frac{1}{2}$ and the $G$ matrix includes all four scalars of the theory as\footnote{The parameters appears in $G$ sticking to the notation of \cite{drukker2020bps}, so that unbarred (barred) parameters accompany (anti-)chiral fields in the chiral decomposition of the theory in $\cN=2$ language. We stress that barred/unbarred parameters are not complex conjugates of each others.}
\beq
\label{eqn:1/24-G}
G = \begin{pmatrix}
0 && \bar\alpha^1 C_1 + \bar\alpha^2 C_2 + e^{-i\tau}(\beta^3 C_3 + \beta^4 C_4) \\
\alpha_1 \bar{C}^1 + \alpha_2 \bar{C}^2 + e^{i\tau} (\bar\beta_3 \bar{C}^3 + \bar\beta_4 \bar{C}^4) && 0
\end{pmatrix}\,.
\eeq
Plugging this in \eqref{eqn:deformation} we find that the resulting superconnection $\cL$ can be explicitly written as\footnote{To write $\cL$ we suitably scaled couplings so to recover the $1/2$ BPS loop of \cite{Drukker:2009hy} when $\alpha^2=\bar\alpha_2=-1$ and all other parameters are zero. Also, whenever omitted, spinorial indices are meant to be contracted up-down, {\it i.e.} $\lambda \chi \equiv \lambda^\alpha \chi_\alpha$.}
\beq
\label{eqn:1/24-superconnection}
\cL = \begin{pmatrix}
\cA'+ \frac{1}{2} && \eta \, (\bar\alpha^1\bar{\psi}^2 - \bar\alpha^2 \bar{\psi}^1) + e^{-i\tau} \xi \, (\beta^3 \bar\psi^4 - \beta^4 \bar\psi^3) \\
\xi \, (\alpha_1 \psi_2 - \alpha_2 \psi_1) + e^{i\tau} \eta \, (\bar\beta_3 \psi_4 - \bar\beta_4 \psi_3) && \hat{\cA}' 
\end{pmatrix}\,,
\eeq
where the commuting spinors $\eta$ and $\xi$ are
\beq\label{eq:spinors}
\eta^\alpha = \sqrt{\frac{2\pi i}{k}}(1,-ie^{-i\tau})^\alpha \,,\qquad \xi^\alpha = \sqrt{\frac{2\pi i}{k}}(-ie^{i\tau},1)^\alpha \,.
\eeq
The diagonal entries are primed because now the scalar coupling matrix $M_J^{\ I}$ is such that it receives the contribution coming from $G^2$, {\it i.e.} it is 
\beq
\label{eqn:M124bos}
M_J^{\ I} = \begin{pmatrix}
-1 + 2\bar\alpha^1 \alpha_1 && 2\bar\alpha^1 \alpha_2 && 2e^{i\tau} \bar\alpha^1 \bar\beta_3 && 2e^{i\tau} \bar\alpha^1 \bar\beta_4 \\
2\bar\alpha^2\alpha_1 && -1 + 2\bar\alpha^2 \alpha_2 && 2e^{i\tau}\bar\alpha^2 \bar\beta_3 && 2e^{i\tau}\bar\alpha^2 \bar\beta_4 \\
2e^{-i\tau} \beta^3 \alpha_1 && 2e^{-i\tau} \beta^3 \alpha_2 && 1 + 2\beta^3 \bar\beta_3 && 2\beta^3 \bar\beta_4 \\
2e^{-i\tau} \beta^4 \alpha_1 && 2e^{-i\tau} \beta^4 \alpha_2 && 2\beta^4 \bar\beta_3 && 1+ 2\beta^4 \bar\beta_4
\end{pmatrix}\,.
\eeq
The resulting operator,
\beq
\label{def-W}
{\cal W}=
\sTr \cP \exp\left(-i\oint \cL \,d\tau\right)\,,
\eeq
preserves $\cQ$. Following the proposal of \cite{drukker2020bps}, it can be represented in terms of a quiver diagram as the one shown in figure \ref{fig:1/24-quiver}. 

As to the best of our knowledge this is the first time that such an operator is presented,\footnote{The quiver representation of figure \ref{fig:1/24-quiver} has already appeared in \cite{drukker2020bps}, see their figure 7, but the corresponding operator was not written down explicitly.} we find it enlightening to stop and make a few comments about it before proceeding. First of all we notice that in the quiver of figure \ref{fig:1/24-quiver} solid arrows point both into and out of the squiggly node. From the analysis in \cite{Drukker:2020dvr}, one can conclude that only $\cQ$ is preserved, generically, and the loop is therefore 1/24 BPS. 

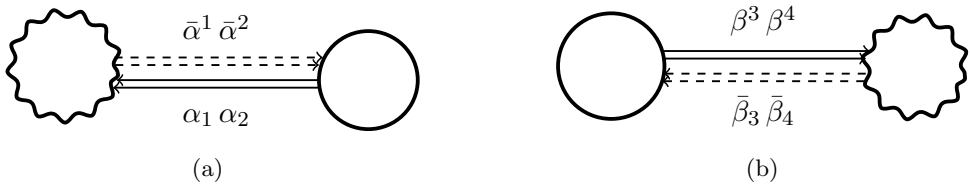
\begin{figure}[h]
    \centering
    \subfigure[]{
    \begin{tikzpicture}
\draw[decoration={snake,amplitude = .5mm,segment length=3.46mm},decorate,line width=.5mm] (6,2) circle (7mm);
\draw[line width=.5mm] (10,1.9) circle (6.5mm);
\draw[line width=.25mm,->,dashed] (6.66,2.2) to (9.4,2.2);
\draw[line width=.25mm,->,dashed] (6.67,2.1) to (9.35,2.1);
\draw[line width=.25mm, <-] (6.69,1.9) to (9.34,1.9);
\draw[line width=.25mm, <-] (6.64,1.8) to (9.34,1.8);
\draw (6,2) node  []  {};
\draw (10,2) node  []  {};
\draw (8,2.6) node  []  {${\bar\alpha}^{1}\,{\bar\alpha}^2$};
\draw (8,1.4) node  []  {$\alpha_{1}\,\alpha_2$};
\end{tikzpicture}
\label{subfig:1/6-quiver1}} \qquad\qquad
    \subfigure[]{\begin{tikzpicture}
\draw[line width=.5mm] (6,2) circle (7mm);
\draw[decoration={snake,amplitude = .5mm,segment length=3.46mm},decorate, line width=.5mm] (10,1.9) circle (6.5mm);
\draw[line width=.25mm,dashed,<-] (6.69,1.9) to (9.4,1.9);
\draw[line width=.25mm,dashed,<-] (6.67,1.8) to (9.35,1.8);
\draw[line width=.25mm,->] (6.66,2.2) to (9.4,2.2);
\draw[line width=.25mm,->] (6.67,2.1) to (9.35,2.1);
\draw (6,2) node  []  {};
\draw (10,2) node  []  {};
%
\draw (8,1.4) node  []  {${\bar\beta}_{3}\,\bar\beta_4$};
\draw (8,2.6) node  []  {$\beta^3\,\beta^4$};
\end{tikzpicture}
\label{subfig:1/6-quiver2}}
    \caption{Branches of $1/6$ BPS loops breaking different $SU(2)$ R-symmetries of $\cL_0$.}
    \label{fig:1/6-quivers}
\end{figure}

Particular subcases of supersymmetry enhancement can be read off directly from the quiver structure. We know \cite{Drukker:2020dvr} that when solid arrows point only into the squiggly node, all supercharges originally preserved by $\cL_0$ are preserved by $\cL$ and the resulting operator is $1/6$ BPS. To be explicit, when only $\alpha_i,\bar\alpha^i$ parameters appear in $G$, see \eqref{eqn:1/24-G}, the corresponding operator can be depicted in terms of a quiver diagram as in figure \ref{subfig:1/6-quiver1}. In this case $G$ breaks only one $SU(2)$ R-symmetry subgroup of $\cL_0$. Moreover, at the particular point $\bar\alpha^i\alpha_i=1$ the resulting loop enjoys extra $SU(3)$ symmetry and becomes $1/2$ BPS. On the other hand, for the case where only the $\beta^j,\bar\beta_j$ parameters appear in $G$, it is useful to employ the gauge where the constant shift (and therefore the squigglyness of the corresponding quiver diagram) lies in the second node. In this case $G$ loses the awkward $e^{\pm i\tau}$ phases and the resulting operator can be depicted as in figure \ref{subfig:1/6-quiver2}. This corresponds to $G$ breaking the other $SU(2)$ R-symmetry subgroup of $\cL_0$ and at the particular point where $\beta^j\bar\beta_j=-1$, $SU(3)$ symmetry is restored and the loop is $1/2$ BPS. This is summarized in figure \ref{fig:net}.

The particular cases outlined above recover the original analysis proposed in the second chapter of \cite{Drukker:2019bev}, where the authors propose $G$'s that can be comprised of $\{C_1,\bar{C}^1,C_2,\bar{C}^2\}$ or of $\{C_3,\bar{C}^3,C_4,\bar{C}^4\}$. Our construction is therefore a generalization of that and corresponds to the most generic BPS operator one can build out of $\cL_0$. All previously known examples can be derived from it through appropriate choices of the parameters.

Finally, seen from a different perspective, by viewing ABJ(M) as the orbifold of $\cN=4$, the operator outlined above corresponds to the $1/16$ BPS operator appearing in figure 5 of \cite{Drukker:2020dvr}, now specialized to ABJ(M).


\subsection{1/12 BPS latitude Wilson loops}
\label{sec:parametriclatitude}

The latitude operators are supported along
\beq 
x^\mu=(\sin\theta,\cos\theta\cos\tau,\cos\theta\sin\tau)\,.
\eeq
Their bosonic representatives are still written as \eqref{eqn:bosonicconnections}, but this time with
\beq
M_J^{\ I} = \begin{pmatrix} -\cos\theta && 0 && e^{-i\tau}\sin\theta && 0 \\
0 && -1 && 0 && 0 \\
e^{i\tau}\sin\theta && 0 && \cos\theta && 0 \\
0 && 0 && 0 && 1
\end{pmatrix}\,.
\eeq
This form of $M$ is such that the resulting loops are invariant under
\bal
\label{eqn:latitudeQs}
\cos\frac{\theta}{2}\bigg(Q_{12}^+ - i e^{-i\theta} S_{12}^+\bigg) &- i\sin\frac{\theta}{2}\bigg(Q_{23}^-+ie^{-i\theta}S_{23}^-\bigg)\,,\\
\cos\frac{\theta}{2}\bigg(Q_{34+} - i e^{i\theta} S_{34+}\bigg) &- i\sin\frac{\theta}{2}\bigg(Q_{14-}+ie^{i\theta}S_{14-}\bigg)\,.
\eal

As before, we follow the prescription \eqref{eqn:deformation} to construct the fermionic counterparts. We take the supercharge $\cQ$ to be given by the sum of the supercharges above. Then the analysis of possible scalars to include in $G$ is parallel to the $\theta\neq0$ discussion of \cite{Drukker:2020dvr}. We find that $C_2,\bar{C}^2$ and $C_4,\bar{C}^4$ can not be included simultaneously due to the non-periodicity of boundary conditions that can not be fixed by means of a gauge transformation. To be precise, the superconnection would transform as the supercovariant derivative of
\beq
\begin{pmatrix}
0 && \bar\alpha^2 C_2 + e^{-i\tau\cos\theta} \beta^4 C_4 \\
\alpha_2 \bar{C}^2 + e^{i\tau\cos\theta} \bar\beta_4 \bar{C}^4 && 0
\end{pmatrix}\,,
\eeq
which does not have well-behaved boundary conditions. As for the inclusion of $C_1,\bar{C}^1$ and $C_3,\bar{C}^3$, we find that it requires promoting the superconnection to a $3\times 3$ supermatrix and taking a cover of the quiver of the theory. Since this goes beyond the scope of our present discussion, we leave such possibility to the future.

\begin{figure}[h]
    \centering
    \subfigure[]{
    \begin{tikzpicture}
\draw[decoration={snake,amplitude = .5mm,segment length=3.46mm},decorate, line width=.5mm] (6,2) circle (7mm);
\draw[line width=.5mm] (10,1.9) circle (6.5mm);
\draw[line width=.25mm,->,dashed] (6.66,2.1) to (9.37,2.1);
\draw[line width=.25mm,<-] (6.68,1.9) to (9.35,1.9);
\draw (6,2) node  []  {};
\draw (10,2) node  []  {};
\draw (8,2.6) node  []  {$\bar\alpha^2$};
\draw (8,1.4) node  []  {$\alpha_2$};
\end{tikzpicture}
\label{subfig:thetaneq0loops1}} \qquad\qquad
    \subfigure[]{\begin{tikzpicture}
\draw[line width=.5mm] (6,2) circle (7mm);
\draw[decoration={snake,amplitude = .5mm,segment length=3.46mm},decorate, line width=.5mm] (10,1.9) circle (6.5mm);
%
\draw[line width=.25mm, dashed,<-] (6.69,1.9) to (9.4,1.9);
\draw[line width=.25mm,->] (6.67,2.1) to (9.3,2.1);
\draw (6,2) node  []  {};
\draw (10,2) node  []  {};
%
\draw (8,1.4) node  []  {$\bar\beta_4$};
\draw (8,2.6) node  []  {$\beta^4$};
\end{tikzpicture}
\label{subfig:thetaneq0loops2}}
    \caption{Branches of $1/12$ BPS latitude loops. Points where supersymmetry is enhanced correspond to $\alpha^2\bar\alpha_2=-\bar\beta^4\beta_4=1$, where an $SU(2)$ subgroup of R-symmetry is restored and the operators become $1/6$ BPS.}
    \label{fig:thetaneq0loops}
\end{figure}
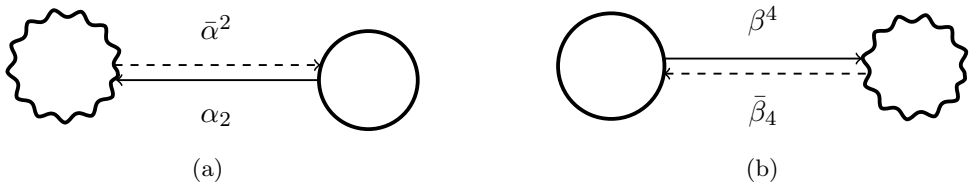

We consider, therefore, two possible loops built out of $G$ coupling either to $C_2,\bar{C}^2$ or to $C_4,\bar{C}^4$. Both options are represented in figure \ref{fig:thetaneq0loops}, where the squigglyness of the nodes now stands for a constant shift of $c=\frac{\cos\theta}{2}$. 

For brevity, we focus here on the explicit construction of the first branch. Its composite bosonic connection has a constant shift lying in the first node and the final form of the superconnection is
\beq
\cL^\theta = \begin{pmatrix}
\cA^{'} + \frac{\cos\theta}{2} && -\bar\alpha^2 \eta\left(\cos\frac{\theta}{2}\bar\psi^1-\sin\frac{\theta}{2}e^{i\tau}\bar\psi^3\right) \\
-\alpha_2 \xi\left(\cos\frac{\theta}{2}\psi_1-\sin\frac{\theta}{2}e^{-i\tau}\psi_3\right) && \hat\cA^{'}
\end{pmatrix}\,,
\eeq
with the scalar coupling now given by
\beq
M_J^{\ I} = \begin{pmatrix} -\cos\theta && 0 && e^{-i\tau}\sin\theta && 0 \\
0 && -1+ 2\bar\alpha^2\alpha_2 && 0 && 0 \\
e^{i\tau}\sin\theta && 0 && \cos\theta && 0 \\
0 && 0 && 0 && 1
\end{pmatrix}\,.
\eeq
At the particular point where $\bar\alpha^2\alpha_2=1$, an $SU(2)$ subgroup of R-symmetry is preserved. The loop is invariant under the supercharges \eqref{eqn:latitudeQs} and the ones obtained by swapping the $2\leftrightarrow 4$ indices. This is the $1/6$ BPS latitude operator introduced in \cite{Bianchi:2014laa} and further studied in \cite{Bianchi:2018bke,Griguolo_2021}.


\subsection{Removing the constant shift}
\label{sec:constantshift}

The constant shift $c$ in \eqref{eqn:superconnection_bos} is useful in the definition of the operators (see chapter 2 of \cite{Drukker:2019bev}). Its presence gives rise to a manifestly reparametrisation invariant operator. Moreover, Wilson loops with this shift are (super)gauge invariant without the need for an additional twist matrix \cite{Cardinali:2012ru} and can be defined as the supertrace of a superconnection, as in \eqref{def-W}, rather than with a trace, as in the original construction of \cite{Drukker:2009hy}. However, the presence of this shift makes the perturbative calculation more intricate (see chapter 5 of \cite{Drukker:2019bev}), so we find it helpful to remove it before proceeding to the next section.

To illustrate the procedure we will take the latitude operator. The $\theta\rightarrow 0$ limit of the analysis below reproduces the circular case. We make a $U(N_1)$ gauge transformation in order to remove the constant shift from the first node,
\begin{equation}
     A_{\mu}\dot x^{\mu} + \frac{\cos\theta}{2} \quad\longrightarrow\quad A_{\mu}\dot x^{\mu} + \frac{d\Lambda}{d\tau} + \frac{\cos\theta}{2} = A_{\mu} \dot{x}^\mu + \text{boundary terms}\,,
\end{equation}
where boundary terms may arise from the discontinuity of $\Lambda$ on the circle. Precisely, we choose 
\begin{equation}
    \Lambda = -\frac{\cos\theta}{2} \tau + \Delta \sum_{n\in \mathbb{Z}}\theta(\tau-2\pi n)\,,
\end{equation}
where a constant $\Delta$ has been introduced, so to insure that $\int_0^{2\pi} d\tau  \frac{d\Lambda}{d\tau}$ vanishes. Requiring
\begin{equation}
    0=\int_0^{2\pi} d\tau \ \frac{d\Lambda}{d\tau}= \int_0^{2\pi}d\tau \left(-\frac{\cos\theta}{2}+ \Delta \,\delta(\tau-2\pi) \right) = -\pi\cos\theta + \Delta\,,
\end{equation}
we obtain $\Delta = \pi\cos\theta$. Therefore, the original gauge term $(A_{\mu}\dot x^{\mu} + \frac{\cos\theta}{2})$ in the superconnection is now replaced by $(A_{\mu}\dot x^{\mu} + \pi \cos{\theta} \, \delta(\tau - 2\pi))$.

Taking this delta function contribution into account, we recover the twist matrix $\cT$
\begin{equation}
    \cP \exp\left( -i\int_{2\pi-\epsilon}^{2\pi+\epsilon} \cL\, d\tau \right) \rightarrow \exp\left( -i\begin{pmatrix} \pi\cos\theta&0 \\ 0&0 \end{pmatrix} \right) = \begin{pmatrix} e^{-i\pi\cos\theta} & 0 \\ 0 & 1\end{pmatrix} \equiv \cT\,.
\end{equation}
In the circular case this is simply $\cT=\diag(-1,1)$. In the latitude case, in order to follow the same conventions of \cite{Bianchi:2014laa}, we rescale it such that
\begin{equation}
\label{eqn:twistmatrix}
    \cT \equiv \begin{pmatrix}
        e^{-i\pi (\cos\theta)/2} & 0\\ 0 & e^{i\pi(\cos\theta)/2}
    \end{pmatrix}\,.
\end{equation}

The gauge transformation we have performed in order to remove the constant shift acts on the matter fields as
\begin{equation}
\begin{split}
    & \bar\psi^I  \to  \bar\psi^I e^{-i\Lambda} = \bar\psi^I e^{i\cos\theta \,\tau/2}\,, \qquad\qquad \psi_I  \to  \psi_I e^{i\Lambda} = \psi_I e^{-i\cos\theta \,\tau/2}\,, \\
    & \bar C^I  \to  \bar C^I e^{-i\Lambda} = \bar C^I e^{i\cos\theta\,\tau/2}\,, \qquad\qquad C_I  \to  C_I e^{i\Lambda} = C_I e^{-i\cos\theta \,\tau/2}\,.
\end{split}
\end{equation}
The diagonal elements of the superconnection remain unchanged, while the fermionic entries gain extra phases. For the circular $1/24$ BPS operator these are
\begin{equation}\label{eq:ffbar}
\begin{split}
    \bar f &= e^{i\tau /2}\,\eta\left( \bar\alpha^1 \bar\psi^2 - \bar\alpha^2\bar\psi^1 \right) + e^{-i\tau /2}\xi \, (\beta^3 \bar\psi^4 - \beta^4\bar\psi^3)  \,,\\
     f &= e^{-i\tau /2}\,\xi \, \left( \alpha_1 \psi_2 - \alpha_2\psi_1 \right) + e^{i\tau /2}\eta \, (\bar\beta_3 \psi_4 - \bar\beta_4\psi_3) \,,
\end{split}
\end{equation}
while for the latitude $1/12$ BPS operator they are
\bal\label{eq:ffbar2}
\bar{f}^\theta &= -e^{i\cos\theta\,\tau/2}\,\eta\,\bar\alpha^2 \left(\cos\frac{\theta}{2}\bar\psi^1-\sin\frac{\theta}{2}e^{i\tau}\bar\psi^3\right),\\
f^\theta &= -e^{-i\cos\theta\,\tau/2}\,\xi\,\alpha_2 \left(\cos\frac{\theta}{2}\psi_1-\sin\frac{\theta}{2}e^{-i\tau}\psi_3\right)\,.
\eal
Therefore, the final form of the superconnection is
\begin{equation}\label{eq:superconnection}
   \cL = \begin{pmatrix} \cA & \bar f \\ f & \hat \cA \end{pmatrix} \,,
\end{equation}
without the constant shift in the first diagonal block, unlike \eqref{eqn:superconnection_bos}, and similarly for the case with $\theta\neq 0$.\footnote{We keep the same symbol ${\cal L}$ for this superconnection without the shift, hoping that it will not be confusing. From now on, ${\cal L}$ will refer to this expression.}

The Wilson loop operator is now written as
\begin{equation}
\label{eqn:cW}
    \cW 
    = \cR^{-1}\, 
    \sTr{\cal P}\left(e^{-i\oint {\cal L}d\tau} \cT\right) 
    \,,
\end{equation}
where we have introduced the normalization factor $\cR = \sTr(\cT)$. In particular, from now on, we will refer to the circular Wilson loop as
\begin{equation}
    \cW 
    = \frac{
   W
    }{N_1+N_2}, \qquad W\equiv  \text{Tr}\,{\cal P}\exp \left(-i \oint {\cal L}\,d\tau\right),
\end{equation}
with the ${\cal L}$ in \eqref{eq:superconnection}.
\section{Renormalization}
\label{sec:renormalization}

\subsection{1D effective field theory for the Wilson loop VEV}\label{sec:1dtheory}

At weak coupling, the standard procedure for computing the vacuum expectation value of a Wilson loop is ordinary perturbation theory. In the functional approach, this amounts to expanding the exponential of the interaction part of the bulk action in powers of the coupling constant and performing contractions with the Wilson loop expansion using Feynman rules for the bulk theory. 

In the QCD context, in the 80's Samuel \cite{Samuel:1978iy},  Gervais and Neveu \cite{Gervais:1979fv} proposed an alternative method to study Wilson loop operators, based on the formulation of a one-dimensional effective field theory. Subsequently, this method was further developed and heavily exploited to study the renormalization of composite operators \cite{Arefeva:1980zd,CRAIGIE1981204,Dorn:1986dt}.

The method makes use of auxiliary one-dimensional fermions and can be briefly summarized as follows. 
Suppose that in a given gauge theory one wants to evaluate a generic Wilson loop supported along a contour $\cC$,
\begin{equation}
    W[\cC] = \Tr \cP \exp\left( -i\int_\cC \cL\,d\tau \right)\,.
\end{equation}
In this expression $\cL$ may be the ordinary gauge connection $A$ or one of the bosonic connections $\cA$ given in \eqref{eqn:bosonicconnections}. In any case, one can write the perturbative expansion of the operator as 
\begin{equation}
\begin{split}
     W[{\cal C}] &=\Tr\left(1+\sum_{k=1}^{\infty}\frac{(-i)^k}{k!}\cP\int_{\cC}d\tau_1d\tau_2\dots d\tau_k\,\cL(\tau_1)\cL(\tau_2)\dots \cL(\tau_k)\right) \\ &=\Tr\left( 1+\sum_{k=1}^{\infty}(-i)^k\int_{\cC}d\tau_1d\tau_2\dots d\tau_k\, \theta(\tau_k-\tau_{k-1})\dots \theta(\tau_2-\tau_1)\cL(\tau_1)\cL(\tau_2)\dots\cL(\tau_k)\right)\,.
\end{split}
\end{equation}
The idea is to interpret $\theta(\tau_i-\tau_j)$ as the propagator of an  auxiliary fermionic field $z$ living on the Wilson loop contour, whose interaction with the rest of the fields is dictated by $\cL$. Taking the $z$ field in the fundamental representation of the gauge group, its action is chosen to be 
\begin{equation}
\label{Seff}
    S_\textrm{eff}=S + \int d\tau\, \bar z(\tau)\left[ \partial_{\tau} +i\cL \right]z(\tau)\,,
\end{equation}
where $S$ is the action of the underlying gauge theory. Performing the  Gaussian $z$-integral in the generating functional, it can be shown that for a contour $\cC_{12}$ connecting two points parametrized by $\tau_1, \tau_2$ one has \cite{CRAIGIE1981204,Dorn:1986dt}
\begin{equation}
    \langle W[\cC_{12}] \rangle =  \left\langle \Tr \cP \exp\left(-i \int_{\tau_2}^{\tau_1} \cL\, d\tau\right)\right\rangle = \langle z(\tau_2)\bar z(\tau_1) \rangle \,,
\end{equation}
where
\begin{equation}
 \langle z(\tau_2)\bar z(\tau_1) \rangle = \int [{\mathcal D} z {\mathcal D} \bar{z}] \, z(\tau_2)\bar z(\tau_1) \, e^{-S_\textrm{eff}}.
\end{equation}
Therefore, the expectation value of  $W$ is nothing but the two-point function of the one-dimensional theory defined on it. 

This method could be naturally generalized to study renormalization properties of Wilson loops in supersymmetric theories. 
Here, we propose a generalization that captures the expectation value of operators in the ABJ(M) theory.

Since in this theory Wilson loops are defined in terms of supermatrices, the natural way to proceed is to replace the one-dimensional auxiliary $z$ fermion with a Grassmann odd supermatrix
\begin{equation}\label{eq:oddmatrix}
    \Psi = \begin{pmatrix} z & \varphi \\ \tilde \varphi& \tilde z\end{pmatrix}\,, \qquad \bar\Psi = \begin{pmatrix} \bar z & \bar{\tilde\varphi} \\ \bar{\varphi}& \bar{\tilde z}\end{pmatrix}\,,
\end{equation}
where $z$ ($\tilde z$) and $\varphi$ ($\tilde \varphi$) are a spinor and a scalar, respectively, in the fundamental representation of $U(N_1)$ ($U(N_2)$). We then look for an effective theory such that the ABJ(M) Wilson loop VEV \eqref{eqn:cW} can be computed as a two-point function of the one-dimensional fields. 
To this end, we consider the following action
\begin{equation}\label{eq:effaction}
    S_\textrm{eff}= S_\textrm{ABJ(M)} +\int d\tau\,
\Tr\left( \bar\Psi \cD_{\tau}\Psi \right)\,,
\end{equation}
where $S_\textrm{ABJ(M)}$ is the ABJ(M) action (see \eqref{eq:ABJMaction}) and $\cD_{\tau} = \partial_{\tau} + i\cL(\tau)$, $\cL$ being the Wilson loop superconnection. It is then easy to prove that 
\begin{equation}\label{eq:2pt}
    \langle W[\cC_{12}] \rangle = \, \langle \Tr \Psi(\tau_2) \bar\Psi(\tau_1) \rangle\,,
\end{equation}
where the vacuum functional on the right-hand side includes the integrations over both the bulk fields and the one-dimensional $\Psi, \bar\Psi$ supermatrices, weighted by the action \eqref{eq:effaction}.\footnote{Here it is sufficient to assume that a consistent definition of integration over supermatrices exists, which leads to well-defined, finite and non-vanishing results for Gaussian integrals.}  We provide more details about this derivation in appendix \ref{sec:apxDorn}.

We focus here on the circular Wilson loop defined in section \ref{sec:parametricWL}, while postponing the investigation of the latitude operator of section \ref{sec:parametriclatitude} to a future publication \cite{CPTT}. Expanding the matrix product and defining for simplicity $g=\sqrt{\frac{2\pi}{k}}$, the effective action can be written explicitly as
\begin{equation}
\hspace{-0.5cm}  
\label{eqn:effectiveaction}
\begin{split}
   S_\textrm{eff} = S_\textrm{ABJ(M)} +\int d\tau \Big[& \bar\varphi D_{\tau}\varphi + \bar{\tilde\varphi} \hat D_{\tau} \tilde \varphi  + \bar z D_{\tau} z + \bar{\tilde z} \hat D_{\tau} \tilde z  \\ & + i(\bar{\tilde z} f \varphi + \bar\varphi \bar f \tilde z + \bar{\tilde\varphi} f z+\bar z\bar f \tilde \varphi)\Big]\,,
\end{split}
\end{equation}
where we have defined $D_{\tau}=\partial_{\tau} +i\cA$ and $\hat D_{\tau}=\partial_{\tau}+i\hat \cA$. $\cA$, $\hat{\cA}$, $f$ and $\bar f$ are the even and odd elements of the Wilson loop superconnection, see \eqref{eq:superconnection}. The covariant $\tau$-derivatives give rise to the usual minimal coupling between the one-dimensional fields and the bulk gauge vectors, plus quartic interactions with bulk scalar bilinears. We have not inserted the explicit expressions of $f, \bar{f}$, which can be found in \eqref{eq:ffbar2}. At this stage, it is only important to take into account that these couplings are proportional to one power of $g$. The tree-level propagators of the one-dimensional fields are
\begin{equation}
\begin{alignedat}{3}
    & \includegraphics[width=0.2\textwidth]{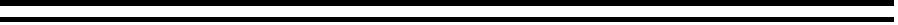} &&= \langle z^i(\tau_1)\bar z_j(\tau_2) \rangle &&= \delta_j^i \, \theta(\tau_1-\tau_2)\,, \\
    & \includegraphics[width=0.2\textwidth]{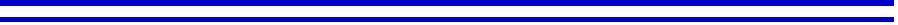} &&= \langle {\tilde z}^{\hat i}(\tau_1)\bar{\tilde z}_{\hat j}(\tau_2) \rangle &&= \delta_{\hat j}^{\hat i} \,  \theta(\tau_1-\tau_2)\,, \\
    & \includegraphics[width=0.2\textwidth]{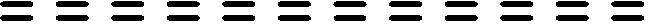} &&= \langle \varphi^i(\tau_1)\bar{\varphi}_j(\tau_2) \rangle &&= \delta_j^i \,  \theta(\tau_1-\tau_2)\,, \\
    & \includegraphics[width=0.2\textwidth]{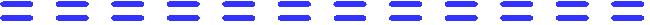} &&= \langle \tilde \varphi^{\hat i}(\tau_1)\bar{\tilde \varphi}_{\hat j}(\tau_2) \rangle &&= \delta_{\hat j}^{\hat i} \,  \theta(\tau_1-\tau_2)\,.
\end{alignedat}
\end{equation}

\vspace{0.2cm}


\subsection{Renormalization scheme}

We now focus on the perturbative evaluation of the two-point function \eqref{eq:2pt} for the one-dimensional theory. This first requires investigating whether the one-dimensional fields and the couplings  undergo a non-trivial renormalization, due to short distance divergences arising on the loop. 

For each one-dimensional field $\phi=\{\varphi,\tilde{\varphi},z,\tilde{z}\}$ the corresponding renormalization functions are defined as $\phi=Z_\phi^{-\frac{1}{2}}\phi_0$,
where $\phi_0$ stands for the bare quantity. We note that since the action \eqref{eqn:effectiveaction} is invariant under the formal exchanges $z \leftrightarrow \varphi$ and $\tilde{z} \leftrightarrow \tilde\varphi$, we can set $Z_z = Z_\varphi$ and $Z_{\tilde{z}} = Z_{\tilde\varphi}$. 
As we are going to prove, the field function renormalization  is sufficient to cancel UV divergent contributions to both the kinetic terms and the interaction vertices between auxiliary fields and gauge connections, {\it i.e.} the $\bar\phi A_\mu \dot{x}^\mu \phi$ vertices. This is consistent with the expectation that the addition of the auxiliary action in \eqref{eq:effaction} does not affect the  UV finiteness of the ABJ(M) theory ($A_\mu$ does not renormalize).

As follows from the definition of $f, \bar{f}$ and $M_I^{\ J}$ in (\ref{eq:ffbar}) and \eqref{eqn:M124bos}, respectively, the fermionic interactions (as for instance $\bar{\tilde z}f \varphi$) and the quartic couplings with the scalar bilinears contain the $g$ coupling and the $\alpha_i, \bar\alpha^i, \beta^j, \bar\beta_j$ parameters as further couplings. 

For the renormalization of the fermionic interactions we define
\begin{equation}
\begin{split}\label{eq:parren}
   & (\bar\alpha^i)_0 \, Z^{1/2}_{\tilde z}Z_{\varphi}^{1/2} =(\bar\alpha^i)_0 \, Z^{1/2}_{z}Z_{\tilde\varphi}^{1/2}= Z_{\bar\alpha^i} \, \bar\alpha^i\,,\\
   & (\alpha_i)_0 \,  Z^{1/2}_{\tilde z}Z_{\varphi}^{1/2} =(\alpha_i)_0 \, Z^{1/2}_{z}Z_{\tilde\varphi}^{1/2}= Z_{\alpha_i} \, \alpha_i\,,\\
   & (\beta^j)_0 \,  Z^{1/2}_{\tilde z}Z_{\varphi}^{1/2} =(\beta^j)_0 \,  Z^{1/2}_{z}Z_{\tilde\varphi}^{1/2}= Z_{\beta^j} \, \beta^j\,,\\
   & (\bar\beta_j)_0 \, Z^{1/2}_{\tilde z}Z_{\varphi}^{1/2} = (\bar\beta_j)_0 \,  Z^{1/2}_{z}Z_{\tilde\varphi}^{1/2}= Z_{\bar\beta_j} \, \bar\beta_j\,,
\end{split}
\end{equation}
where the subscript 0 denotes bare parameters and we have used that the ABJ(M) coupling does not renormalize, {\it i.e.} $g_0=g$.
The scalar vertices of the form $M_I^{\ J}\bar\phi C_J\bar C^I\phi$ deserve more attention since the parametric dependence is hidden inside the scalar coupling matrix $M_I^{\ J}$. We set 
\begin{equation}
\begin{alignedat}{2}
\label{eq:Mren}
    &Z_{\varphi}({M}_I^{\ J})_0 =Z_{\varphi C} M_I^{\ J}\,, \qquad &&Z_{\tilde\varphi}({M}_I^{\ J})_0=Z_{\tilde\varphi C} M_I^{\ J}\,,\\ &Z_{z}({M}_I^{\ J})_0=Z_{z C} M_I^{\ J}\,, \qquad &&Z_{\tilde z}\,({M}_I^{\ J})_0=Z_{\tilde z C}M_I^{\ J}\,,
\end{alignedat}
\end{equation}
where $({M}_I^{\ J})_0$ is the scalar coupling matrix expressed in terms of the bare parameters.

Using the standard BPHZ renormalization procedure, we write all renormalization functions as $Z = 1 + \delta$, where $\delta$ are the corresponding countertems. We then extract the Feynman rules from the one-dimensional Lagrangian written as the sum of a renormalized Lagrangian plus the counterterm part
\begin{equation}\label{eq:Ltot}
    \cL_\textrm{1D}=\cL^\textrm{ren}_\textrm{1D} + \cL_\textrm{1D}^{\text{ct}}\,,
\end{equation}
where $\cL^\textrm{ren}_\textrm{1D}$ is given by \eqref{eqn:effectiveaction} written in terms of renormalized quantities and the counterterms read
\begin{equation}\label{eq:counterterms}
\begin{split}
     \cL^{\text{ct}}_\textrm{1D} &= \sum_{\phi=\varphi,z}\left(\delta_{\phi}\, \bar\phi \partial_{\tau}\phi + \delta_{\phi}\, ig \bar\phi A_{\mu}\dot x^{\mu}\phi + \delta_{\phi C} \, g^2|\dot x|  M_I^{\ J} \bar \phi C_J \bar C^I \phi \right)\\ & + \sum_{\tilde\phi=\tilde\varphi,\tilde z}\left(\delta_{\tilde\phi} \, \bar{\tilde\phi} \partial_{\tau}\tilde\phi + \delta_{\tilde\phi} \, ig \bar{\tilde\phi} \hat A_{\mu}\dot x^{\mu}\tilde\phi + \delta_{\tilde\phi C} \, g^2|\dot x| M_I^{\ J}  \bar{\tilde\phi}  \bar C^I C_J \tilde\phi \right)\\ &+ i\bar{\tilde z}  \left( e^{-\frac{i\tau}{2}}\xi\left( \delta_{\alpha_1} \, \alpha_1 \psi^2 - \delta_{\alpha_2} \, \alpha_2\psi^1 \right) + e^{\frac{i\tau}{2}}\eta \, (\delta_{\bar\beta_3} \, \bar\beta_3 \psi^4 - \delta_{\bar\beta_4} \, \bar\beta_4\psi^3) \right) \varphi\\ &+ i\bar{\tilde\varphi} \left( e^{-\frac{i\tau}{2}}\xi\left( \delta_{\alpha_1} \, \alpha_1 \psi^2 - \delta_{\alpha_2} \, \alpha_2\psi^1 \right) + e^{\frac{i\tau}{2}}\eta \, (\delta_{\bar\beta_3} \, \bar\beta_3 \psi^4 - \delta_{\bar\beta_4} \, \bar\beta_4\psi^3) \right) z\\ &+ i\bar\varphi  \left(e^{\frac{i\tau}{2}}\eta  \left( \delta_{\bar\alpha^1} \, \bar\alpha^1 \bar\psi^2 - \delta_{\bar\alpha^2} \, \bar\alpha^2\bar\psi^1 \right) + e^{-\frac{i\tau}{2}}\xi \, (\delta_{\beta^3} \, \beta^3 \bar\psi^4 - \delta_{\beta^4} \, \beta^4\bar\psi^3) \right) \tilde z\\ &+ i\bar z  \left( e^{\frac{i\tau}{2}}\eta\left( \delta_{\bar\alpha^1} \, \bar\alpha^1 \bar\psi^2 - \delta_{\bar\alpha^2} \, \bar\alpha^2\bar\psi^1 \right) + e^{-\frac{i\tau}{2}}\xi \, (\delta_{\beta^3}\beta^3 \bar\psi^4 - \delta_{\beta^4} \, \beta^4\bar\psi^3) \right) \tilde\varphi\,,
\end{split}
\end{equation}
with obvious meanings of the $\delta$'s.

\vspace{0.2cm}


\subsection{Evaluation of one-loop counterterms}
\label{sec:counterterms}

We begin by investigating the structure of the counterterms at one loop. We tame short distance divergences arising from the evaluation of Feynman integrals by using dimensional regularization in $D = 1-2\epsilon$ and a minimal subtraction scheme. We work in the large $N_1, N_2$ limit.

Since we want to study the UV behavior of our one-dimensional theory, we work in the $\tau_2\to\tau_1$ limit, where $\tau$ parameterizes the curve on which the theory is defined. Therefore, any regular contour can be approximated, around a point, by a straight segment, such that $|\dot x|=1$ and $\dot x \cdot \ddot x=0 $. In this limit, for a generic one-dimensional field $\phi$ we use the following approximation\footnote{We use the notation $\phi(\tau_i)\equiv\phi_i$ and $x_i \equiv x(\tau_i)$.} 
\begin{equation}\label{eq:fieldexp}
    \phi_2 \simeq \phi_1 + (\tau_2 - \tau_1) \dot{\phi}_1\,,
\end{equation}
as well as the following expansion for the coordinates on the contour
\begin{equation}\label{eq:coordexp}
\begin{split}
    & x^{\mu}_2 \simeq x^{\mu}_1 +(\tau_2-\tau_1)\dot x^{\mu}_1\,,\qquad
    \dot x^{\mu}_2 \simeq \dot x^{\mu}_1 +(\tau_2-\tau_1)\ddot x^{\mu}_1\,, \\
    &\left( x_2-x_1 \right)^2\simeq(\tau_2-\tau_1)^2 \, \dot{x}_1^2 \, \left( 1+(\tau_2-\tau_1)\frac{\dot{x}_1 \cdot \ddot{x}_1}{\dot{x}_1^2} \right)\,,
\end{split}
\end{equation}

To keep the discussion as clear as possible, we provide here details for the first few diagrams and collect the rest of the calculations in appendix \ref{app:renormalization}.

\vspace{0.2cm}
\subsubsection*{Corrections to the kinetic term}

\begin{figure}[]
    \centering
    \subfigure[]{
    \includegraphics[width=0.25\textwidth]{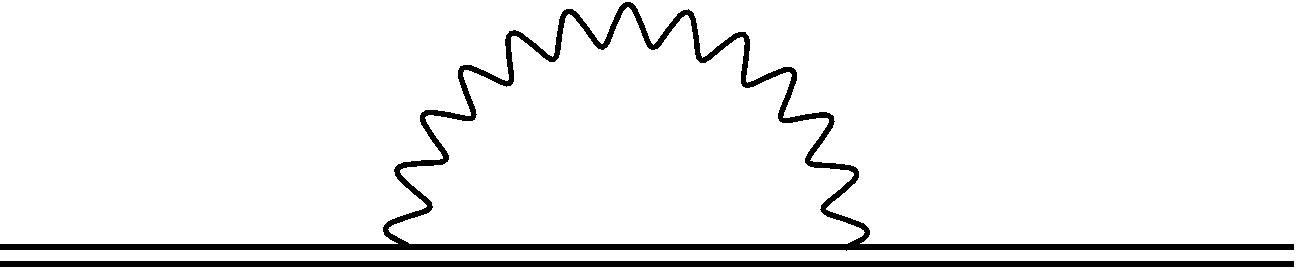}
    \label{subfig:vertexcorrectiona}} \qquad
    \subfigure[]{
    \includegraphics[width=0.25\textwidth]{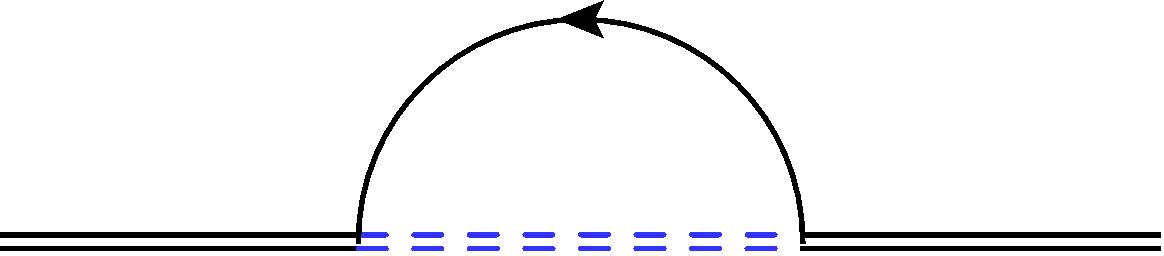}
    \label{subfig:vertexcorrectionb}}  \qquad
    \subfigure[]{
    \includegraphics[width=0.25\textwidth]{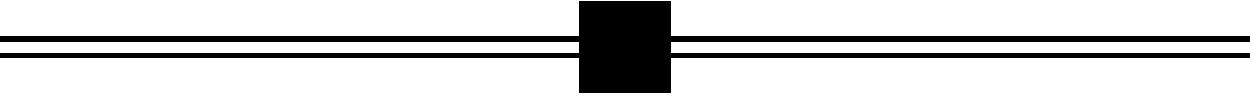}
    \label{subfig:vertexcorrectiond}} 
    \caption{One-loop corrections to the fermionic propagator $\langle\bar z z\rangle$. Double straight lines represent $z$ and $\bar{z}$, blue double dashed lines are the one-dimensional $\tilde\varphi$ scalars, simple straight lines are ABJ(M) fermions, whereas wavy lines describe the ABJ(M) $U(N_1)$ gauge field. Diagram (c) is the $\delta_z$ counterterm in \eqref{eq:counterterms}.}
    \label{fig:kinetic1}
\end{figure}

We begin by considering one-loop self-energy corrections to the $\langle\bar z z\rangle$ propagator of the one-dimensional theory. The contributing diagrams are drawn in figure \ref{fig:kinetic1} (we neglect tadpole diagrams, as they vanish in dimensional regularization).

The first diagram is the gauge field correction and gives rise to the following contribution
\begin{equation}
    \Sigma^{\rm \subref{subfig:vertexcorrectiona}}_z = \int d\tau_1 \int d\tau_2 \,\bar z_1\, z_2 \,\dot x_1^{\mu}\, \dot x_2^{\nu} \, \theta(\tau_1-\tau_2)\langle A_{\mu}(\tau_1) A_{\nu}(\tau_2) \rangle \,.
\end{equation}
However, inserting the explicit expression \eqref{eqn:propagator} for the gauge propagator and using the expansions \eqref{eq:coordexp}, it is easy to see that in dimensional regularization this integral vanishes, due to the antisymmetry of the $\epsilon_{\mu\nu \rho}$ tensor. 

The second diagram gives (we define $\tau_{12}\equiv \tau_1-\tau_2$) 
\begin{equation}
\begin{split}
    \Sigma^{\rm \subref{subfig:vertexcorrectionb}}_z &= \int d\tau_1 \int d\tau_2 \left(i \bar z \bar f \tilde \varphi \right)(\tau_1)\left(i \bar{\tilde\varphi} f z \right)(\tau_2) \\
    &=\bar\alpha^i\alpha_i \int d\tau_1 \int^{\tau_1} d\tau_2 \,\bar z_1\, z_2\,  e^{\frac{i\tau_{12}}{2}}\eta_{\alpha}(\tau_1)\xi^{\beta}(\tau_2)\langle \bar\psi^{\alpha}(\tau_1)\psi_{\beta}(\tau_2) \rangle\\
    & +\beta^j\bar\beta_j \int d\tau_1 \int^{\tau_1} d\tau_2\, \bar z_1\, z_2\,  e^{-\frac{i\tau_{12}}{2}}\xi_{\alpha}(\tau_1)\eta^{\beta}(\tau_2)\langle \bar\psi^{\alpha}(\tau_1)\psi_{\beta}(\tau_2) \rangle\,,
\end{split}
\end{equation}
where we have already used $\langle \tilde\varphi_1 \bar{\tilde \varphi}_2 \rangle = \theta(\tau_1 - \tau_2)$. Inserting the fermionic propagator \eqref{eqn:propagator}, it explicitly reads
\begin{equation}\label{eq:5b}
\begin{split}
    \Sigma^{\rm \subref{subfig:vertexcorrectionb}}_z 
    & =i \bar\alpha^i\alpha_i \frac{\Gamma(\frac32 - \epsilon)}{2\pi^{\frac32 - \epsilon}}
    \int d\tau_1 \int^{\tau_1} d\tau_2 \,\bar z_1\, z_2\, e^{i\frac{\tau_{12}}{2}}\xi^{\alpha}(\tau_2)(\gamma_{\mu})_{\alpha}^{\ \beta}\eta_{\beta}(\tau_1) \frac{(x_2-x_1)^{\mu}}{|x_2 - x_1|^{3-2\epsilon}} \\
    & +i\beta^j\bar\beta_j \frac{\Gamma(\frac32 - \epsilon)}{2\pi^{\frac32 - \epsilon}} \int d\tau_1 \int^{\tau_1} d\tau_2\, \bar z_1\, z_2\,  e^{-i\frac{\tau_{12}}{2}}\eta^{\alpha}(\tau_2)(\gamma_{\mu})_{\alpha}^{\ \beta}\xi_{\beta}(\tau_1)\frac{(x_2-x_1)^{\mu}}{|x_2 - x_1|^{3-2\epsilon}}  \,.
\end{split}
\end{equation}
In the $\tau_2\to \tau_1$ limit, using the explicit expression for the $\xi, \eta$ spinors and the gamma matrices in \eqref{eq:gamma}, we can write 
\begin{equation}\label{eq:xietaid}
    \begin{split}
    e^{i\frac{\tau_{12}}{2}}\xi^{\alpha}(\tau_2)\, (\gamma_{\mu})_{\alpha}^{\ \beta}\, \eta_{\beta}(\tau_1)\, (x_2-x_1)^{\mu} &= -4i g^2  \sin\frac{\tau_{12}}{2}\, \simeq \, -2i g^2 (\tau_1-\tau_2)\,,\\ e^{-i\frac{\tau_{12}}{2}}\eta^{\alpha}(\tau_2)\, (\gamma_{\mu})_{\alpha}^{\ \beta} \, \xi_{\beta}(\tau_1)\, (x_2-x_1)^{\mu} &= -4i g^2 \sin\frac{\tau_{12}}{2} \, \simeq \, -2i g^2(\tau_1-\tau_2) \,.
    \end{split}
\end{equation}
Expanding the rest of the integrand with \eqref{eq:fieldexp}, \eqref{eq:coordexp}, the integral reduces to
\begin{equation}
\begin{split}
\Sigma^{\rm \subref{subfig:vertexcorrectionb}}_z & = -g^2 N_2 \, (\bar\alpha^i\alpha_i + \beta^j\bar\beta_j)\frac{\Gamma(\frac{3}{2}-\epsilon)}{\pi^{\frac{3}{2}-\epsilon}}\int d\tau_1  \bar z_1\, \dot z_1\int^{\tau_1} d\tau_2\,  (\tau_1-\tau_2)^{-1+2\epsilon} + \cdots \\
 & = -g^2\frac{N_2}{4\pi\epsilon}(\bar\alpha^i\alpha_i +  \beta^j\bar\beta_j)\int d\tau \, \bar z\, \partial_{\tau}z + \text{finite terms}\,, 
\end{split} 
\end{equation}
where in the first line dots indicate terms of the expansion which give rise to finite integrals, and in the second line we have extracted the divergent part for $\epsilon \to 0$.

Finally, the counterterm contribution is 
\begin{equation}
    \Sigma^{\rm \subref{subfig:vertexcorrectiond}}_z = -\delta_z \int d\tau\, \bar z \partial_{\tau} z\,.
\end{equation}
Therefore the total correction to the $z,\bar z$ propagator $\Sigma_z$, given by the sum of all diagrams above, is
\begin{equation}
    \Sigma_z = \left( -g^2\frac{N_2}{4\pi\epsilon}(\bar\alpha^i\alpha_i + \beta^j\bar\beta_j)-\delta_z \right)\int d\tau \bar z \partial_{\tau} z\,.
\end{equation}
Requiring the counterterm to cancel the divergence, we eventually find
\begin{equation}
\label{eqn:zren}
    Z_z= Z_\varphi 
    =1-g^2\frac{N_2}{4\pi\epsilon}(\bar\alpha^i\alpha_i + \beta^j\bar\beta_j)\,.
\end{equation}
The same procedure can be applied to the tilde fields, obtaining  similar contributions
\beq \label{eq:ztilderen}
 Z_{\tilde z} = Z_{\tilde\varphi}
 =1-g^2\frac{N_1}{4\pi\epsilon}(\bar\alpha^i\alpha_i + \beta^j\bar\beta_j)\,.
\eeq

\vspace{0.2cm}


\subsubsection*{Corrections to the gauge-fermion vertex}

\begin{figure}[]
    \centering
    \subfigure[]{
    \includegraphics[width=0.25\textwidth]{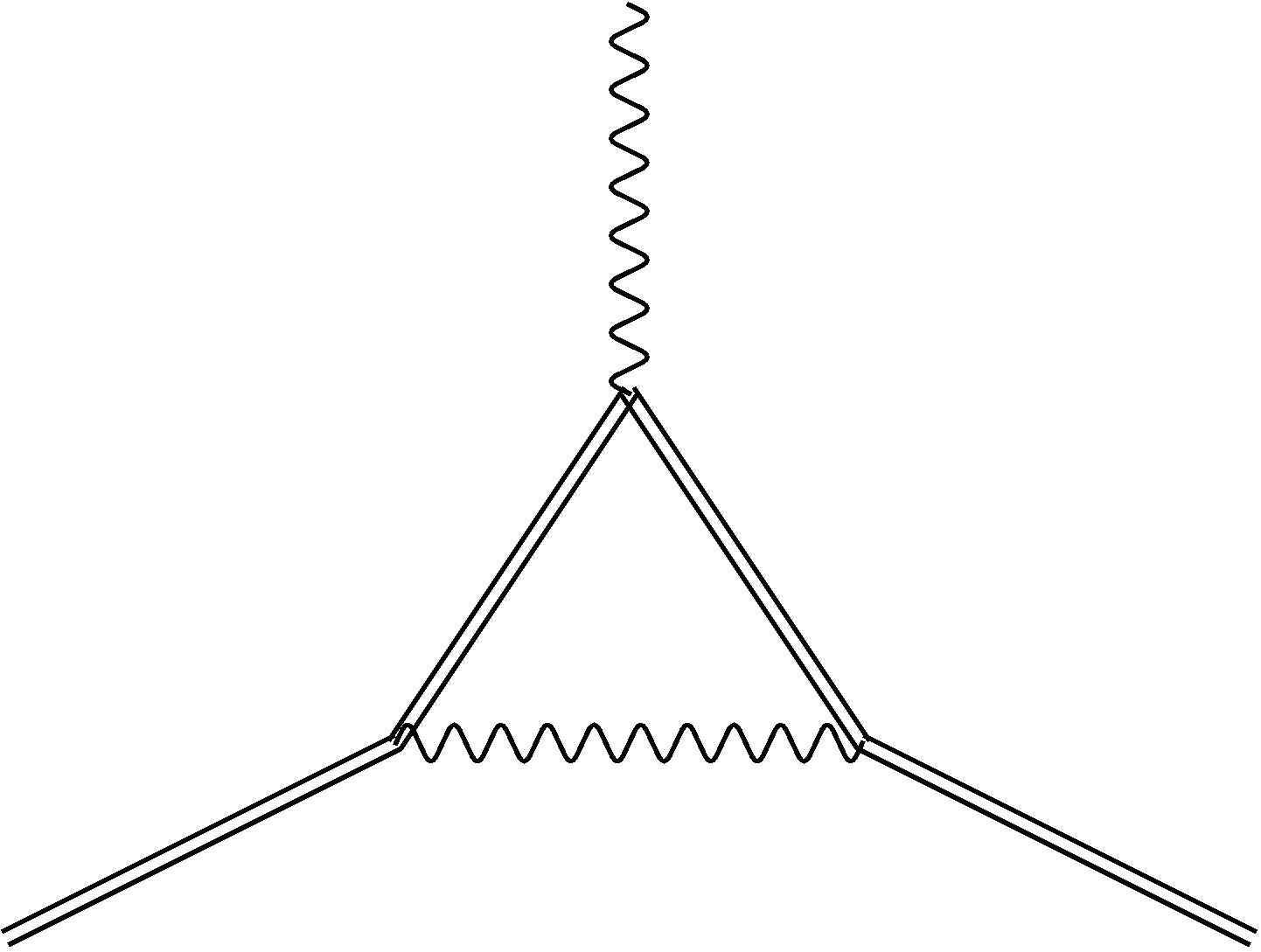}
    \label{subfig:gauge1a}} \qquad
    \subfigure[]{
    \includegraphics[width=0.25\textwidth]{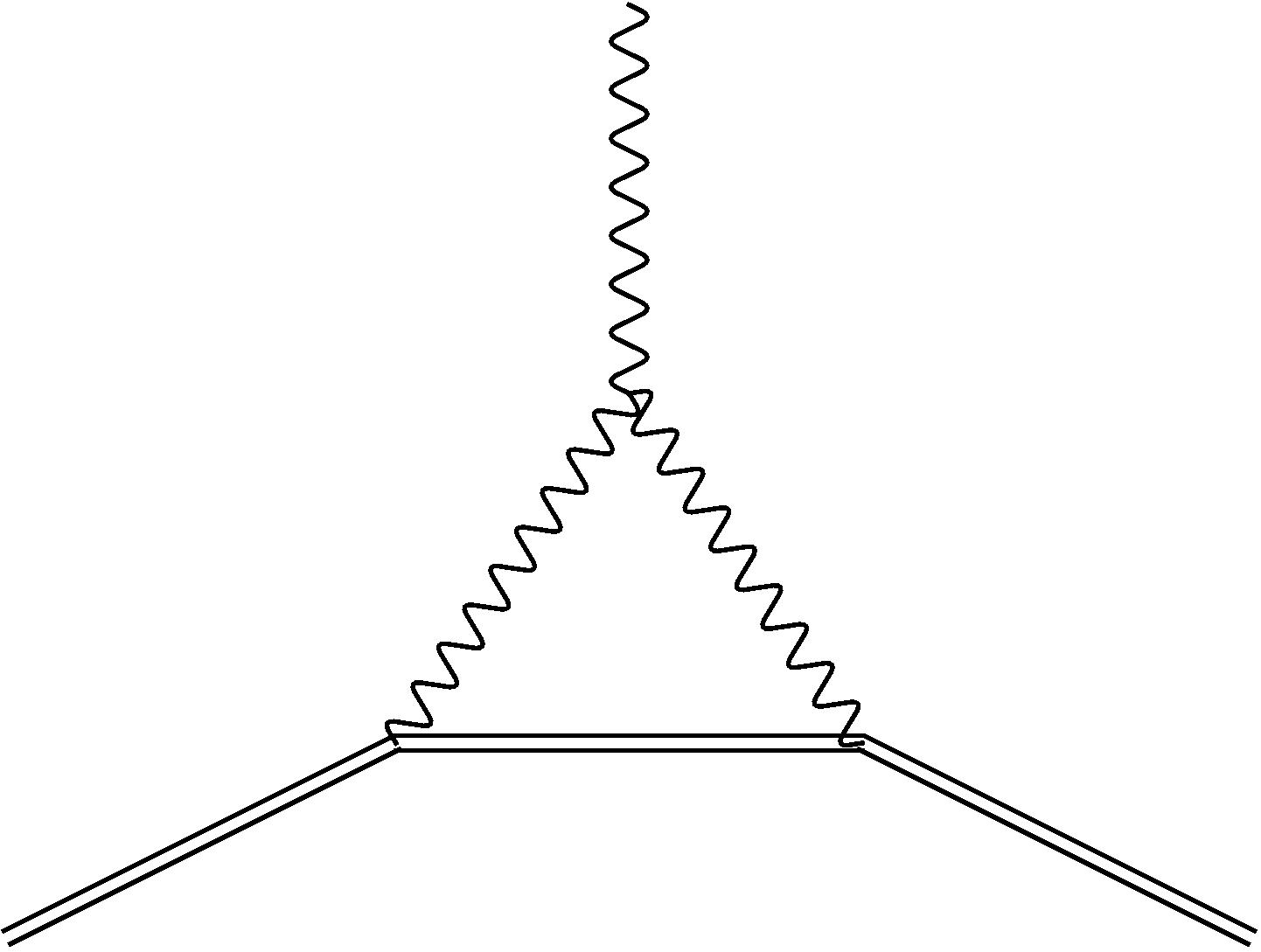}
    \label{subfig:gauge1b}}
    \qquad
    \subfigure[]{
    \includegraphics[width=0.25\textwidth]{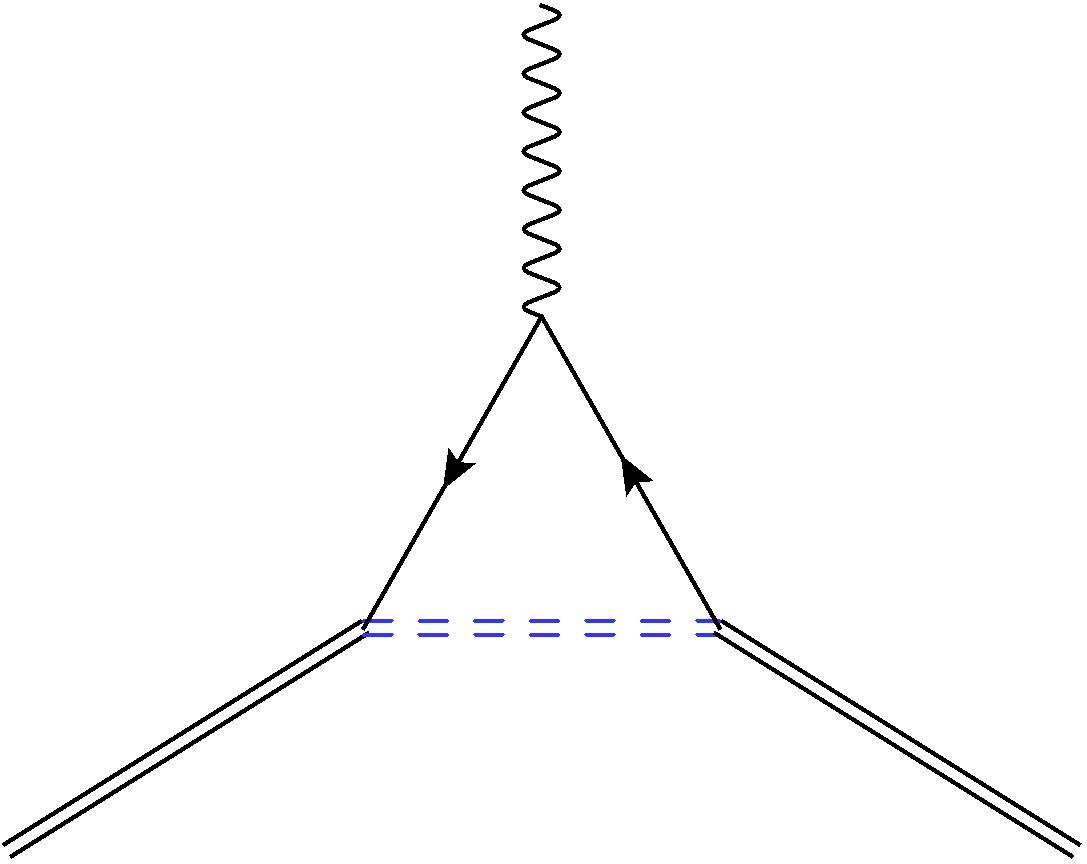}
    \label{subfig:gauge1c}} 
    
    \subfigure[]{
    \includegraphics[width=0.20\textwidth]{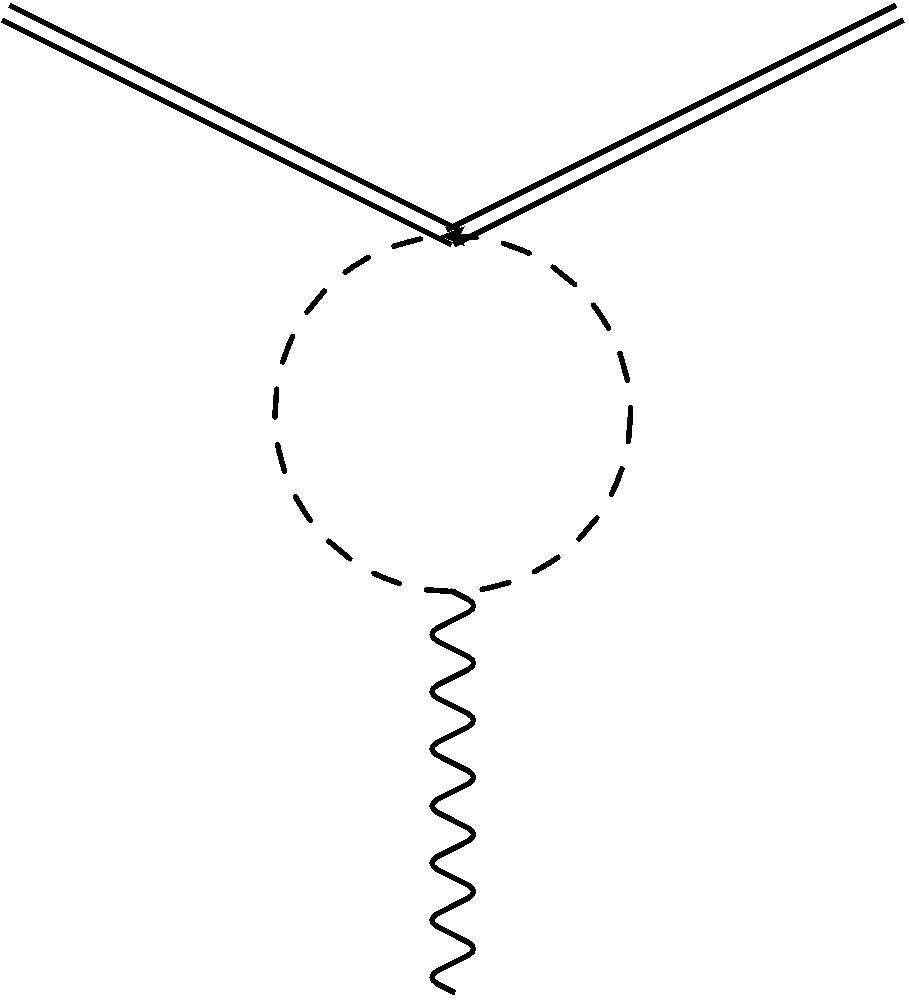}
    \label{subfig:gauge1d}} \qquad\qquad
    \subfigure[]{
    \includegraphics[width=0.20\textwidth]{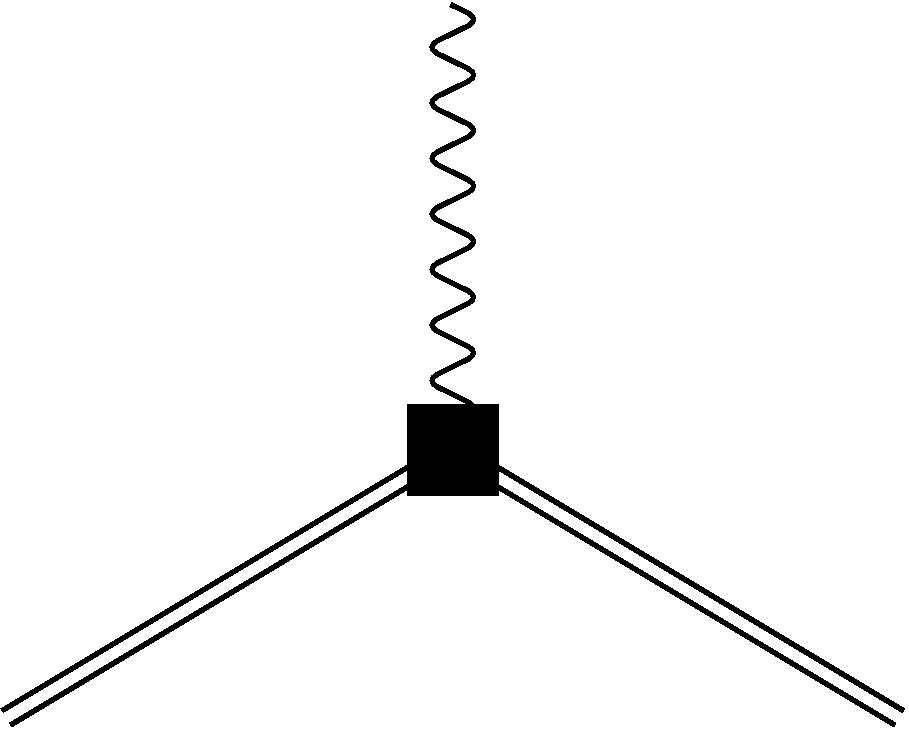}
    \label{subfig:gauge1e}} 
    \caption{One-loop corrections of the $V_{\bar z A z}$ vertex. The blue double-dashed line describes the $\tilde{\varphi}$ propagator, while the single-dashed line describes the ABJM scalar fields $C_I$. The last diagram represents the $\delta_z$ counterterm. }
    \label{fig:gauge1}
\end{figure}

We now consider one-loop corrections to the gauge-fermion vertex $S_{\bar z A z}=i\int d\tau\,\bar z A_{\mu} \dot{x}^{\mu} z$. The corresponding diagrams are summarized in figure \ref{fig:gauge1}. In the following we simply list the final result of each integral, referring to appendix \ref{sec:gamma} for the details of the computation.

To begin with, it is easy to see that diagram \ref{subfig:gauge1a} does not contribute, due to planarity. In fact, using the expansions \eqref{eq:coordexp}, the associated divergence turns out to be proportional to $\epsilon_{\mu \nu \rho} \dot{x}_1^\mu \dot{x}_1^\nu$.

For the same reason, as discussed in appendix \ref{sec:gamma}, the divergent contribution of diagram \ref{subfig:gauge1b} also vanishes. This diagram contains the three gauge field vertex coming from the ABJ(M) action, $\frac{1}{3g^2}\epsilon^{\mu\nu\rho}\int d^d x\,\Tr (A_{\mu} A_{\nu} A_{\rho})$ (here $d=3-2\epsilon$). Therefore, it is  proportional to the product of three epsilon tensors, one from the vertex and two from the gauge propagators. Using ordinary epsilon tensor algebra, this product can be reduced to a single epsilon, but eventually the remaining tensor is contracted with the same vector twice. 

Diagram \ref{subfig:gauge1c} is built using the gauge-fermion-fermion vertex from the original ABJ(M) action $-\int d^dx\Tr(g\bar\psi^J\gamma^{\mu}A_{\mu}\psi_J)$. Its divergent contribution reads 
\begin{equation}
\label{eqn:vertex2}
    \Gamma^{\rm \subref{subfig:gauge1c}}_{\text{gauge}} = -\frac{g^2N_2}{4\pi \epsilon} (\bar\alpha^i\alpha_i+\beta^j\bar\beta_j)\int d\tau \ i\bar z A_{\mu} \dot x^{\mu} z\,.
\end{equation}

Finally, diagram \ref{subfig:gauge1d} contains the gauge-scalar vertex coming from minimal coupling in the ABJ(M) action, $i \int d^d x\, (A^\mu C_I \partial_{\mu}\bar{C}^I - \partial_\mu  C_I \, \bar{C}^I A^\mu)$. In Lorentz gauge this diagram turns out to be equal to zero, as shown in appendix \ref{sec:gamma}. 

Summing all the contributions, the correction to the gauge vertex $\Gamma_\text{gauge}$ is eventually given by
\begin{equation}
    \Gamma_\text{gauge} = \left( -\delta_{z} - \frac{g^2 N_2}{4\pi\epsilon}(\bar\alpha^i\alpha_i + \beta^j\bar\beta_j) \right)  \int d\tau\, i \bar z A_{\mu} \dot x^{\mu} z\,.
\end{equation}
Comparing with \eqref{eqn:zren}, we see that $\delta_z$ cancels exactly  the divergence. 

Following the same procedure for the $\bar{\tilde z} \hat A_{\mu} \tilde{z}$ vertex, we find that the result changes only by a color factor. Precisely, we obtain 
\begin{equation}
    {\hat\Gamma}_\text{gauge} = \left( -\delta_{\tilde z} - \frac{g^2 N_1}{4\pi\epsilon}(\bar\alpha^i\alpha_i + \beta^j\bar\beta_j) \right)  \int d\tau\, i \bar{\tilde z} \hat{A}_{\mu} \dot x^{\mu} {\tilde z}\,,
\end{equation}
and $\delta_{\tilde z}$ in \eqref{eq:ztilderen} cancels exactly this vertex divergence. 

The same pattern holds also for the remaining gauge-boson vertices, {\it i.e.} $\bar{\varphi} A_\mu \dot{x}^\mu \varphi$ and $\bar{\tilde \varphi} A_\mu \dot{x}^\mu {\tilde \varphi}$.

\vspace{0.2cm}


\subsubsection*{Fermion vertex corrections}

To compute the counterterm associated with the fermion vertex correction (last four lines in \eqref{eq:counterterms}), we first consider the coupling $i\bar{\tilde z} f \varphi$. Inserting the explicit expression \eqref{eq:ffbar} for $f$, this amounts to evaluating four different vertex structures, precisely
\begin{equation}\label{eq:fermionvertices}
i \alpha_1 e^{-i\tau/2} \xi \, \bar{\tilde z} \psi^2 \varphi - i \alpha_2 e^{-i\tau/2} \xi \, \bar{\tilde z} \psi^1 \varphi + i \bar{\beta}_3 e^{i\tau/2} \eta \, \bar{\tilde z} \psi^4 \varphi - i \bar{\beta}_4 e^{i\tau/2} \eta \,\bar{\tilde z} \psi^3 \varphi \, ,
\end{equation}
with $\xi, \eta$ given in \eqref{eq:spinors}.

\begin{figure}[H]
    \centering
    \subfigure[]{
    \includegraphics[width=0.25\textwidth]{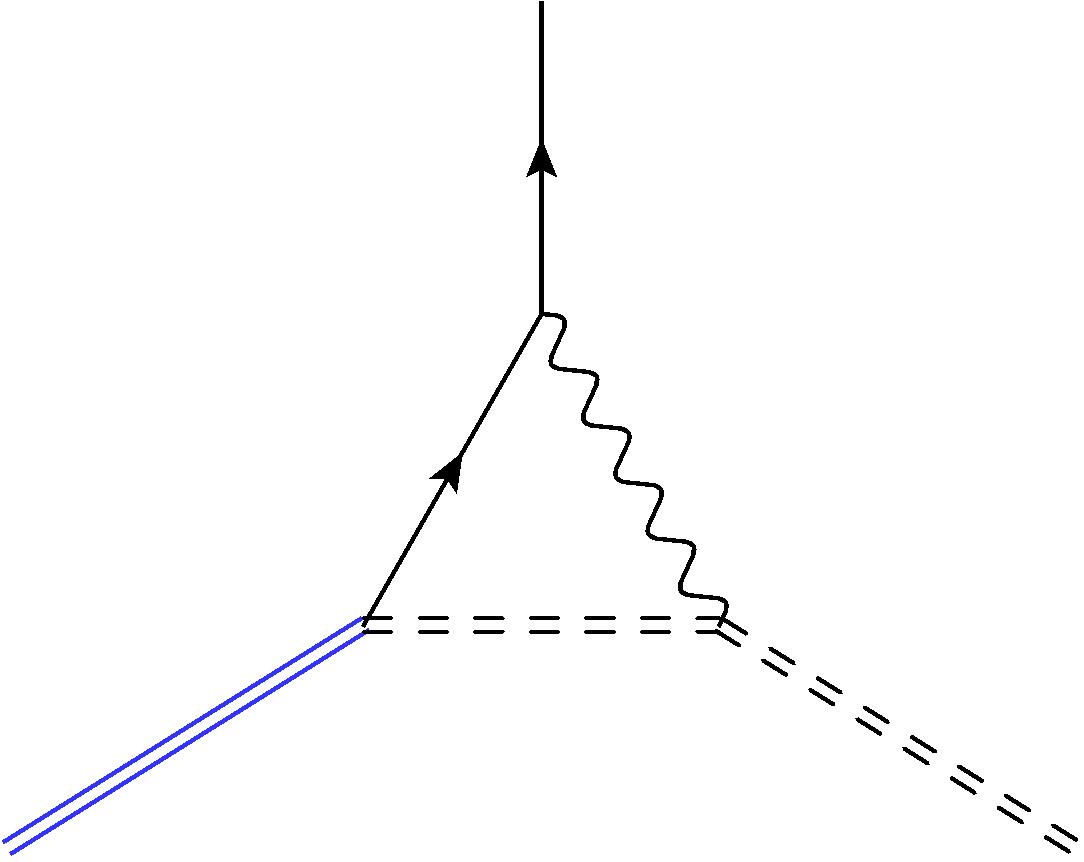}
    \label{subfig:fermion1a}} \qquad
    \subfigure[]{
    \includegraphics[width=0.25\textwidth]{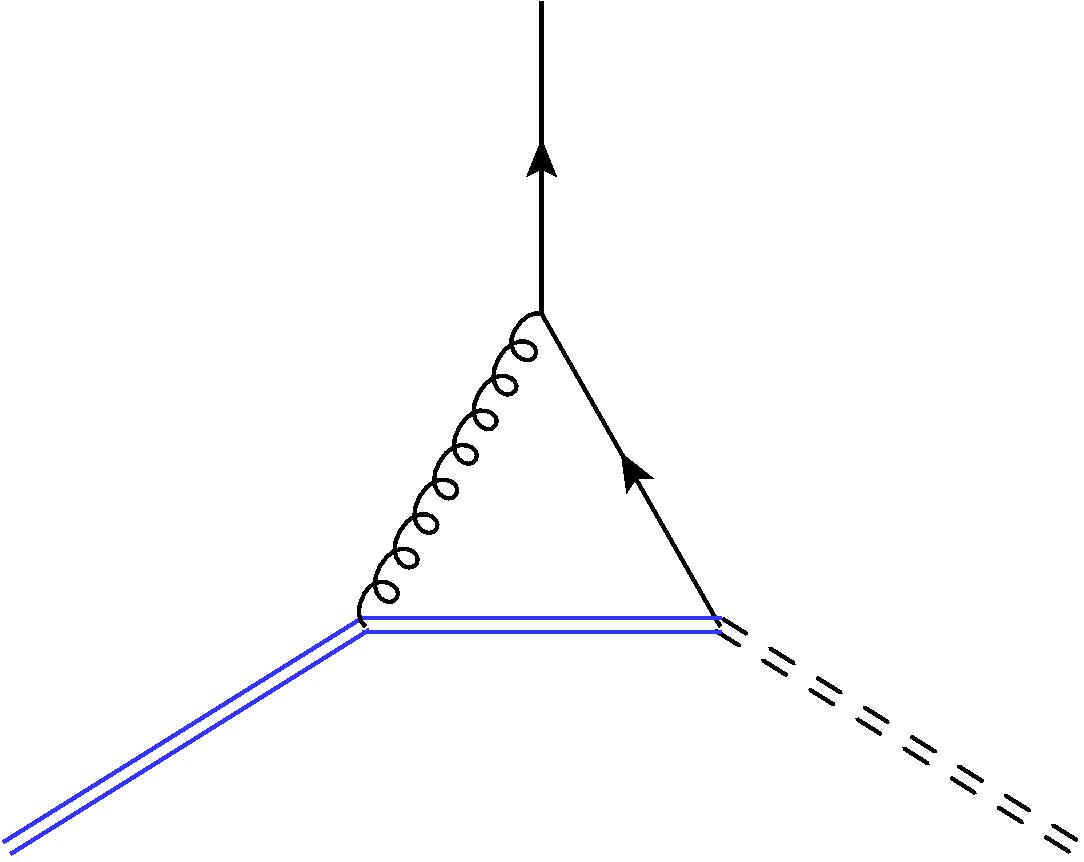}
    \label{subfig:fermion1b}}
    \qquad
    \subfigure[]{
    \includegraphics[width=0.20\textwidth]{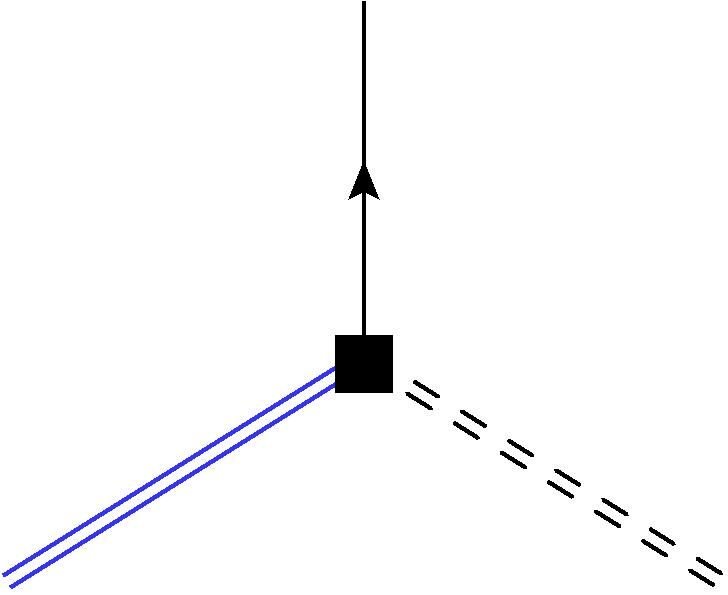}
    \label{subfig:fermion1c}} 
    \caption{One-loop corrections to the $V_{\bar{\tilde z} f \varphi }$ vertex. Wavy lines correspond to $A_\mu$ propagators, whereas wiggly lines represent $\hat{A}_\mu$ propagators.}
    \label{fig:fermion1}
\end{figure}

For all the structures, the typologies of diagrams to be considered are shown in figure \ref{fig:fermion1}. 
The ABJ(M) vertices $-\bar\psi^I \gamma^{\mu} \psi_I A_{\mu}$ and $\bar\psi^I \gamma^{\mu} \hat{A}_{\mu}\psi_I$ appear in  \ref{subfig:fermion1a} and \ref{subfig:fermion1b}, respectively. Since these vertices are diagonal in the fermion colors, the correction to the $\alpha_1$ and $\alpha_2$ vertices will be the same, as well as the ones for $\bar{\beta}_3$ and $\bar{\beta}_4$.  

Considering first the corrections to the $\alpha_i$ vertices, from diagram \ref{subfig:fermion1a} we obtain (the details are in appendix \ref{sec:gamma_af})
\begin{equation}
\label{eqn:gamma_af}
    \Gamma^{\rm \subref{subfig:fermion1a}}_{\text{fermion}} = \frac{g^2 N_1}{8\pi\epsilon}i \int d\tau \bar{\tilde z} \left( \alpha_1 e^{-i\tau/2} \xi \,  \psi^2  - \alpha_2 \, e^{-i\tau/2} \xi \,  \psi^1 \right) \varphi  \,.
\end{equation}
For diagram \ref{subfig:fermion1b} we find the same contribution with $N_1$ replaced by $N_2$.

Summing up the three diagrams, the one-loop correction to the $\alpha_i$ fermion vertices is given by
\begin{equation}
   i \int d\tau \bar{\tilde z} \left[ \left( \delta_{\alpha_1} + \frac{g^2(N_1+N_2)}{8\pi\epsilon} \right)\alpha_1 e^{-i\tau/2} \xi \, \psi^2  - \left( \delta_{\alpha_2} + \frac{g^2(N_1+N_2)}{8\pi\epsilon} \right)\alpha_2 \, e^{-i\tau/2} \xi \, \psi^1 \right] \varphi \,.
\end{equation}

It is easy to check that performing the same computation for the  fermionic vertices proportional to $\bar{f}$ in \eqref{eq:ffbar}, we obtain the same corrections to the $\bar\alpha^i$ couplings.
Consequently, we find
\begin{equation}\label{eq:Zalfa}
    Z_{\bar{\alpha}^i} = Z_{\alpha_i}
    =1-\frac{g^2(N_1+N_2)}{8\pi\epsilon}, \qquad \quad i=1,2 \, .
\end{equation}
Similarly, for the $\beta^j,\bar\beta_j$ couplings in \eqref{eq:fermionvertices} we obtain
\begin{equation}
    Z_{\beta^j} = Z_{\bar\beta_j} 
    =  1+\frac{g^2(N_1+N_2)}{8\pi\epsilon},\qquad \quad j=3,4 \,.
\end{equation}
The different sign compared with \eqref{eq:Zalfa} comes from the different couplings accompanying $\bar\alpha^i,\alpha_i$ and $\beta^j,\bar\beta_j$ parameters.

\vspace{0.2cm}


\subsubsection*{Scalar vertex corrections}

Now we study the corrections to the scalar vertex $S_{\varphi C}=g^2M_I^{\ J}\int d\tau \bar \varphi C_J\bar C^I \varphi$, as the prototype of the four-point vertices in \eqref{eqn:effectiveaction}. This vertex requires particular attention since the components of $M_I^{\ J}$ are functions of  the  $\alpha_i,\bar{\alpha}^i,\beta^j,\bar{\beta}_j$ parameters. In the most general case, the $1/24$ BPS matrix of \eqref{eqn:M124bos}, the parameters appear in all the components, and this renders the computation rather involved. However, considering the particular case $\bar\alpha^2\alpha_2=\beta^3\bar\beta_3=\beta^4\bar\beta_4=0$ is sufficient to compute the desired corrections, while simplifying considerably the calculations. 
We will then stick to this case.

\begin{figure}[h]
    \centering
    \subfigure[]{
    \includegraphics[width=0.30\textwidth]{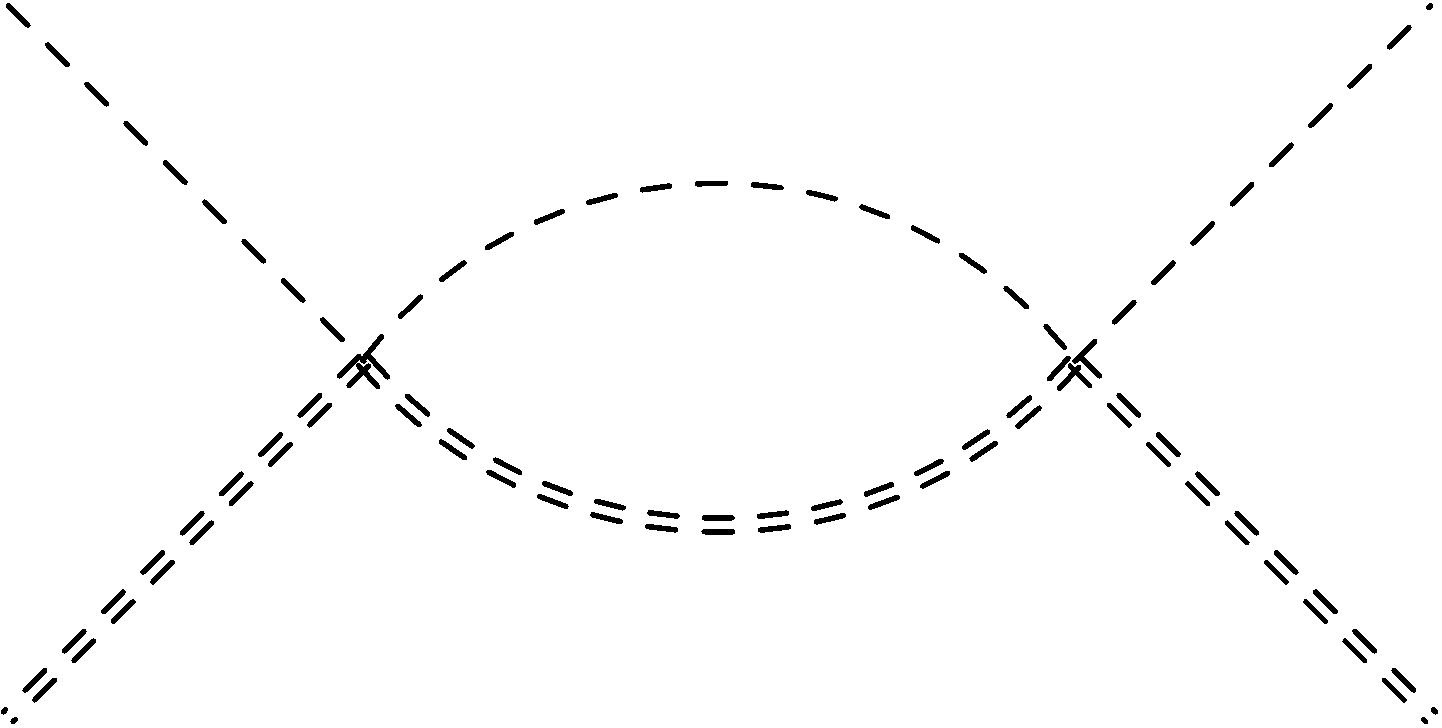}
    \label{subfig:scalarvertexa}} \qquad
    \subfigure[]{
    \includegraphics[width=0.20\textwidth]{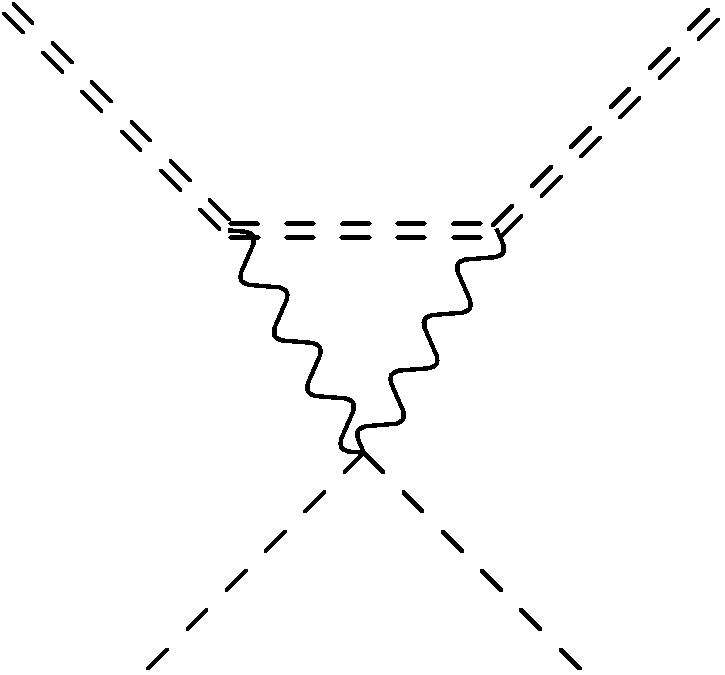}
    \label{subfig:scalarvertexb}}\qquad
    \subfigure[]{
    \includegraphics[width=0.20\textwidth]{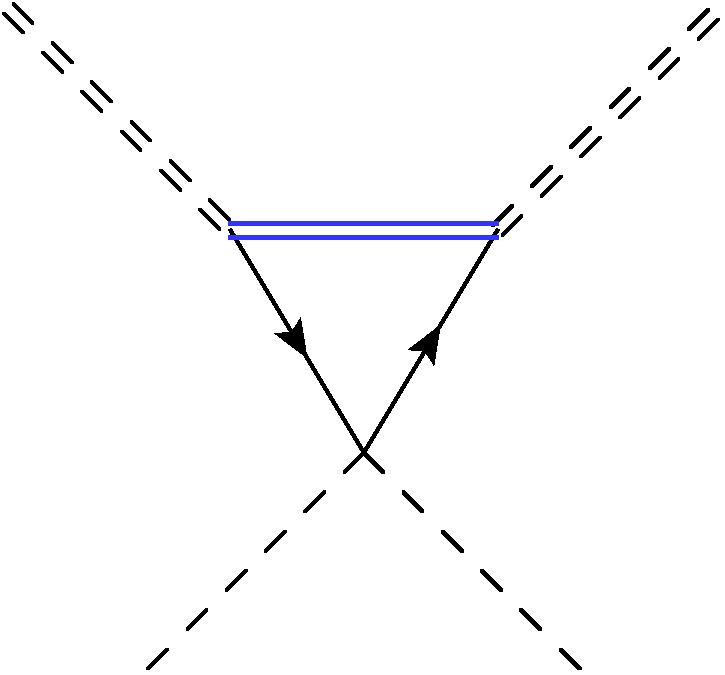}
    \label{subfig:scalarvertexc}} \qquad
    
    \subfigure[]{
    \includegraphics[width=0.20\textwidth]{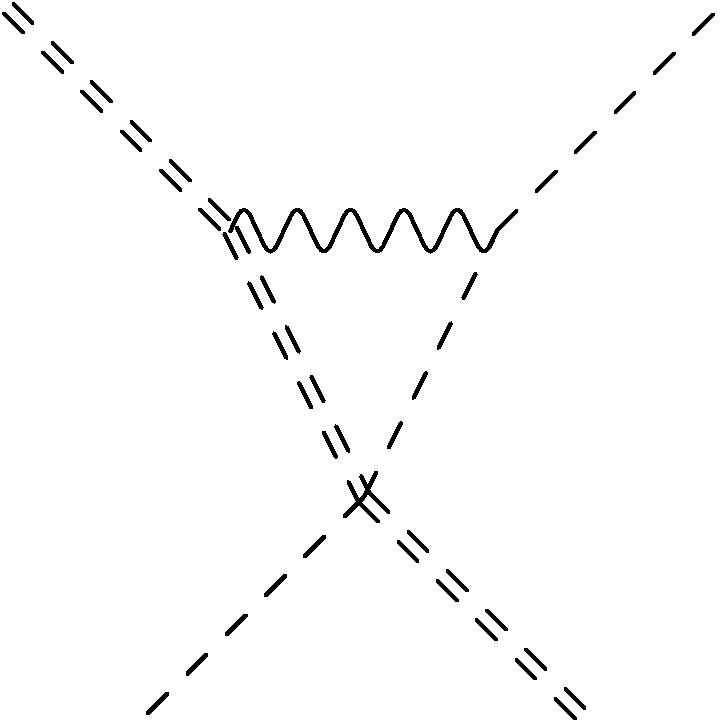}
    \label{subfig:scalarvertexf}} \qquad
    \subfigure[]{
    \includegraphics[width=0.20\textwidth]{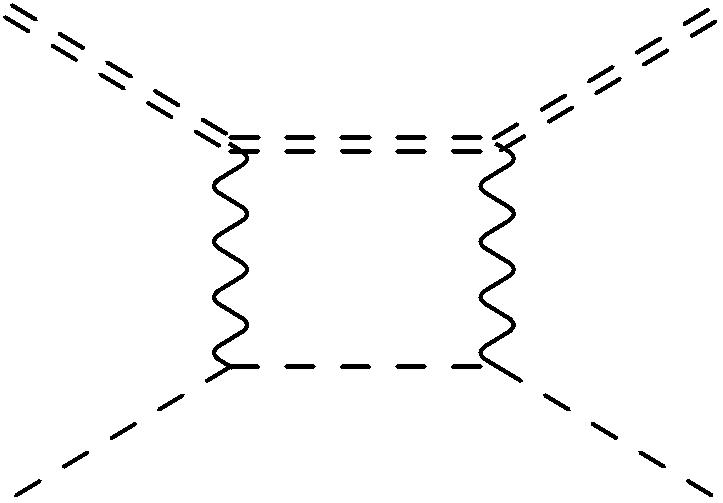}
    \label{subfig:scalarvertexe}} \qquad
    \subfigure[]{
    \includegraphics[width=0.20\textwidth]{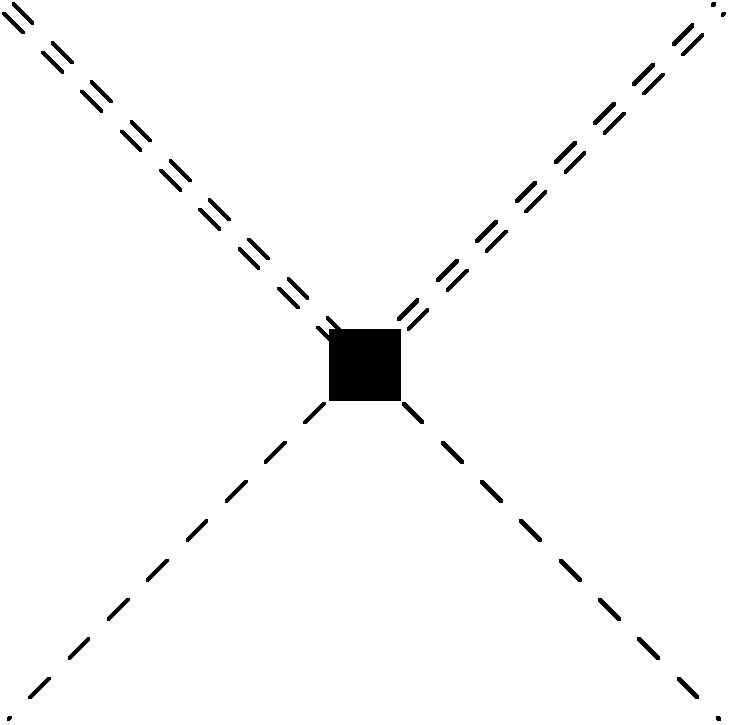}
    \label{subfig:scalarvertexd}}
    \caption{Leading one-loop corrections of the $V_{\bar \varphi C \bar C \varphi }$ vertex. The last diagram corresponds to the $\delta_{\varphi C}$ counterterm.}
    \label{fig:scalarvertex}
\end{figure}

At leading order in the gauge colors, the diagrams that contribute to the four-point vertex are depicted in figure \ref{fig:scalarvertex}. Further non-vanishing diagrams could be drawn, which however lead to subleading corrections proportional to double-trace vertices. Since we work at large $N_1, N_2$, we neglect them. 

Details on the computation of each diagram are presented separately in appendix \ref{sec:gamma_f}. Here, we list only the results. 

From the first two diagrams we obtain
\begin{equation}\label{eqn:gammabscalar}
\begin{split}
    & \Gamma^{\rm \subref{subfig:scalarvertexa}}_{\text{scalar}} =  -g^4 \frac{N_1}{8\pi\epsilon} M_{I}^{ \ K} M_{K}^{\ J} \int d\tau\,  \bar\varphi \, C_J \bar C^I \varphi\,, \\
    &\Gamma^{\rm \subref{subfig:scalarvertexb}}_{\text{scalar}} =  g^4 \frac{N_1}{8\pi\epsilon} \int d\tau\,  \bar\varphi \, C_I \bar C^I \varphi\,.
\end{split}
\end{equation}

Diagram \ref{subfig:scalarvertexc} involves the Yukawa couplings appearing in the last two lines of the ABJ(M) action in \eqref{eq:ABJMaction}. It can be built either using the $2g^2C_I\bar C^J\bar\psi^I\psi_J$ vertex or the $-g^2C_I\bar C^I\bar\psi^J\psi_J$ one. The two corresponding contributions read respectively 
\begin{equation}\label{eqn:Gammascalarc1}
\begin{split}
    & \Gamma^{\rm \subref{subfig:scalarvertexc},1}_{\text{scalar}}= -g^4\frac{N_2}{2\pi\epsilon}\bar\alpha^1\alpha_1\int d\tau \bar\varphi \, C_2 \bar C^2 \varphi\,, \\
    & \Gamma^{\rm \subref{subfig:scalarvertexc},2}_{\text{scalar}}  =g^4\frac{N_2}{4\pi\epsilon}\bar\alpha^1\alpha_1\int d\tau \bar\varphi \, C_I \bar C^I \varphi \,.
    \end{split}
\end{equation}
Therefore, we can summarize the result from this diagram as
\begin{equation}
    \Gamma^{\rm \subref{subfig:scalarvertexc}}_{\text{scalar}} =  g^4\frac{N_2}{4\pi\epsilon}\bar\alpha^1\alpha_1 \begin{pmatrix} 1 &0&0&0 \\ 0 & -1 &0 &0 \\ 0 & 0 & 1 &0\\ 0&0&0&1\end{pmatrix}_{\! \! \, I}^{\ J}\int d\tau \, \bar\varphi C_J \bar C^I \varphi\,.\, 
\end{equation}

Finally, we move on to diagrams \ref{subfig:scalarvertexf} and \ref{subfig:scalarvertexe}. Working in Lorentz gauge, it is easy to see that the corresponding contributions are not divergent. In fact, we can always integrate by parts the $\partial_\mu$ derivatives coming from the ABJ(M) vertices on the external $C_I, \bar{C}^I$ lines. As a consequence, the integrand is finite for dimensional reasons. 

Summing all the contributions, we eventually obtain
\begin{equation}
    0=g^2\left[ \delta_{\varphi C}M_I^{\ J} - g^2\frac{N_1}{8\pi\epsilon}\left( M_I^{\ K}M_{K}^{\ J}-\delta_I^J \right) +g^2\frac{N_2}{4\pi\epsilon}\bar\alpha^1\alpha_1(\delta_I^J-2\delta_I^2\delta^J_2) \right]\int d\tau \, \bar\varphi C_J \bar C^I \varphi\,.
\end{equation}
This implies that
\begin{equation}
    Z_{\varphi C}M_I^{\ J} = (1 + \delta_{\varphi C} ) M_I^{\ J} = M_I^{\ J} +\frac{g^2}{4\pi\epsilon} \bar\alpha^1\alpha_1 \left[ -N_2(\delta_I^J-2\delta_I^2\delta^J_2) + 2N_1( \bar\alpha^1\alpha_1-1)\delta_I^1\delta^J_1 \right] \,.
\end{equation}
From the definition \eqref{eq:Mren}, for the scalar coupling renormalization we can write 
\begin{equation}
    ({M}_I^{\ J})_0 = \frac{Z_{\varphi C}}{Z_{\varphi}}M_I^{\ J}= M_I^{\ J} + \frac{g^2}{4\pi\epsilon}\bar\alpha^1\alpha_1\left[ -N_2(\delta_I^J-2\delta_I^2\delta^J_2) + 2N_1( \bar\alpha^1\alpha_1-1)\delta_I^1\delta^J_1 + N_2 M_I^{\ J} \right]\,.
\end{equation}
The term at order $g^2$ on the right-hand side is zero for $I=J=2,3,4$, whereas for $I=J=1$ we obtain
\begin{equation}\label{eq:aabarren}
    (\bar\alpha^1\alpha_1)_0=\bar\alpha^1\alpha_1\left[ 1+\frac{g^2}{4\pi\epsilon}(N_1+N_2)(\bar\alpha^1\alpha_1-1) \right]\,,
\end{equation}
where on the left-hand side the subscript indicates the product of the two bare parameters. This result is consistent with the $\alpha_1, \bar{\alpha}^1$ renormalization that we have already discussed.


\subsection{$\beta$-functions}
\label{sec:betafunction}

Having determined the renormalization functions, we can now compute the one-loop $\beta$-functions for the parameters. 
To this end, we first recall that the definition of the bare parameters are given in \eqref{eq:parren}.
Collecting the results for the renormalization functions found in the previous sections,
\begin{equation}
\begin{split}
    Z_{z}^{1/2}&=1-g^2\frac{N_2}{8\pi\epsilon}(\bar\alpha^i\alpha_i + \beta^j\bar\beta_j)\,,\\
    Z_{\tilde \varphi}^{1/2} &= 1-g^2\frac{N_1}{8\pi\epsilon}(\bar\alpha^i\alpha_i + \beta^j\bar\beta_j)\,,\\
    Z_{\bar\alpha^i}=Z_{\alpha_i}&= 1-\frac{g^2(N_1+N_2)}{8\pi\epsilon}\,,\\ Z_{\beta^j}=Z_{\bar\beta_j}&= 1+\frac{g^2(N_1+N_2)}{8\pi\epsilon} \,,
\end{split}
\end{equation}
and plugging them there, we find
\begin{equation}
\label{eqn:renzeta}
\begin{split}
    &(\alpha_k)_0 = \left( 1+\frac{g^2}{8\pi\epsilon} \, (N_1+N_2)(\bar\alpha^i\alpha_i + \beta^j\bar\beta_j-1) \right)\alpha_k \qquad k=1,2 \,, \\ &  (\bar\alpha^{k})_0 = \left( 1+\frac{g^2}{8\pi\epsilon} \,(N_1+N_2) (\bar\alpha^i\alpha_i + \beta^j\bar\beta_j-1)\right)\bar\alpha^k \, , \\ &(\bar\beta_l)_0 = \left( 1+\frac{g^2}{8\pi\epsilon} \,(N_1+N_2) (\bar\alpha^i\alpha_i + \beta^j\bar\beta_j+1) \right)\bar\beta_l \; \, \qquad l = 3,4  \, , \\ &(\beta^l)_0 = \left( 1+\frac{g^2}{8\pi\epsilon} \, (N_1+N_2)(\bar\alpha^i\alpha_i + \beta^j\bar\beta_j+1) \right)\beta^l\, .
\end{split}
\end{equation}
As already mentioned, if we set $\bar\alpha^2=\alpha_2=\bar\beta_j =\beta^j =0$ and consider the product $(\bar{\alpha}^1 \alpha_1)_0$, we obtain exactly the expression \eqref{eq:aabarren} coming from the renormalization of the four-point scalar vertices. This is a non-trivial check of our renormalization procedure.  

The one-dimensional theory under investigation possesses nine dimensionless coupling constants $g_a=(g^2,\alpha_i,\bar\alpha^i,\beta^j,\bar\beta_j)$, with the new indices $a, b, \ldots$ running over these nine couplings. In dimensional regularization with $d=3-2\epsilon$, the $\alpha_i,\bar\alpha^i,\beta^j,\bar\beta_j$ parameters remain dimensionless, while $g^2$ acquires dimension $\Delta_{g^2} = 2\epsilon$. 

Expressing the bare coupling constants $(g_{a})_0$ as a function of the renormalized ones as
\begin{equation}
     (g_{a})_0= \mu^{u_a\epsilon}\left[ g_a + \frac{1}{\epsilon}K_{a} + \cO\left(\tfrac{1}{\epsilon^2}\right) \right]\,, 
\end{equation}
with $u_{g^2}=2$ and the others vanishing, the corresponding $\beta$-functions are given by
\begin{equation}
    \beta_{a} = \mu \frac{d g_a}{d\mu}=-\epsilon u_a g_a - u_a K_{a} + \sum_b u_b g_b \frac{\partial K_{a}}{\partial g_b}\,.
\end{equation}
Specializing this to the $\alpha_k$ parameters, we find 
\begin{equation}
    (\alpha_k)_0 = \alpha_k + \frac{1}{\epsilon}K_{\alpha_k}\,, \qquad \qquad \beta_{\alpha_k} = 2 g^2 \frac{\partial K_{\alpha_k}}{\partial g^2}\,,
\end{equation}
and similarly for the other parameters.

From \eqref{eqn:renzeta} we can read off the explicit expressions of the $K$'s, which lead to the following one-loop $\beta$-functions
\begin{equation}
\label{eqn:betafunction}
\begin{split}
    \beta_{\alpha_k}&= \frac{g^2}{4\pi}(N_1+N_2)\, (\bar\alpha^i\alpha_i + \beta^j\bar\beta_j-1)\alpha_k\,,  \qquad 
    \beta_{\bar\alpha^k}= \frac{g^2}{4\pi}(N_1+N_2)\, (\bar\alpha^i\alpha_i + \beta^j\bar\beta_j-1)\bar\alpha^k,\\ \beta_{\bar\beta_l}&= \frac{g^2}{4\pi}(N_1+N_2)\, (\bar\alpha^i\alpha_i + \beta^j\bar\beta_j+1)\bar\beta_l \,,  \qquad\,\,\, \beta_{\beta^l}= \frac{g^2}{4\pi}(N_1+N_2)\, (\bar\alpha^i\alpha_i + \beta^j\bar\beta_j+1)\beta^l \,.
\end{split}
\end{equation}
These are the analogues of the  Polchinski-Sully $\beta$-functions for the parameter $\zeta$ of the interpolating Wilson loop in $\cN=4$ super Yang-Mills theory \cite{Polchinski_2011}. 

To conclude this section we observe that the results we have obtained for the renormalization functions and the $\beta$-functions are path independent, since short distance divergences should be blind to the actual form of the Wilson loop contour. Therefore, we expect them to be valid also for the renormalization of the parametric latitude Wilson loops of section \ref{sec:parametriclatitude}. In fact, as it will be discussed in \cite{CPTT}, the renormalization functions that remove UV divergences in that case are independent of the latitude angle and coincide with the present ones. 


\section{Wilson loop expectation value}
\label{sec:WLVEV}

In this section we compute the two-loop VEV for the circular $1/24$ BPS Wilson loop. In the auxiliary field approach this is given by (see appendix \ref{sec:apxDorn} for the proof of this identity)
\begin{align}
\label{eqn:WLDorn}
   \langle W_{1/24}(\bar\alpha^i,\alpha_i,\beta^j,\bar\beta_j) \rangle 
   &= \frac{1}{2}\langle \Tr\left( \Psi_0(2\pi)\bar\Psi_0(0)   \right) \rangle\nonumber \\
    &=\frac{1}{2}\Big( \langle z_0(2\pi) \bar{z}_0(0) \rangle + \langle \varphi_0(2\pi)\bar\varphi_0(0) \rangle + \langle \tilde{z}_0(2\pi) \bar{\tilde z}_0(0) \rangle +\langle \tilde\varphi_0(2\pi)  \bar{\tilde\varphi}_0(0) \rangle \Big)\nonumber \\
    & = (1+ \delta_z) \langle z(2\pi) {\bar z}(0) \rangle + (1+ \delta_{\tilde z}) \langle {\tilde z}(2\pi) \bar{\tilde z}(0) \rangle  \, .
\end{align}
where in the last line we have taken into account the relation between bare and renormalized fields, and the fact that $z$ and $\varphi$ in the auxiliary matrix \eqref{eq:oddmatrix} have the same two-point function, as well as $\tilde{z}$ and $\tilde\varphi$. We recall that the two counterterms $\delta_z, \delta_{\tilde z}$ can be read off from \eqref{eqn:zren} and \eqref{eq:ztilderen}, respectively. 

Since the one-dimensional auxiliary field method is analogous to the conventional way of computing Wilson loops VEVs, there are straightforward relations between diagrams of the one-dimensional theory and diagrams coming from the perturbative expansion of the Wilson loop. In fact, if in the diagrams contributing to the two-point functions we identify the end points, and identify the one-dimensional propagators with the Wilson loop contour, we formally reproduce the one- and two-loop diagrams from the expansion of the Wilson loop. 
It then follows that the typologies of integrals are the same in the two cases, so we can exploit the results already present in the literature for two-loop integrals of Wilson loops. We refer in particular to \cite{Bianchi:2013zda,Bianchi:2013rma} for details on the evaluation of the integrals in the same set of conventions. 

We evaluate the $\langle z(2\pi) {\bar z}(0) \rangle$, $\langle {\tilde z}(2\pi) \bar{\tilde z}(0) \rangle$ correlators at two loops using the Lagrangian \eqref{eq:Ltot}, that is using the Feynman rules for renormalized quantities. According to \eqref{eqn:WLDorn} the result for the Wilson loop VEV is then obtained by multiplying by the renormalization factors $(1 + \delta)$ and keeping the correct order in loops. 

For instance, focusing on the $\langle z {\bar z} \rangle$ correlator, we organize the perturbative expansion as
\begin{equation}
\begin{split}\label{eq:2ptexp}
    \langle W \rangle &= (1 + \delta_z^{(1)} + \delta_z^{(2)} + \cdots) \left( \langle z(2\pi) {\bar z}(0) \rangle^{(0)} + \langle z(2\pi) {\bar z}(0) \rangle^{(1)} + \langle z(2\pi) {\bar z}(0) \rangle^{(2)} + \cdots \right) \\
    & = 1 + \left[ \delta_z^{(1)}  + \langle z(2\pi) {\bar z}(0) \rangle^{(1)} \right] 
    + \left[ \delta_z^{(2)}  + \delta_z^{(1)} \langle z(2\pi) {\bar z}(0) \rangle^{(1)} + \langle z(2\pi) {\bar z}(0) \rangle^{(2)} \right] + \cdots
\end{split}
\end{equation}  
where $\delta^{(L)}_z$ indicates the counterterm at order $L$. The presence of counterterms properly grouped according to their loop order is crucial to remove short distance divergences from the integrals and make the expansion order by order finite. In the next two sections we evaluate the finite contributions corresponding to the two square brackets in \eqref{eq:2ptexp}.


\subsection{One-loop analysis}

At one-loop, the diagrams contributing to the two-point functions are the ones depicted in figures \ref{subfig:vertexcorrectiona} and \ref{subfig:vertexcorrectionb}. We can then exploit part of the previous calculations, except that now we have to evaluate the finite part of the integrals, having removed already the short distance divergence. 

Diagram \ref{subfig:vertexcorrectiona} still vanishes for planarity, as the epsilon tensor coming from the vector propagator is contracted with three vectors lying on the plane of the circular contour.

The contribution from \ref{subfig:vertexcorrectionb} is given in \eqref{eq:5b}. The integrals appearing there were computed in dimensional regularization in \cite{Bianchi:2013zda,Bianchi:2013rma,Griguolo_2013a}. Using those results and taking into account that at this order we find $\langle z \bar z \rangle = \langle \tilde z \bar{\tilde z} \rangle$, the one-loop expectation value of the 1/24 BPS operator reads
\begin{equation}
\label{eqn:1l-circular}
    \langle W_{1/24}(\bar\alpha^i,\alpha_i,\beta^j,\bar\beta_j) \rangle^{(1)}=-(\bar\alpha^i\alpha_i+\beta^j\bar\beta_j)g^2N_1N_2\frac{4^{\epsilon}\pi^{\epsilon+1}\sec\pi\epsilon}{\Gamma(\epsilon)}=-(\bar\alpha^i\alpha_i+\beta^j\bar\beta_j)g^2 N_1N_2 \pi\epsilon.
\end{equation}
In the $\epsilon \to 0$ limit this contribution vanishes. However, since it will enter later at two loops multiplied by the counterterms, it is necessary to keep it for finite $\epsilon$. 


\subsection{Two-loop analysis}

We now move on to the evaluation of the two-point functions in \eqref{eqn:WLDorn} at two loops. In what follows we focus separately on bosonic and fermionic diagrams, as well as on contributions due to the counterterms of the one-dimensional theory.

\begin{figure}[H]
    \centering
    \subfigure[]{
    \includegraphics[width=0.25\textwidth]{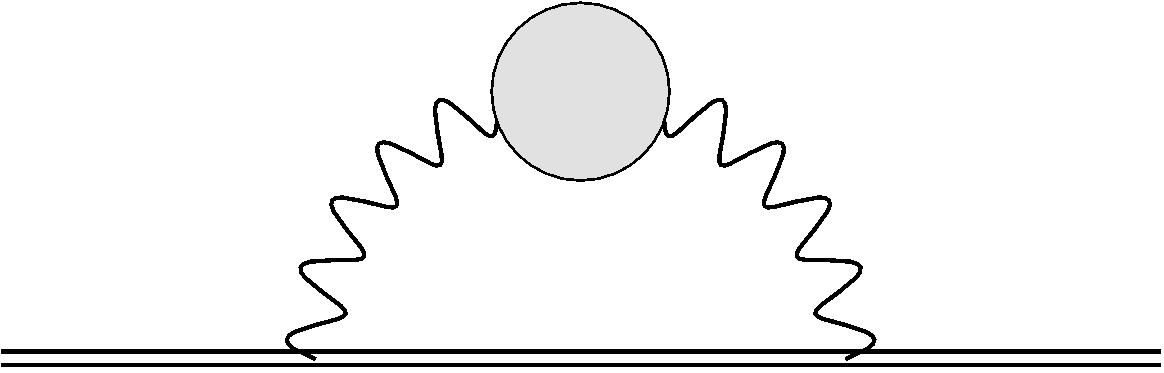}
    \label{subfig:2-loop_bosa}} \qquad
    \subfigure[]{
    \includegraphics[width=0.25\textwidth]{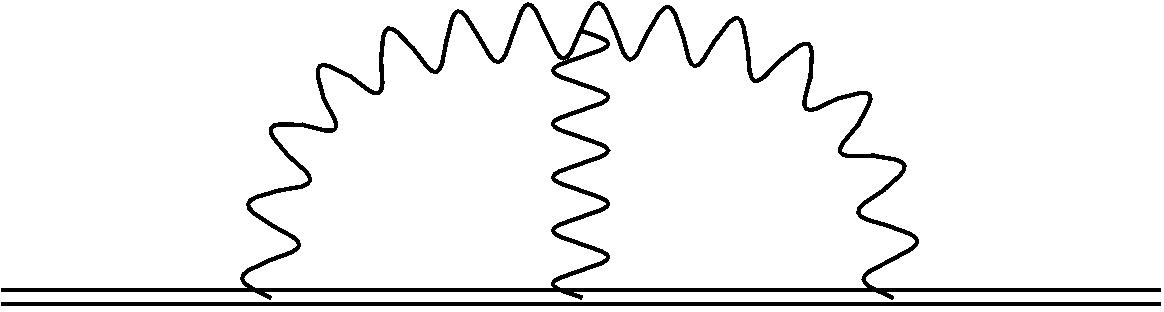}
    \label{subfig:2-loop_bosb}} \qquad
    \subfigure[]{
    \includegraphics[width=0.25\textwidth]{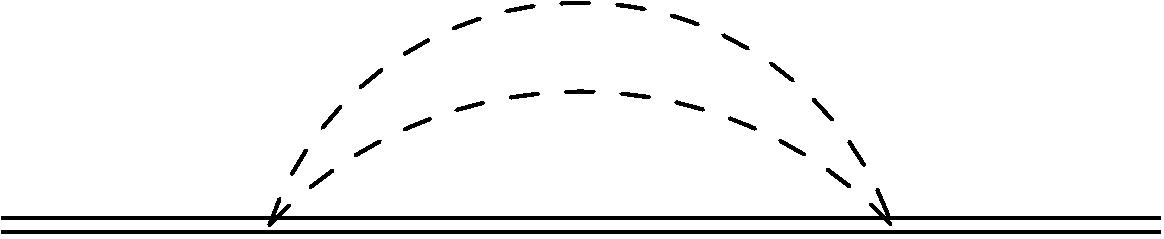}
    \label{subfig:2-loop_bosc}} 
    
    \subfigure[]{
    \includegraphics[width=0.25\textwidth]{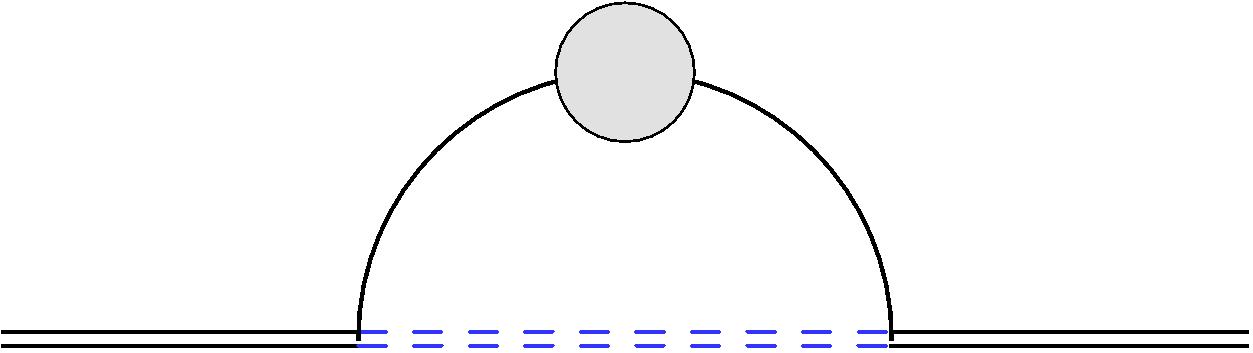}
    \label{fig:2-loop_fer1}}\qquad
    \subfigure[]{
    \includegraphics[width=0.25\textwidth]{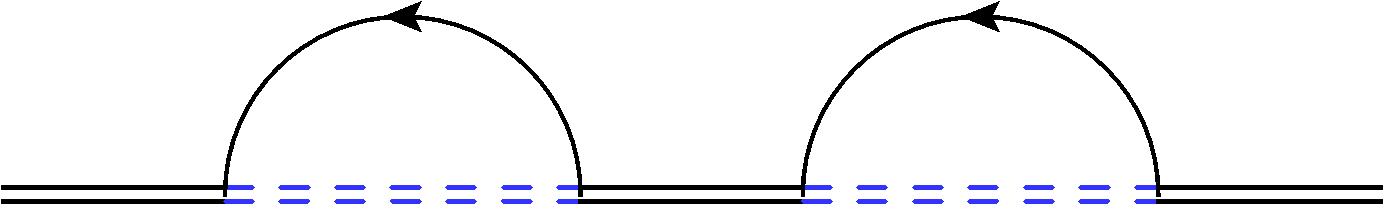}
    \label{subfig:2-loop_fer2a}} \qquad
    \subfigure[]{
    \includegraphics[width=0.25\textwidth]{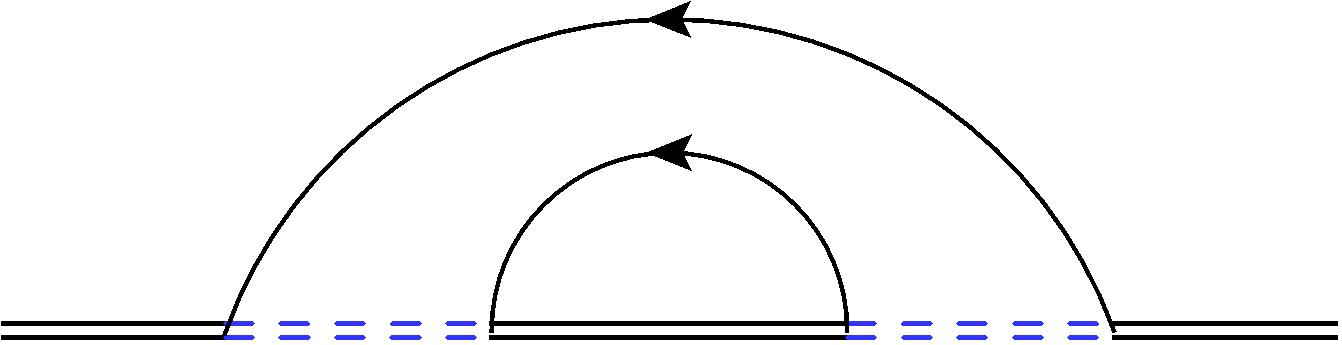}
    \label{subfig:2-loop_fer2b}}
    
    \subfigure[]{
    \includegraphics[width=0.25\textwidth]{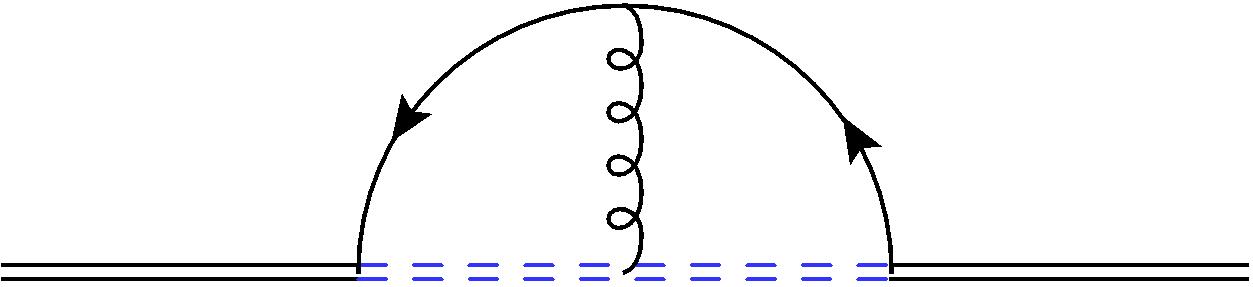}
    \label{subfig:2-loop_fer3a}} \quad
    \subfigure[]{
    \includegraphics[width=0.25\textwidth]{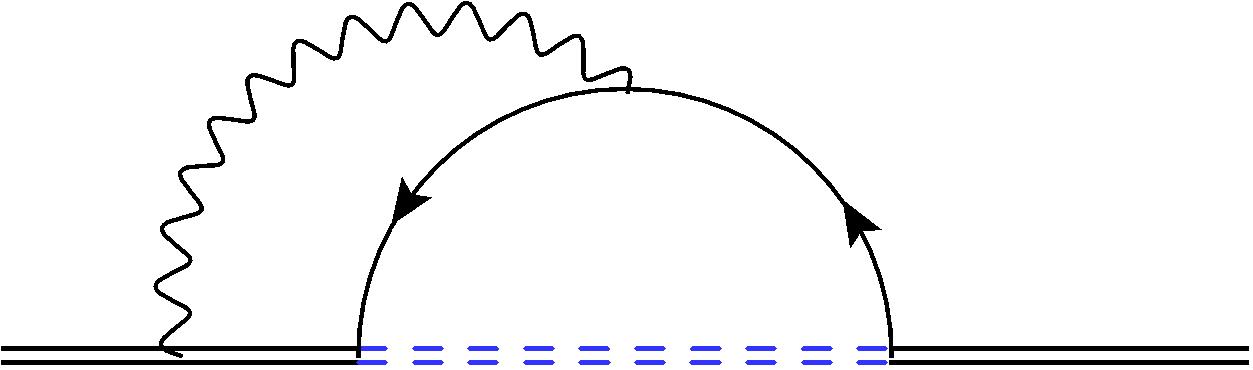}
    \label{subfig:2-loop_fer3b}}\quad
    \subfigure[]{
    \includegraphics[width=0.25\textwidth]{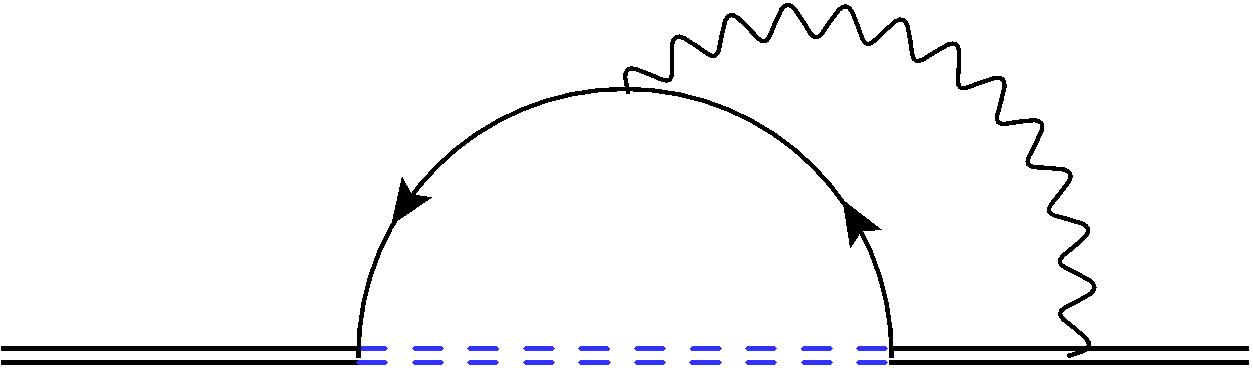}
    \label{subfig:2-loop_fer3c}}
    \caption{Two-loop corrections to the one-dimensional fermionic propagator $\langle z \bar z\rangle$.}
    \label{fig:2-loop_bos}
\end{figure}

\subsubsection*{Bosonic diagrams}

The bosonic diagrams which contribute non-trivially are reported in the first line of figure \ref{fig:2-loop_bos}. 

Diagram \ref{subfig:2-loop_bosa} contains the gauge propagator corrected at one loop. Using its explicit expression \eqref{eqn:onelooppropagator}, the corresponding contribution to $\langle z\bar z \rangle^{(2)}$ reads
\begin{equation}
    {\rm \ref{subfig:2-loop_bosa}}=-\int_0^{2\pi}d\tau_1\int_0^{\tau_1}d\tau_2 \langle A_{\mu}(\tau_1)A_{\nu}(\tau_2) \rangle^{(1)} \dot x_1^{\mu}\dot x^{\nu}_2 = \frac{g^4}{4}N^2_1N_2\,.
\end{equation}
The result for $\langle \tilde z \bar{\tilde z} \rangle^{(2)}$ is the same with $N_1$ and $N_2$ exchanged. 

Diagram \ref{subfig:2-loop_bosb} contains the ABJ(M) pure gauge vertex. Exploiting the results in \cite{Bianchi:2013zda,Bianchi:2013rma} for the corresponding integral, this gives
\begin{align}
    {\rm{\ref{subfig:2-loop_bosb}}} =-\frac{i}{3g^2}\int_0^{2\pi}d\tau_1 \int_0^{\tau_1}&d\tau_2\int_0^{\tau_2}d\tau_3 \, \epsilon^{\alpha\beta\gamma}\dot x_1^{\mu}\,\dot x_2^{\nu}\,\dot x_3^{\rho}\\ & \times\langle A_{\mu}(\tau_1)A_{\nu}(\tau_2)\rangle \langle A_{\rho}(\tau_3) A_{\alpha}(x) \rangle \langle A_{\beta}(x)A_{\gamma}(x) \rangle\nonumber = -\frac{g^4}{24}N_1^3\,.
\end{align}
The result for $\langle \tilde z\bar{\tilde z}\rangle^{(2)}$ is the same with $N_1^3$ replaced by $N_2^3$.

Diagram \ref{subfig:2-loop_bosc} deserves more attention, since it is the only diagram which contributes to the 1/24 BPS operator, but is absent in the more supersymmetric cases. Its contribution to $\langle z\bar z \rangle^{(2)}$ is
\begin{equation}
    {\rm \ref{subfig:2-loop_bosc}}=g^4N^2_1N_2\frac{\Gamma^2(\frac{1}{2}-\epsilon)}{16\pi^{3-2\epsilon}}\int_0^{2\pi} d\tau_1\int_0^{\tau_1} d\tau_2\, \Tr\left( M(\tau_1)M(\tau_2) \right) \frac{1}{|x_1-x_2|^{2-4\epsilon}}\,.
\end{equation}
As long as the trace of two $M$ matrices is $\tau$-independent, this integral is identically zero (see for instance \cite{Bianchi:2013rma}). This is what happens in the 1/6 and 1/2 BPS cases.  
However, in the present case 
this trace acquires a non-trivial $\tau$-dependence proportional to the loop parameters, 
\begin{equation}
    \Tr\left(M(\tau_1)M(\tau_2)  \right) \, \to \, 8(\bar\alpha^i\alpha_i)(\beta^j\bar\beta_j)\cos\tau_{12}\,.
\end{equation}
This modifies the nature of the integral 
leading to a non-vanishing result. In fact, the resulting integral is the same as the one-loop correction to the gauge field propagator \ref{subfig:2-loop_bosa}. Exploiting that result, we obtain
\begin{equation}
\begin{split}\label{eq:10cbos}
    {\rm \ref{subfig:2-loop_bosc}}&=g^4N_1N_2(N_1+N_2)(\bar\alpha^i\alpha_i)(\beta^j\bar\beta_j)\frac{\Gamma^2(\frac{1}{2}-\epsilon)}{2\pi^{3-2\epsilon}}\int_0^{2\pi}d\tau_1\int_0^{\tau_1}d\tau_2\, \frac{\cos \tau_{12}}{|x_{12}|^{2-4\epsilon}}\\ &=- \frac{g^4}{2}(\bar\alpha^i\alpha_i)(\beta^j\bar\beta_j)N_1N_2(N_1+N_2) \,.
\end{split}
\end{equation}
Summarizing, the bosonic contribution to  the $1/24$ BPS Wilson loop in \eqref{eqn:WLDorn} is 
\begin{equation}
    \cB = \frac{g^4}{4}N_1N_2(N_1+N_2)-\frac{g^4}{24}\left( N_1^3+N_2^3 \right)- \frac{g^4}{2}(\bar\alpha^i\alpha_i)(\beta^j\bar\beta_j)N_1N_2(N_1+N_2)\,.
\end{equation}

\vspace{0.2cm}

\subsubsection*{Fermionic diagrams}

Fermionic diagrams contributing to the two-point functions are depicted in the second and third lines of figure \ref{fig:2-loop_bos}. 
The first diagram \ref{fig:2-loop_fer1} contains the one-loop corrected fermion propagator given in \eqref{eqn:onelooppropagator}. Since this is proportional to $(N_1-N_2)$, when in \eqref{eqn:WLDorn} we sum up the contribution of $\langle z(2\pi) \bar z(0)\rangle$ with the one from $\langle \tilde z (2\pi)\bar{\tilde z} (0)\rangle$ obtained by exchanging $N_1$ with $N_2$, they cancel each other. Therefore, this diagram does not contribute to the Wilson loop VEV.

Moving to the double fermion-exchange diagrams illustrated in figures \ref{subfig:2-loop_fer2a}-\ref{subfig:2-loop_fer2b}, using known integrals from the literature \cite{Bianchi:2013zda,Bianchi:2013rma}, we obtain 
\begin{equation}
\begin{split}
    {\rm\ref{subfig:2-loop_fer2a}}+{\rm\ref{subfig:2-loop_fer2b}}&=\frac{1}{24}\int d\tau_1d\tau_2d\tau_3d\tau_4 \, \langle z(2\pi)\bar z(0) \big( \bar z \bar f \tilde \varphi \big)(\tau_1)\big( \bar z \bar f \tilde \varphi \big)(\tau_2)\\ & \qquad \qquad \times\big( \bar{\tilde\varphi} f z \big)(\tau_3)\big( \bar{\tilde \varphi} f z \big)(\tau_4)\rangle + (\tau_1,\tau_2,\tau_3,\tau_4 \text{ permutations}) \\
    & = \frac{3g^4}{8}N_1N_2(N_1+N_2) (\bar\alpha^i\alpha_i+\beta^j\bar\beta_j)^2
\end{split}
\end{equation}

Finally, the three diagrams in the last line of figure \ref{fig:2-loop_bos} correspond to the three different ways of contracting the fields that exit the fermion-vector vertex.
Their sum reads 
\begin{equation}
\begin{split}
    &{\rm\ref{subfig:2-loop_fer3a}}+{\rm\ref{subfig:2-loop_fer3b}} +{\rm\ref{subfig:2-loop_fer3c}} \\
    &= i\int_0^{2\pi}d\tau_1 \int_0^{\tau_1}d\tau_2\int_0^{\tau_2}d\tau_3 \Big[ \langle \bar f(\tau_1) \hat A_{\mu}(\tau_2)\dot x^{\mu}_2 f(\tau_3) \rangle \\
     & \hspace{5.5cm} \langle \bar f(\tau_1) f(\tau_2) A_{\mu}(\tau_3)\dot x^{\mu}_3\rangle + \langle A_{\mu}(\tau_1)\dot x^{\mu}_1 \bar f(\tau_2) f(\tau_3) \rangle  \Big]\,.
\end{split}
\end{equation}
Inserting the explicit expressions \eqref{eq:ffbar} for the $f, \bar{f}$ functions, we obtain a linear combination of integrals which are the same ones appearing in the ordinary Wilson loop expansion. Therefore, exploiting known results in the literature \cite{Bianchi:2013zda,Bianchi:2013rma} and combining the contributions from $\langle z(2\pi) \bar z(0)\rangle$ and $\langle \tilde z(2\pi)\bar{\tilde z}(0)\rangle$, we eventually obtain
\begin{equation}
\label{eqn:vertexresult}
    {\rm\ref{subfig:2-loop_fer3a}}+{\rm\ref{subfig:2-loop_fer3b}} +{\rm\ref{subfig:2-loop_fer3c}} = -\frac{g^4}{2}N_1N_2(N_1+N_2)(\bar\alpha^i\alpha_i-\beta^j\bar\beta_j)\,.
\end{equation}

In conclusion, the total sum of fermionic diagrams reads
\begin{equation}
{\mathcal F} = \frac{3g^4}{8}N_1N_2(N_1+N_2) (\bar\alpha^i\alpha_i+\beta^j\bar\beta_j)^2 -\frac{g^4}{2}N_1N_2(N_1+N_2)(\bar\alpha^i\alpha_i-\beta^j\bar\beta_j)\,.   
\end{equation}

\vspace{0.2cm}

\subsubsection*{Counterterms}
\begin{figure}[h]
    \centering
    \subfigure[]{
    \includegraphics[width=0.30\textwidth]{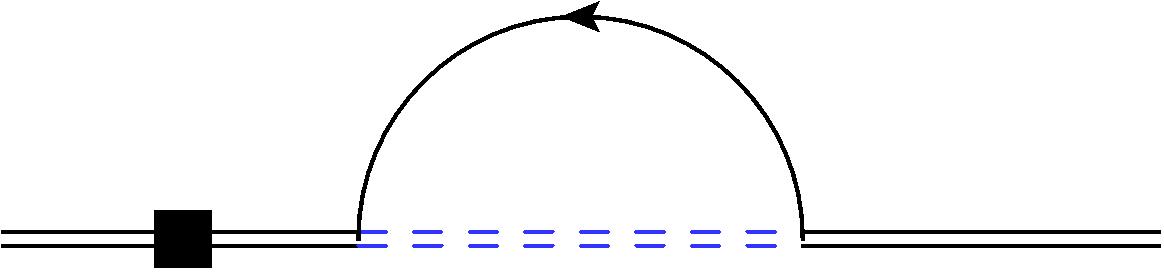}
    \label{subfig:diagcounta}} \qquad\qquad
    \subfigure[]{
    \includegraphics[width=0.30\textwidth]{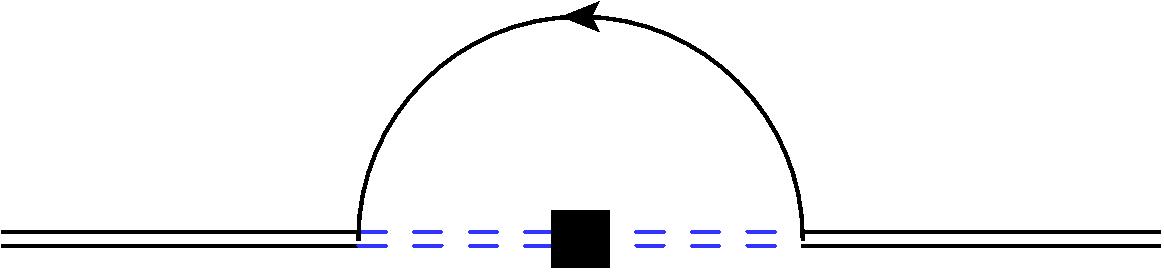}
    \label{subfig:diagcountb}} 
    
    \subfigure[]{
    \includegraphics[width=0.30\textwidth]{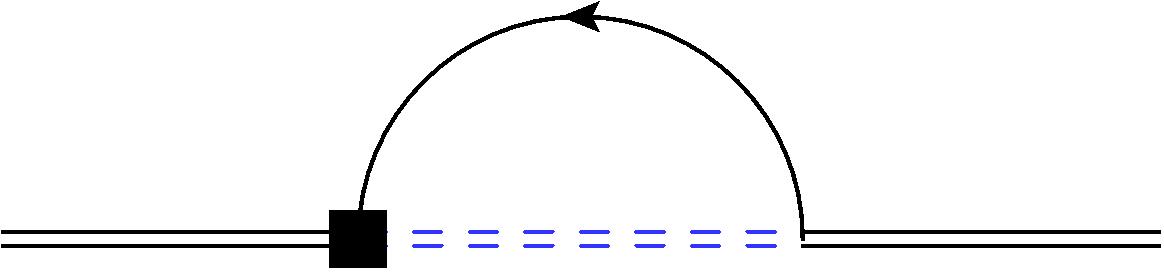}
    \label{subfig:diagcountc}} \qquad\qquad
    \subfigure[]{
    \includegraphics[width=0.30\textwidth]{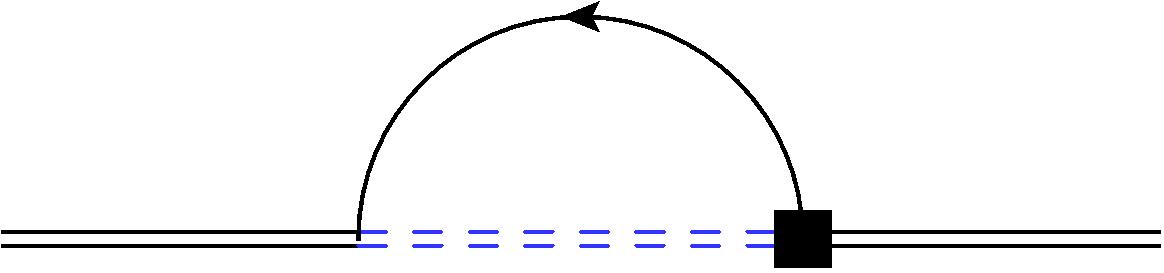}
    \label{subfig:diagcountd}} 
    \caption{Counterterm contributions to the self-energy of the one-dimensional fermion $z$.}
    \label{fig:diagcount}
\end{figure}

As seen in section \ref{sec:counterterms}, we also need to include diagrams with vertices coming from the counterterms. In particular, for the $\langle z(2\pi)\bar z(0) \rangle$ two-loop correction we obtain the four diagrams in figure \ref{fig:diagcount}. For all the other one-dimensional fields we have analogous diagrams and the results extend straightforwardly. 

We start from diagram \ref{subfig:diagcounta}, which corresponds to the insertion of a one-loop $\delta_z$ counterterm. We obtain
\begin{equation}
\begin{split}
    {\rm \ref{subfig:diagcounta}} &= \delta_z\int d\tau\int d\tau_1\int d\tau_2\langle z(2\pi)\bar z(0) \big( \bar z \bar f \bar{\tilde \varphi} \big)(\tau_1)  \big( \varphi f z \big)(\tau_2) \big(\bar z \partial_{\tau} z\big)(\tau) \rangle + (\tau_1 \leftrightarrow \tau_2)\\
    &=2\delta_z \int_0^{2\pi}d\tau_1\int_0^{\tau_1}d\tau_2 \langle \bar f(\tau_1)f(\tau_2) \rangle\,,
\end{split}
\end{equation}
where we have used $\partial_{\tau}\theta(\tau-x)=\delta(\tau-x)$. Similarly, from \ref{subfig:diagcountb} we have
\begin{equation}
    {\rm \ref{subfig:diagcountb}}=\delta_{\tilde\varphi} \int_0^{2\pi}d\tau_1\int_0^{\tau_1}d\tau_2 \langle \bar f(\tau_1)f(\tau_2) \rangle\,.
\end{equation}
Diagrams \ref{subfig:diagcountc} and \ref{subfig:diagcountd} correspond to insertions of a fermionic counterterm vertex 
\begin{equation}
\begin{aligned}
    {\rm\ref{subfig:diagcountc}} + {\rm\ref{subfig:diagcountd}} &= -\frac12\int d\tau_1\int d\tau_2\langle z(2\pi)\bar z(0) \big( \bar z \bar f \bar{\tilde \varphi} \big)(\tau_1)  \big( \varphi \, (\delta_f f) z \big)(\tau_2) \rangle \\
    &\quad -\frac12\int d\tau_1\int d\tau_2\langle z(2\pi)\bar z(0) \big( \bar z \, (\delta_{\bar f} \bar f) \bar{\tilde \varphi} \big)(\tau_1)  \big( \varphi f z \big)(\tau_2) \rangle + (\tau_1 \leftrightarrow \tau_2)
    \\&=
    -\int_0^{2\pi}d\tau_1 \int_0^{\tau_1}d\tau_2 \bigg(\langle \bar f(\tau_1) \delta_f f(\tau_2) \rangle  
+ \langle \delta_{\bar f} \bar f(\tau_1) f(\tau_2) \rangle\bigg) \,,
\end{aligned}
\end{equation}
where we have defined
\begin{equation}
    \delta_ff =e^{-\frac{i\tau}{2}}\xi\left( \delta_{\bar\alpha^1}\bar\alpha^1 \psi^2 - \delta_{\bar\alpha^2}\bar\alpha^2\psi^1 \right) + e^{\frac{i\tau}{2}}\eta(\delta_{\beta^3}\beta^3 \psi^4 - \delta_{\beta^4}\beta^4\psi^3)\,.
\end{equation}
and similarly for $\delta_{\bar f} \bar f$.

The same calculation can be reproduced for the $\langle \tilde z \bar{\tilde z}\rangle$ two-point function. Taking into account that $\delta_{\tilde \varphi}=\delta_{\tilde z}$ and $\delta_{\varphi}=\delta_{z}$ the total contribution is
\begin{equation}
\label{eqn:counterresult1}
     \left( \langle z(2\pi)\bar z(0) \rangle+\langle \tilde z(2\pi) \bar{\tilde z}(0)\rangle\right)^{\text{ct}}=\left(\delta_f+\delta_{\bar f} -  3\delta_z -3\delta_{\tilde z} \right)\langle W \rangle^{(1)}\,,
\end{equation}
where $\langle W \rangle^{(1)}$ is the one-loop contribution to circular Wilson loops \eqref{eqn:1l-circular}.

According to expansion \eqref{eq:2ptexp}, at two loops we have extra finite contributions coming from the product of the one-loop counterterms (set $\delta^{(1)}_z \equiv \delta_z, \delta_{\tilde z}$ there) multiplying the one-loop ${\cal O}(\epsilon)$ two-point functions 
\begin{equation}\label{eqn:counterresult2}
   \delta_z \langle z(2\pi)\bar z(0) \rangle^{(1)} + \delta_{\tilde z}\langle \tilde z(2\pi) \bar{\tilde z}(0) \rangle^{(1)} = \left( \delta_z +\delta_{\tilde z} \right)\langle W \rangle^{(1)} \,.
\end{equation}
Therefore, summing \eqref{eqn:counterresult1} and \eqref{eqn:counterresult2}, the final contribution to the Wilson loop VEV from the counterterms is
\begin{equation}
    \cC= -g^4\frac{N_1N_2}{4}(N_1+N_2)\left[ (\bar\alpha^i\alpha_i+\beta^j\bar\beta_j)^2-\bar\alpha^i\alpha_i+\beta^j\bar\beta_j \right]\,.
\end{equation}


\subsection{The final result for the Wilson loop VEV}
\label{sec:Wresults}

Combining all the previous results, ${\cal B}+{\cal F}+{\cal C}$, the large $N_1,N_2$ expectation value for the parametric circular $1/24$ BPS Wilson loop at two loops reads
\begin{equation}
\label{eqn:result1}
    \langle \cW_{1/24}(\alpha_i,\bar\alpha^i,\beta^j,\bar\beta_j)  \rangle = 1 -\frac{g^4}{24}\left[ N_1^2+N_2^2-4N_1N_2-3N_1N_2(\bar\alpha^i\alpha_i-\beta^j\bar\beta_j -1)^2 \right] + \cO(g^6) \,,
\end{equation}
where we recall that we have defined $g=\sqrt{2\pi/k}$. 

By setting $\beta^j=\bar\beta_j=0$ or $\bar\alpha^i=\alpha_i=0$, one recovers two branches of interpolating $1/6$ BPS fermionic Wilson loops, which have the following VEVs
\begin{equation}
\label{eqn:1/6result}
\begin{split}
    \langle \cW_{1/6}^\textrm{I}(\alpha_i,\bar\alpha^i) \rangle &= 1 -\frac{g^4}{24}\left[ N_1^2+N_2^2-4N_1N_2-3N_1N_2(\bar\alpha^i\alpha_i -1)^2 \right] + \cO(g^6) \,,\\
    \langle \cW_{1/6}^\textrm{II}(\beta^j,\bar\beta_j )\rangle &= 1 -\frac{g^4}{24}\left[ N_1^2+N_2^2-4N_1N_2-3N_1N_2(\beta^j\bar\beta_j+ 1)^2 \right] + \cO(g^6) \,.
\end{split}
\end{equation}
We recall that, according to the classification in \cite{Ouyang:2015iza,Ouyang:2015bmy}, ``type I" and ``type II" $1/6$ BPS fermionic Wilson loops differ by the preserved $SU(2) \subset SU(4)$ R-symmetry group. 

If in particular we choose $\bar{\alpha}^i \alpha_i = 1$ in $\cW_{1/6}^\textrm{I}(\alpha_i,\bar\alpha^i)$ or $\beta^j \bar{\beta}_j =-1$ in $\cW_{1/6}^\textrm{II}(\beta^j,\bar\beta_j )$, we recover the known result for  1/2 BPS operators \cite{Bianchi:2013zda,Bianchi:2013rma,Griguolo_2013a}, at two loops and in the large $N_1, N_2$ limit:
\begin{equation}\label{eq:W12}
\begin{split}
\langle \cW^\textrm{I}_{1/2}\rangle =
\langle\cW^\textrm{II}_{1/2}\rangle 
= 1 -\frac{g^4}{24}\left( N_1^2+N_2^2-4N_1N_2\right) + \cO(g^6)\,.
\end{split}
\end{equation}

Finally, if we set all the parameters to zero we obtain the two-loop expectation value of the bosonic operator \cite{Drukker:2008zx,Chen:2008bp,Kluson:2008zrv,Rey:2008bh}
\begin{equation}\label{eq:W+}
   \langle \cW_{1/6}^{\rm bos}\rangle \equiv 
   \left\langle  \frac{N_1 {\cal W}^{\rm bos} + N_2 \hat{\cal W}^{\rm bos}}{N_1+N_2} \right\rangle = 1 - \frac{g^4}{24} \left( N_1^2+N_2^2-7N_1N_2 \right) + \cO(g^6).
\end{equation}
where ${\cal W}^{\rm bos}, \hat{\cal W}^{\rm bos}$ are the bosonic operators defined in \eqref{eqn:bosonicconnections}.

We have  evaluated the Wilson loop VEVs exploiting the one-dimensional auxiliary field formulation. Alternatively, one could use the ordinary procedure of expanding $W$ in powers of the superconnection  and evaluate correlation functions of ${\cal L}$. We have checked that proceeding in this way, once we replace bare parameters with their renormalized expressions found above, the final result
coincides with \eqref{eqn:result1}. This is a non-trivial check of our procedure. 

\vspace{0.2cm}
\section{Discussion}
\label{sec:results}

\subsection{Renormalization Group flows}

In section \ref{sec:betafunction} we have shown that the introduction of the weakly relevant couplings $\alpha_i,\bar\alpha^i,\beta^j,\bar\beta_j$ triggers a RG flow driven by the eight $\beta$-functions \eqref{eqn:betafunction}. Here we study this flow by focusing on the ``type I" operators defined above. The study of RG flows involving ``type II" Wilson loops will be presented elsewhere \cite{CPTT}.

\begin{figure}[H]
        \centering
    \includegraphics[width=0.50\textwidth]{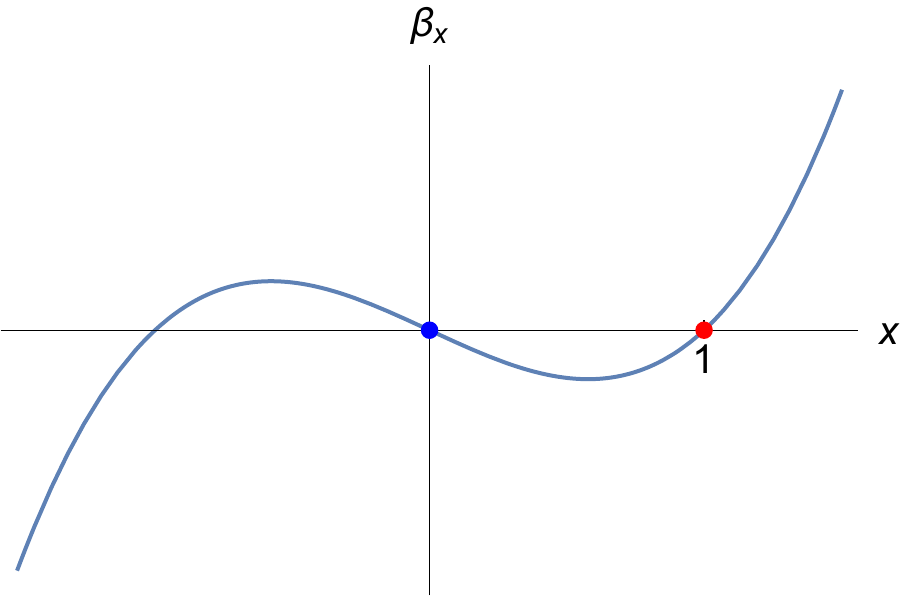}
    \caption{Plot of $\beta_x$ when $y=0$. To make contact with figure \ref{fig:RGflow}, we have highlighted the fixed points.}
    \label{fig:beta}    
\end{figure}

In order to give an intuitive visual description of the RG flow we define $\bar\alpha^1=\alpha_1=x$ and $\beta^3=\bar\beta_3=y$ and set the other parameters to zero. The relevant $\beta$-functions are then
\begin{equation}
\begin{split}\label{eq:betaxy}
    \beta_{x}(x,y) &= \mu\frac{\partial x}{\partial\mu} = \frac{g^2}{4\pi}(N_1+N_2)(x^2+y^2-1) \, x\,, \\
    \beta_{y}(x,y) &= \mu\frac{\partial y}{\partial\mu} = \frac{g^2}{4\pi}(N_1+N_2)(x^2+y^2+1) \, y \,.
\end{split}
\end{equation}
In figure  \ref{fig:beta} we plot $\beta_x(x,y=0)$ and highlight the zeros of the $\beta_x$-function.

\begin{figure}[h]
    \centering
    \includegraphics[width=0.5\textwidth]{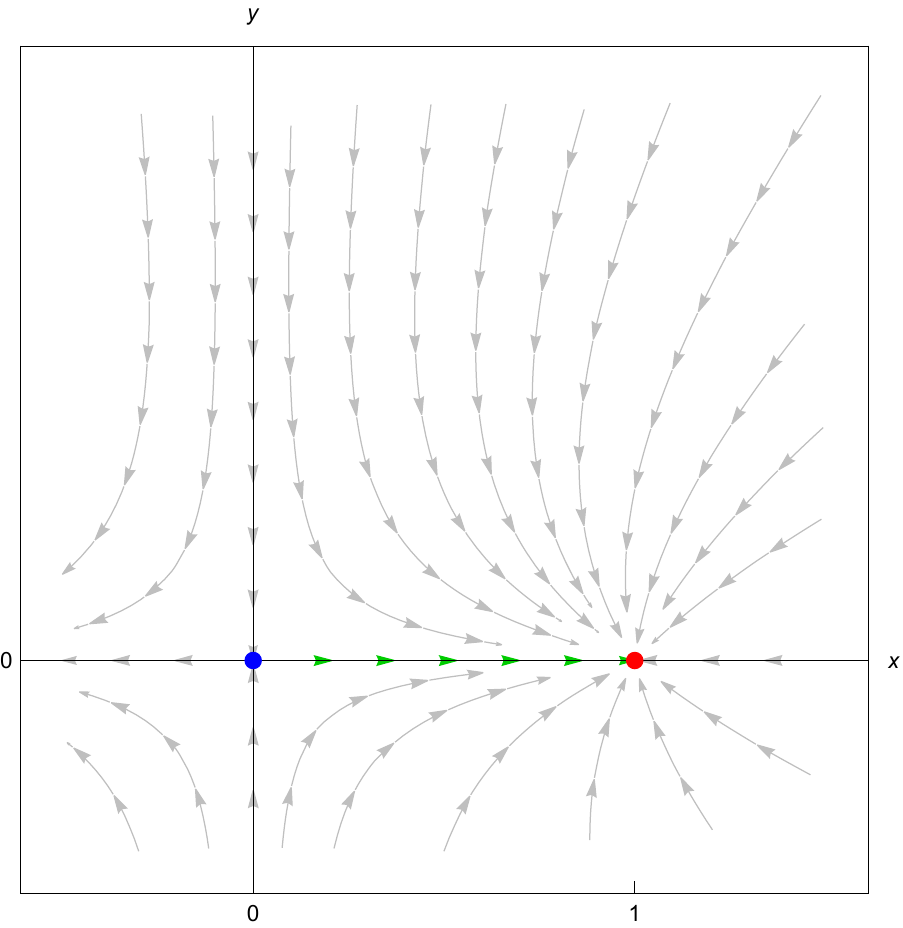}
    \caption{The RG flow in the $(x,y)$ plane. Arrows go from the UV to the IR. Arrows on the $x,y$ axes correspond to $1/6$ BPS flows, while arrows outside the $x,y$ axes correspond to $1/24$ BPS flows. Horizontal green arrows correspond to the RG flow between the bosonic $1/6$ BPS Wilson loop (blue dot) and ``type I" fermionic $1/2$ BPS Wilson loop described by $\langle \cW^\textrm{I}_{1/6}(\alpha_i,\bar\alpha^i) \rangle$ at $\bar\alpha^i\alpha_i=1$ (red dot). }
    \label{fig:RGflow}
\end{figure}

In figure \ref{fig:RGflow} we draw the RG flow trajectories in the $(x,y)$ plane, for $x, y$ real and non-negative. For negative $x$ we would obtain an equivalent fixed point. The blue point in figures \ref{fig:beta}, \ref{fig:RGflow} corresponds to the $x=y=0$ fixed point where the associated operator is the bosonic $1/6$ BPS Wilson loop ${\cal W}_{1/6}^{\rm bos}$ defined in \eqref{eq:W+}. This is a UV fixed point where the parameters trigger an outgoing flow. Moving along the horizontal green line in figure \ref{fig:RGflow} we reach an IR fixed point (highlighted in red) corresponding to ``type I" fermionic $1/2$ BPS Wilson loop obtained from $\cW^\textrm{I}_{1/6}$ by selecting $\bar\alpha^i\alpha_i=1$.

The green line describes an enriched RG flow between UV and IR fixed points, along which supersymmetry is partially preserved. In fact, all the points on the two axes, even those not highlighted in green, correspond to operators which preserve four supercharges.

Similarly, flows along a generic direction in the plane preserve one supercharge, corresponding to the $1/24$ BPS operator, whose VEV is given in \eqref{eqn:result1}. In this sense, they can still be interpreted as enriched RG flows.

\begin{figure}[h]
    \centering
    \includegraphics[width=0.5\textwidth]{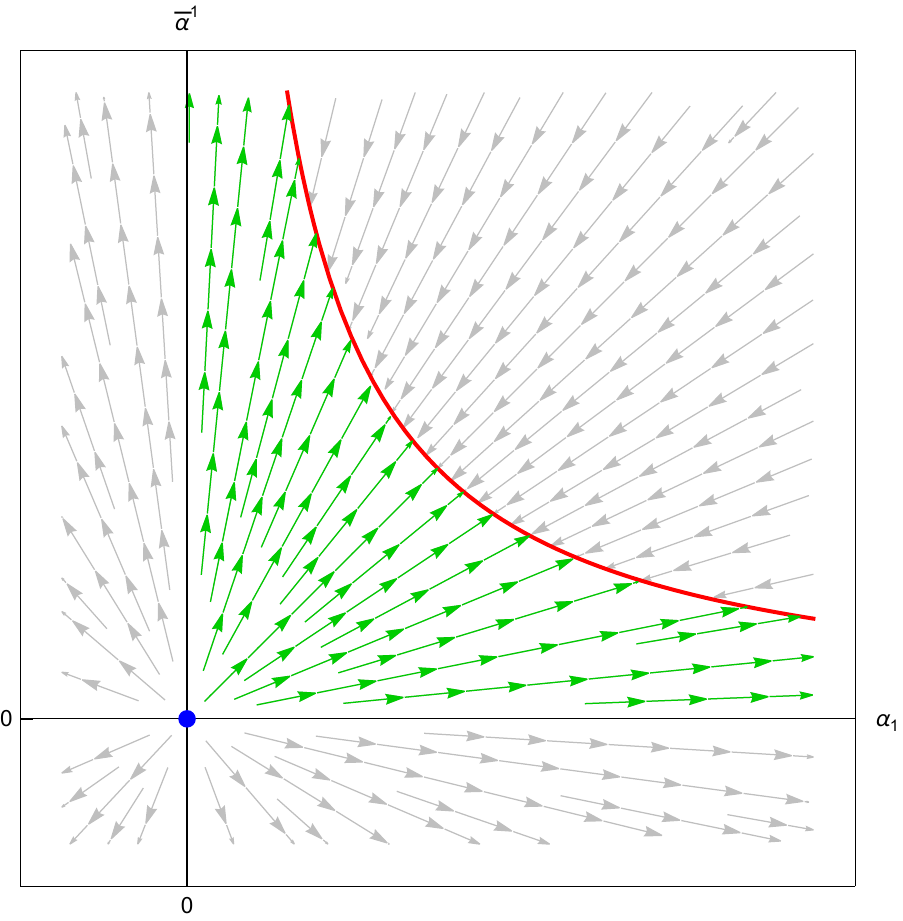}
    \caption{The RG flow in the $(\alpha_1,\bar\alpha^1)$ plane. Arrows go from the UV to the IR. The red curve corresponds to $\alpha_1\bar\alpha^1=1$ and the blue dot is the bosonic $1/6$ BPS Wilson loop.}
    \label{fig:RGflowAlpha}
\end{figure}

More generally, still setting $y=0$, we relax the condition $\bar\alpha^1=\alpha_1=x$, and consider the flow in the $(\alpha_1,\bar\alpha^1)$ plane, as presented in figure \ref{fig:RGflowAlpha}. As expected, the $1/2$ BPS curve $\alpha_1\bar\alpha^1=1$ corresponds to a set of attractive points. This is in agreement with the analysis of figure \ref{fig:RGflow} where the red dot is also attractive and green trajectories connect fixed points.

\vspace{0.2cm}


\subsection{The defect SQFT}

Non-local operators like Wilson loops can be used to describe one-dimensional defect quantum field theories (dQFTs).\footnote{For a quite exhaustive list of references on linear defects, see for instance \cite{Penati:2021tfj}.} In particular, if the operator preserves the one-dimensional conformal algebra $sl(2,{\mathbb R})$, it defines a defect conformal field theory (dCFT). In addition, if the operator is BPS and preserves a sufficient amount of supersymmetry, the corresponding defect is a superconformal field theory (dSCFT). 

Focusing on the set of ABJ(M) circular Wilson loops considered in this paper, it is well known that the 1/6 and 1/2 BPS fermionic operators preserve the $su(1,1|1)$ and $su(1,1|3)$ one-dimensional superconformal algebras, respectively. Therefore, they describe superconformal defects. Instead, the new 1/24 BPS operator defined in (\ref{eqn:1/24-superconnection})-(\ref{eqn:M124bos}) describes a supersymmetric, but not (super)conformal defect, as it preserves only one supercharge and the dependence of the scalar couplings on the contour coordinate breaks conformal invariance.

Regardless of their superconformal or only supersymmetric nature, the dQFTs  supported by ABJ(M) fermionic Wilson loops are generated by local operators defined by $U(N_1|N_2)$ supermatrices localized on the Wilson loop. 
The dQFT is featured by the set of correlation functions of these local operators, inserted on the Wilson loop vacuum. Precisely, the defect $n$-point function of a set of local operators inserted at points $\tau_1, \ldots , \tau_n$ along the circle is defined as 
\begin{equation}
    \llangle \Tr\left( O_n O_{n-1}\ldots O_1 \right) \rrangle \equiv \frac{\left\langle \Tr\cP\left( e^{-i\int_{\tau_n}^{2\pi}d\tau \cL(\tau)} O_n e^{-i\int_{\tau_{n-1}}^{\tau_n}d\tau\cL(\tau)} O_{n-1}\ldots  O_1 e^{-i\int_{0}^{\tau_1}d\tau\cL(\tau)} \right)\right\rangle}{\langle {\cal W} \rangle}\,,
\end{equation}
where in the right hand side the expectation value is on the ABJ(M) vacuum. The insertion of Wilson links ensures gauge invariance.
In the one-dimensional auxiliary field formalism introduced in section \ref{sec:1dtheory} the defect $n$-point function can be written entirely in terms of ABJ(M) expectation values of products of supermatrices localized on the contour
\begin{equation}
    \llangle \Tr\left( O_n O_{n-1}\ldots O_1 \right) \rrangle \equiv\frac{1}{2^{n+1}} \frac{\langle \Tr\left( \Psi(2\pi)\bar\Psi(\tau_n) O_n \Psi(\tau_n)\bar\Psi(\tau_{n-1}) O_{n-1}\ldots O_1 \Psi(\tau_1)\bar\Psi(0) \right) \rangle}{\langle \Tr \Psi(2\pi)\bar\Psi(0) \rangle}\,.
\end{equation}

In this context, the $(x, y)$ plane depicted in figure \ref{fig:RGflow} has the nice interpretation of describing different defect theories, with fixed points corresponding to theories at their critical point. The blue point is a UV unstable fixed point corresponding to a one-dimensional ${\cal N}=1$ SCFT. The red point represents an interacting IR critical theory where supersymmetry gets enhanced to ${\cal N}=3$.

The Wilson loop RG flows that we have found are interpreted as flows in the space of one-dimensional defects. The two axes describe a continuum of one-dimensional SCFTs. Along these two directions the enriched flow preserves ${\cal N}=1$ superconformal invariance. As soon as we move out of the two axes, superconformal invariance is broken, although one supercharge is still preserved. Nevertheless  along all of these flows the system is driven towards the IR fixed point corresponding to $\cW_{1/2}^\textrm{I}$.
 
It is important to give a closer look at the weakly relevant operators which perturb the system and drive it away from the UV fixed point. To this end, we recall that the UV fixed point corresponds to the bosonic Wilson operator ${\cal W}_{1/6}^{\rm bos}$ obtained by setting all the parameters to zero.  Moving along the two green lines in figure \ref{fig:RGflow} amounts to adding a deforming operator as ${\cal L}_{1/6}^{\rm bos} \rightarrow {\cal L}_{1/6}^{\rm bos} + {\cal L}^{\rm def}$. For instance, if we move along the horizontal axis, which amounts to setting $\beta^j = \bar\beta_j = 0$, and choose for simplicity $\alpha_1=\bar\alpha^1=0$, we have
\begin{equation}\label{eq:deformation}
  {\cal L}^{\rm def} = \bar\alpha^2 \alpha_2 \begin{pmatrix} -\frac{4\pi i }{k} C_2 \bar{C}^2 & 0 \\ 0 & -\frac{4\pi i }{k} \bar{C}^2 C_2  \end{pmatrix} - \bar\alpha^2\,e^{i\tau/2} \begin{pmatrix} 0 & \eta \bar\psi^1 \\ 0 & 0 \end{pmatrix} - \alpha_2\,e^{-i\tau/2} \begin{pmatrix} 0 & 0 \\ \xi \psi_1 & 0 \end{pmatrix}\,.
 \end{equation}
 
We are interested in computing the anomalous dimensions of these operators. This amounts to computing their two-point functions in the bosonic $1/6$ BPS defect. 

Focusing on the two fermionic operators, it is easy to see that their integrated two-point function can be expressed as 
\begin{multline}
\label{eqn:2pt2}
         \int_0^{2\pi} d\tau_1\int_0^{\tau_1} d\tau_2 \left[ \frac{e^{i\frac{\tau_{12}}{2}}\llangle  (\eta\bar\psi^1)(\tau_1) (\xi\psi_1)(\tau_2)  \rrangle +
     e^{-i\frac{\tau_{12}}{2}}\llangle  (\xi\psi_1)(\tau_1) (\eta\bar\psi^1)(\tau_2)   \rrangle}{N_1+N_2} \right]_{\alpha_2,\bar\alpha^2=0}= \\
      \qquad = - \frac{\partial^2}{\partial \alpha_2\partial\bar\alpha^2}\log\langle {\cal W}^{I}_{1/6}(\alpha_2, \bar\alpha^2) \rangle\Bigg|_{\alpha_2,\bar\alpha^2=0} \,.
\end{multline}
If the operators develop an anomalous dimension $\gamma$, the left hand side of this equation formally becomes 
\begin{equation}
\label{eqn:relation1}
\begin{split}
    -\frac{g^2}{\pi}\frac{N_1N_2}{N_1+N_2}\int_0^{2\pi}\!\! \! \! d\tau_1\int_0^{\tau_1}\! \! \! \! d\tau_2 \, \frac{1}{|4\sin^2\frac{\tau_{12}}{2}|^{1+\gamma}} 
 & =- \frac{\sqrt{\pi}g^2}{2}\frac{ N_1N_2}{N_1+N_2}\frac{\Gamma(-\frac{1}{2}-\gamma)}{\Gamma(-\gamma)} \\&\isEquivTo{\vert\gamma\vert\ll 1} -g^2\pi \frac{N_1N_2}{N_1+N_2}\gamma\,.
\end{split}
\end{equation}
On the other hand, the right hand side of 
\eqref{eqn:2pt2} can be easily evaluated observing that the two-loop result satisfies the following identity
\begin{equation}\label{eq:identity}
    \frac{\partial}{\partial \bar\alpha^2} \log\langle {\cal W}^{I}_{1/6}(\alpha_2, \bar\alpha^2) \rangle = \kappa \,  \beta_{\alpha_2} \, , \qquad \text{with }\qquad \kappa = \pi g^2 \frac{N_1N_2}{N_1+N_2}\,.
\end{equation}
Therefore, comparing the two expressions we finally obtain
\begin{equation}
    \gamma =  \frac{\partial \beta_{\alpha_2}}{\partial \alpha_2}\Bigg|_{\alpha_2,\bar\alpha^2=0} = -\frac{g^2}{4\pi} (N_1+N_2)\,.
\end{equation}
This result can also be checked by explicitly computing the first order correction to the two-point function in \eqref{eqn:2pt2}. 

We have found that the first contribution to the anomalous dimension of the fermionic fields is negative. With a similar reasoning, one can check that also the bi-scalar operator $C_2 \bar{C}^2$ acquires negative anomalous dimension. This confirms that the deformation \eqref{eq:deformation} is a weakly relevant operator. 

More generally, we can compute the $\psi_1, \bar{\psi}^1$ anomalous dimension in the ${\cal W}^I_{1/6}(\alpha_2,\bar{\alpha}^2)$ defect. This amounts to evaluating the derivative of the $\beta$-function without fixing the values of the parameters. We easily find
\begin{equation}
    \gamma(\alpha_2, \bar\alpha^2) =  \frac{\partial \beta_{\alpha_2}}{\partial \alpha_2}= \frac{g^2}{4\pi} (N_1+N_2)(2 \bar\alpha^2\alpha_2  -1)\,.
\end{equation}
This interpolates between the dimension of the weakly relevant operator in the UV and its dimension in the IR.

\vspace{0.2cm}


\subsection{A g-theorem}

Focusing on $1/6$ BPS flow along the green line we now prove the validity of a g-theorem. In order to keep the discussion simpler we again set $\alpha_1=\bar\alpha^1 = x$ and $\beta^3=\bar\beta_3 = y$ with other parameters set to zero.

Referring to the $\beta$-function as written in \eqref{eq:betaxy}, in the $x\in[0,1]$ region we can write
\begin{equation}
    \frac{\partial}{\partial x}\log \langle \cW^\textrm{I}_{1/6}(x) \rangle =2 \kappa \, \beta_x\Big|_{y=0} \,.
\end{equation}
where $\kappa$ has been defined in \eqref{eq:identity}.
First of all, this implies that the blue and the red conformal fixed points in figure \ref{fig:RGflow} are extrema of $\langle \cW_{1/24}(\alpha_i,\bar\alpha^i,\beta^j,\bar\beta_j) \rangle$. Moreover, it is easy to show that the (red) $1/2$ BPS point is a minimum while the (blue) bosonic $1/6$ BPS point is a maximum. Therefore, comparing the two fixed points connected by the horizontal green line in figure \ref{fig:RGflow}, we can write
\begin{equation}\label{eqn:gtheorem}
    \log\langle \cW_{1/6}^\textrm{bos}\rangle\equiv\log\langle \cW_{1/6}^\textrm{I}(x=0) \rangle>\log\langle \cW_{1/6}^\textrm{I}(x=1) \rangle\equiv\log\langle \cW^\textrm{I}_{1/2} \rangle \, .
\end{equation}
This result can be interpreted as a g-theorem \cite{Cuomo:2021rkm} for the one-dimensional defect. In fact, defining the interpolating functions $g^\textrm{I}=\langle \cW^\textrm{I}_{1/6} \rangle$, we find $g^\textrm{I}_{UV}>g^\textrm{I}_{IR}$. Recalling that $\log \langle \cW \rangle$ is nothing but the partition function of the one-dimensional defect, monotonicity is consistent with the decreasing of
degrees of freedom from the UV to the IR fixed point. 
Our result is in line with what has been already found in $\cN=4$ super Yang-Mills \cite{Beccaria:2017rbe,Beccaria:2018ocq}, although in a different setup, as it consists of BPS flows.
\vspace{0.2cm}

\subsection{Comparison with the localization result}

It is well known that in supersymmetric theories defined on compact manifolds BPS Wilson loops can be computed using supersymmetric localization \cite{Pestun_2012}. This provides a representation of the path integral evaluating the Wilson loop VEV  as a matrix integral. 

For the ABJ(M) theory on $S^3$ localization allows to exactly compute the VEV of the bosonic Wilson loops in  \eqref{eqn:bosonicconnections} as the expectation values 
\begin{equation}\label{eq:bosonicVEVloc}
    \langle {\mathcal W}^{\rm bos} \rangle_{1} = \left\langle \frac{1}{N_1}\sum_{i=1}^{N_1} e^{2\pi \lambda_i} \right\rangle_{_\cZ} \,, \qquad \langle \hat{\cal W}^{\rm bos} \rangle_{1} = \left\langle \frac{1}{N_2}\sum_{i=1}^{N_2} e^{2\pi \hat \lambda_i} \right\rangle_{_\cZ} \,,
\end{equation}
where the average is evaluated and normalized using the following non-Gaussian matrix model \cite{Kapustin:2009kz}
\begin{equation}
    \cZ = \int \prod_{i=1}^{N_1} d\lambda_i e^{i\pi k \lambda^2_i} \prod^{N_2}_{j=1} d\hat \lambda_j e^{-i\pi k \hat \lambda^2_j} \frac{\prod_{i<j}^{N_1} \sinh^2(\pi(\lambda_i-\lambda_j))\prod_{i<j}^{N_2} \sinh^2(\pi(\hat\lambda_i-\hat\lambda_j))}{\prod_{i=1}^{N_1}\prod_{j=1}^{N_2} \cosh^2(\pi(\lambda_i-\hat\lambda_j))}\,.
\end{equation}
Here the integrations are on two complete sets of eigenvalues $\{\lambda_i \}$, $\{\hat\lambda_j \}$ of the Cartan subalgebras of $U(N_1)$ and $U(N_2)$, respectively. The ``1'' subscript in \eqref{eq:bosonicVEVloc} indicates that  the matrix model computes the expectation values at framing $f=1$ \cite{Kapustin:2009kz}.\footnote{See also \cite{Bianchi:2013rma,Bianchi:2016yzj} and chapter 6 of \cite{Drukker:2019bev} for an introductory discussion to framing in three-dimensional Chern-Simons-matter theories. As we have emphasized throughout this paper, the dimensional regularization used in section \ref{sec:WLVEV} is alternative to framing regularization, therefore the perturbative results obtained in this paper correspond to the $f=0$ scheme.} 

The main observation is that the prescription \eqref{eq:bosonicVEVloc} automatically provides exact results for the whole class of BPS Wilson loops that we have considered in this paper. This stems from the fact that, classically, the 1/24 BPS, the 1/6 BPS fermionic and the 1/2 BPS Wilson loops are all cohomologically equivalent to the linear combination ${\cal W}_{1/6}^{\rm bos}$ defined in \eqref{eq:W+}. 
In other words, they differ from ${\cal W}_{1/6}^{\rm bos}$ by a ${\cal Q}$-exact term, where ${\cal Q}$ is one of the supercharges preserved by all the operators in the game. Therefore, if cohomological equivalence is preserved at the quantum level, one can in principle use this ${\cal Q}$ to localize the path integral for the Wilson loop VEV. As a consequence, the following identities hold
\begin{equation}\label{eq:Widentities}
    \langle \cW_{1/24}(\alpha_i,\bar\alpha^i,\beta^j,\bar\beta_j) \rangle_1 
    = \langle \cW_{1/6}^\textrm{I}(\alpha_i,\bar\alpha^i)  \rangle_1 = \langle \cW_{1/6}^\textrm{II}(\beta^j,\bar\beta_j)  \rangle_1 = \langle \cW_{1/2}^\textrm{I,II}\rangle_1 = \langle {\cal W}_{1/6}^{\rm bos} \rangle_1\,.
\end{equation}
At framing one, all the VEVs must equal  $\langle {\cal W}_{1/6}^{\rm bos} \rangle_1 = (N_1 \langle {\cal W}^{\rm bos} \rangle + N_2 \langle\hat{\cal W}^{\rm bos}\rangle)/(N_1+N_2)$, which can be easily evaluated from \eqref{eq:bosonicVEVloc}.
At weak coupling, this quantity is known both from the matrix model expansion  \cite{Kapustin:2009kz} and from perturbation theory \cite{Mauri_2018}. Up to two loops it reads
\begin{equation}
\label{eqn:resultR}
    \langle {\cal W}_{1/6}^{\rm bos} \rangle_{1} = 1+\frac{i\pi(N_1-N_2)}{k} -\frac{\pi^2}{6k^2}\left[ 4(N_1^2+N_2^2) -10N_1N_2-1 \right]\,.
\end{equation}

Therefore, as a consequence of identities \eqref{eq:Widentities}, at $f=1$ the VEVs loose any dependence on the alpha and beta parameters.
In other words, all the points of the plot in figure \ref{fig:RGflow} correspond to the same quantum operator. It is then interesting to understand how the parameter dependence arises when the expectation values are evaluated at $f \neq 1$, in particular at framing zero. 

For 1/6 BPS bosonic and 1/2 BPS fermionic Wilson loops, a relation between their expectation values at framing zero computed perturbatively and the ones at framing one coming from the matrix model has been found \cite{Drukker:2009hy, Bianchi_2016}. Up to two loops, these read
\begin{equation}\label{eq:framing10}
    \langle {\cal W}^{\rm bos} \rangle_1 = e^{\frac{i\pi N_1}{k}}\langle {\cal W}^{\rm bos} \rangle_{0}
    \, , \quad 
    \langle \hat {\cal W}^{\rm bos} \rangle_1 = e^{-\frac{i\pi N_2}{k}}\langle \hat {\cal W}^{\rm bos} \rangle_{0} \, , \quad
    \langle {\cal W}^\textrm{I,II}_{1/2} \rangle_{1} = e^{\frac{i\pi (N_1-N_2)}{k}}\langle {\cal W}_{1/2} \rangle_{0}\,.
\end{equation}
They can be generalized to generic (also non-integer) framing $f$ in a rather simple way, see the discussion in \cite{Bianchi:2014laa}. 

Since the interpolating operators under investigation have a non-trivial parametric dependence at framing zero, but loose this dependence at framing one, we expect the parameter dependence to be carried by ``phase'' factors, in analogy with \eqref{eq:framing10}.
Perturbative analysis reveals that up to two loops and in the large $N_1,N_2$ limit, the matrix model results (\ref{eq:Widentities})-(\ref{eqn:resultR}) are related to the perturbative ones in section \ref{sec:Wresults} as follows. For the fermionic 1/6 BPS operators we are led to the following identities
\begin{equation}
\begin{split}\label{eqn:MMrelation2}
   & \langle {\cal W}_{1/6}^\textrm{I}(\bar\alpha^i,\alpha_i) \rangle_1 =\frac{N_1e^{\frac{i\pi}{k}(N_1-\bar\alpha^i\alpha_i N_2)}+N_2\, e^{\frac{i\pi}{k}(\bar\alpha^i\alpha_i N_1-N_2)}}{N_1+N_2} \langle {\cal W}_{1/6}^\textrm{I}(\bar\alpha^i,\alpha_i) \rangle_0\,,\\
    & \langle {\cal W}_{1/6}^\textrm{II}(\beta^j,\bar\beta_j) \rangle_1 =\frac{N_1e^{\frac{i\pi}{k}(N_1+\beta^j\bar\beta_j N_2)}+N_2\, e^{\frac{i\pi}{k}(-\beta^j\bar\beta_j N_1-N_2)}}{N_1+N_2} \langle {\cal W}_{1/6}^\textrm{II}(\beta^j,\bar\beta_j) \rangle_0\,,
\end{split}
\end{equation}
whereas for the more general $1/24$ BPS operator the relation it reads
\bea
\label{eqn:MMrelation}
   \langle {\cal W}_{1/24}
   \rangle_1 = \frac{N_1 e^{\frac{i\pi}{k}(N_1-(\bar\alpha^i\alpha_i-\beta^j\bar\beta_j)N_2)}+N_2 \, e^{\frac{i\pi}{k}((\bar\alpha^i\alpha_i-\beta^j\bar\beta_j)N_1-N_2)}}{N_1+N_2}\langle {\cal W}_{1/24}
   \rangle_0\,.
\eea
In the ABJM limit, $N_1=N_2$, it boils down to 
\begin{equation}
    \langle {\cal W}_{1/24}(\bar\alpha^i,\alpha_i,
    \beta^j,\bar\beta_j) \rangle_1 = \cos\left( \tfrac{\pi N}{k}(1-\bar\alpha^i\alpha_i+\beta^j\bar\beta_j) \right)\langle {\cal W}_{1/24}(\bar\alpha^i,\alpha_i,\beta^j,\bar\beta_j) \rangle_0\,.
\end{equation}
Similar relations come from \eqref{eqn:MMrelation2} for $N_1=N_2$. 

We note that the exponentials carrying the parameter dependence are no longer pure phases as in \eqref{eq:framing10}, since the parameters can be generically complex and the barred parameters are not the complex conjugates. However, for the special values $\bar{\alpha}^i \alpha_i=1$ and $\beta^j \bar{\beta}_j=0$ they reduce to the last phase in \eqref{eq:framing10}.

These identities have been empirically inferred from the two-loop results and are not expected to be true in general. In fact, the parametric exponents are likely to be corrected at higher orders, as already happens at three loops for the ${\cal W}^{\rm bos}, \hat{\cal W}^{\rm bos}$ phases \cite{Bianchi:2016yzj}.

We close this section with a technical observation on the integrals in the two schemes, framing or dimensional regularization ($f=0$). The parameter independence of the framing-one results indicates that a genuine perturbative calculation done at $f=1$ should sensibly differ from our present calculation done using dimensional regularization. In particular, new non-vanishing contributions should arise to compensate the parameter dependence carried by diagrams that are framing independent. For instance, let us focus on the contributions proportional to $\bar\alpha^i\alpha_i\beta^j\bar\beta_j$. At two loops and framing zero, they come from the scalar diagram \ref{subfig:2-loop_bosc} and the two fermion ones, \ref{subfig:2-loop_fer2a} and \ref{subfig:2-loop_fer2b}.  As shown in \cite{Mauri_2018}, the scalar integral is framing independent, thus its dependence on the parameters should survive also at framing one. On the other hand, it has been argued in \cite{Bianchi_2016} that at $f=1$ the fermionic diagrams should be identically vanishing. Therefore, at framing one some new parameter dependent contribution should arise, which eventually cancels the scalar diagram. A proof of this statement would require performing a genuine two-loop calculation at framing one, though this might be obstructed by the difficulty of computing fermionic diagrams at non-trivial framing.

\vspace{0.2cm}

\subsection{Outlook}
\label{sec:conclusion}

There are several natural directions in which these investigations could be extended. 


As stressed repeatedly, the flows considered in this paper are BPS, with at least one supercharge preserved at all points of the flow. It would of course be interesting to introduce a $\zeta$-parameter in the 1/24 BPS Wilson loop, to interpolate to a fully non-supersymmetric limit, like it has been done in \cite{Polchinski_2011} for the 1/2 BPS circular Wilson loop of $\cN=4$ super Yang-Mills in four dimensions. This could be achieved by rescaling the overall couplings to the scalars and the fermions in \eqref{eq:superconnection} and
would make the RG flow space even richer, opening up a new direction corresponding to the renormalization of the new parameter. We plan to address this in a future investigation.

As usual, there is always the question of the holographic dual description in terms of minimal surfaces in M-theory or type IIA superstring theory. The 1/24 BPS Wilson loop considered in this paper should be described by mixed boundary conditions in $AdS_2$, generalizing what has been done in \cite{Correa:2019rdk,Garay:2022szq} for deformations of the 1/6 BPS bosonic operator defined on a straight line. For instance, it would interesting to understand whether the set of boundary conditions that do not preserve conformal invariance -- the reason why they were not further discussed in \cite{Correa:2019rdk,Garay:2022szq} -- may play a role in this context.

We should stress that the generalization of this approach to the present case would require to first adapt it to the circular case, where a conformal anomaly \cite{gross} makes operators, which are cohomologically equivalent at the classical level, no longer equivalent at the quantum level. In particular, as we have discussed, this causes a non-trivial dependence of the VEVs on the deformations. Therefore, in this case the mixing of Neumann and Dirichlet boundary conditions should entail a non-trivial parametric dependence in the interpolating string solutions. More generally, the question of what is, if any, the holographic counterpart of framing is quite important.

Finally, it is interesting to repeat this analysis for the 1/12 BPS latitude Wilson loop of section \ref{sec:parametriclatitude}. This is going to be addressed in \cite{CPTT}. In that case the one-dimensional effective field theory on the Wilson loop is modified by the presence of non-trivial shifts in the superconnection $\cL^\theta$, which result in mass terms for the one-dimensional fields. 

\section*{Acknowledgements}
We are grateful to Luca Griguolo and Domenico Seminara for discussions at the early stage of this work. We also thank Diego Correa, Alberto Faraggi and Guillermo Silva for useful correspondence. LC, SP and MT are partially supported by the INFN grant {\it Gauge Theories, Strings and Supergravity (GSS)}.
DT is supported in part by the INFN grant {\it Gauge and String Theory (GAST)}. DT would like to thank FAPESP’s partial support through the grants 2016/01343-7 and 2019/21281-4.

\newpage

\appendix

\section{Conventions and Feynman rules}
\label{app:abjm}

We follow the conventions in \cite{Bianchi:2014laa}. We work in three-dimensional Euclidean space with coordinates $x^{\mu}=(x^0,x^1,x^2)$. The three-dimensional gamma matrices are defined as
\begin{equation}\label{eq:gamma}
    (\gamma^{\mu})^{ \ \beta}_{ \alpha}=(-\sigma^3,\sigma^1,\sigma^2)_\alpha^{\ \beta}\,,
\end{equation}
with $(\sigma^{i} )^{ \ \beta}_{  \alpha}$ ($\alpha,\beta=1,2$) being the Pauli matrices, such that $\gamma^{\mu}\gamma^{\nu}=\delta^{\mu\nu}+i\epsilon^{\mu\nu\rho}\gamma_{\rho}$, where $\epsilon^{123}=\epsilon_{123}=1$ is totally antisymmetric. Spinorial indices are lowered and raised as $(\gamma^{\mu})^{\alpha}_{\ \beta}=\epsilon^{\alpha\gamma}(\gamma^{\mu})^{\ \delta}_{\gamma} \epsilon_{\beta\delta}$, with $\epsilon_{12}=-\epsilon^{12}=1$. The Euclidean action of $U(N_1)_k\times U(N_2)_{-k}$ ABJ(M) theory is
\begin{equation}\label{eq:ABJMaction}
\begin{split}
    S_{\textrm{ABJ(M)}}=&\frac{k}{4\pi} \int d^3 x\,  \epsilon^{\mu\nu\rho}\Big\{ -i\text{Tr}\left( A_{\mu}\partial_{\nu}A_{\rho} +\frac{2i}{3}A_{\mu}A_{\nu}A_{\rho} \right)+i\text{Tr}\left( \hat A_{\mu}\partial_{\nu}\hat A_{\rho} +\frac{2i}{3}\hat A_{\mu}\hat A_{\nu}\hat A_{\rho} \right) \\ & + \text{Tr}\left[ \frac{1}{\xi}(\partial_{\mu}A^{\mu})^2 - \frac{1}{\xi}(\partial_{\mu}\hat A^{\mu})^2 +\partial_{\mu} \bar c D^{\mu}c-\partial_{\mu} \bar{\hat c}D^{\mu}\hat c\right] \Big\} \\ & + \int d^3x \text{Tr}\left[ D_{\mu} C_I D^{\mu} \bar C^I +i\bar\psi^I \gamma^{\mu}D_{\mu} \psi_I \right]\\ & \begin{split}- \frac{2\pi i}{k}\int d^3 x \text{Tr}\Big[& \bar C^I C_I \psi_J \bar\psi^J - C_I \bar C^I \bar\psi^J \psi_J + 2C_I\bar C^J \bar\psi^I\psi_J \\ &-2\bar C^I C_J \psi_I \bar\psi^J - \epsilon_{IJKL}\bar C^I \bar \psi^J \bar C^K \bar \psi^L +\epsilon^{IJKL} C_I \psi_J C_K \psi_L \Big]+ S^{\text{bos}}_{\text{int}}\,,\end{split}
\end{split}
\end{equation}
with covariant derivatives defined as
\begin{equation}
\begin{split}
    &D_{\mu} C_I = \partial_{\mu} C_I +i A_{\mu} C_I -i C_I \hat A_{\mu}\,, \qquad D_{\mu} \bar C^I=\partial_{\mu} \bar C^I -i \bar C^I A_{\mu} + i\hat A_{\mu} \bar C^I\,, \\ & D_{\mu}\bar\psi^I=\partial_{\mu} \bar\psi^I + iA_{\mu} \bar\psi^I - i\bar\psi^I \hat A_{\mu}\,, \qquad D_{\mu} \psi_I =\partial_{\mu}\psi_I -i\psi_I A_{\mu} +i\hat A_{\mu} \psi_I\,.
\end{split}
\end{equation}
We work in Landau gauge for vector fields and in dimensional regularization with $d=3-2\epsilon$. The tree-level propagators are (with $g=\sqrt{2\pi/k}$)
\begin{equation}
\label{eqn:propagator}
    \begin{split}
        \langle (A_{\mu})_p^{\ q}(x)(A_{\nu})_r^{\ s}(y)\rangle^{(0)} &=\delta_p^s\delta_r^q\, i g^2 \, \frac{\Gamma(\frac{3}{2}-\epsilon)}{2\pi^{\frac{3}{2}-\epsilon}}\frac{\epsilon_{\mu\nu\rho}(x-y)^{\rho}}{|x-y|^{3-2\epsilon}},\\
        \langle (\hat A_{\mu})_{\hat p}^{\ \hat q}(x)(\hat A_{\nu})_{\hat r}^{\ \hat s}(y)\rangle^{(0)} &=-\delta_{\hat p}^{\hat s}\delta_{\hat r}^{\hat q} \, i g^2 \, \frac{\Gamma(\frac{3}{2}-\epsilon)}{2\pi^{\frac{3}{2}-\epsilon}}\frac{\epsilon_{\mu\nu\rho}(x-y)^{\rho}}{|x-y|^{3-2\epsilon}},\\
        \langle  (\psi_I^{\alpha})_{\hat i}^j(x) (\bar\psi_{\beta}^J)_k^{\hat l} (y) \rangle^{(0)} & = -i\delta_I^J\delta_i^{\hat l}\delta_k^{j } \frac{\Gamma(\frac{3}{2}-\epsilon)}{2\pi^{\frac{3}{2}-\epsilon}}\frac{(\gamma_{\mu})^{\alpha}_{\ \beta}(x-y)^{\mu}}{|x-y|^{3-2\epsilon}}\\ & =i\delta_I^J\delta_i^{\hat l}\delta_k^{j } (\gamma_{\mu})^{\alpha}_{\ \beta}\partial_{\mu}\left( \frac{\Gamma(\frac{1}{2}-\epsilon)}{4\pi^{\frac{3}{2}-\epsilon}} \frac{1}{|x-y|^{1-2\epsilon}}\right), \\
        \langle (C_I)_i^{\hat j}(x)(\bar C^J)_{\hat k}^l(y)\rangle^{(0)} &= \delta_I^J \delta_i^l \delta_{\hat k}^{\hat j}  \frac{\Gamma(\frac{1}{2}-\epsilon)}{4\pi^{\frac{3}{2}-\epsilon}}\frac{1}{|x-y|^{1-2\epsilon}} ,
    \end{split}
\end{equation}
while the one-loop propagators are
\begin{equation}
\label{eqn:onelooppropagator}
\begin{split}
    \langle (A_{\mu})_p^{\ q}(x) (A_{\nu})_r^{s}(y) \rangle^{(1)} &= \delta_p^s\delta_r^q \left(\frac{2\pi}{k}\right)^2 N_1 \frac{\Gamma^2(\frac{1}{2}-\epsilon)}{4\pi^{3-2\epsilon}}\left[ \frac{\delta_{\mu\nu}}{|x-y|^{2-4\epsilon}}-\partial_{\mu}\partial_{\nu}\frac{|x-y|^{2\epsilon}}{4\epsilon(1+2\epsilon)} \right] ,\\
    \langle (\hat A_{\mu})_{\hat p}^{\ \hat q}(x) (\hat A_{\nu})_{\hat r}^{ \hat s}(y) \rangle^{(1)} &= \delta_{\hat p}^{\hat s}\delta_{\hat r}^{\hat q} \left(\frac{2\pi}{k}\right)^2 N_2 \frac{\Gamma^2(\frac{1}{2}-\epsilon)}{4\pi^{3-2\epsilon}}\left[ \frac{\delta_{\mu\nu}}{|x-y|^{2-4\epsilon}}-\partial_{\mu}\partial_{\nu}\frac{|x-y|^{2\epsilon}}{4\epsilon(1+2\epsilon)} \right] ,\\
    \langle  (\psi_I^{\alpha})_{\hat i}^j(x) (\bar\psi^J_{\beta})_k^{\hat l}(y) \rangle^{(1)} &= i \delta_I^J \delta_{\hat i}^{\hat l}\delta_{\hat k}^{\hat j} \delta^{\alpha}_{\beta}\left( \frac{2\pi}{k} \right) (N_1-N_2) \frac{\Gamma^2(\frac{1}{2}-\epsilon)}{16\pi^{3-2\epsilon}}\frac{1}{|x-y|^{2-4\epsilon}}.
\end{split}
\end{equation}
The latin indices are color indices. For instance, $(A_{\mu})_p^{\ q} \equiv A_\mu^a (T^a)_p^{\ q}$ where $T^a$ are $U(N_1)$ generators in fundamental representation.


\section{The auxiliary field method for fermionic Wilson loops}
\label{sec:apxDorn}

In this section we prove that the ABJ(M) Wilson loop VEV
\begin{equation}
\label{eqn:ad0}
    \langle W \rangle = \langle \Tr \cP e^{-i\int d\tau \cL} \rangle\,,
\end{equation}
can be written as the two-point function of the one-dimensional field supermatrix, {\it i.e.}
\begin{equation}
\label{eqn:ad1}
    \langle W \rangle = \frac{1}{2}\langle \Tr \Psi_0 \bar\Psi_0 \rangle_{1D}\,,
\end{equation}
where the one-dimensional fields on the r.h.s. of this equation are the bare ones. In order to simplify the notation, in what follows we will neglect the subscript 0 under the assumption that all the fields have to be meant as bare fields. 

We start by defining
\begin{equation}
\label{eqn:ad2}
    \cZ [\eta,\bar\eta]=\int D\Psi D\bar\Psi e^{-\int d\tau \Tr (\bar\Psi D_{\tau}\Psi -\bar\Psi \eta - \bar\eta \Psi)}\,,
\end{equation}
where $\cD_{\tau}= \partial_{\tau}+i\cL$ and
\begin{equation}
    \eta = \begin{pmatrix} \chi & \phi \\ \tilde\phi & \tilde \chi \end{pmatrix}\,,\qquad\qquad \bar\eta = \begin{pmatrix} \bar\chi & \bar{\tilde\phi} \\ \bar\phi & \bar{\tilde \chi} \end{pmatrix}
\end{equation}
are odd supermatrices with $\chi$ ($\tilde \chi$) and $\phi$ ($\tilde\phi$) one-dimensional fermion and scalar fields in the fundamental representation of $U(N_1)$ ($U(N_2)$). The one-dimensional fields two-point function can be written as
\begin{equation}
\label{eqn:ad4}
    \langle \Psi\bar\Psi \rangle_{1D} = \frac{\delta^2 \log\cZ[\eta,\bar\eta]}{\delta^L \bar\eta \, \delta^R \eta}\Big|_{\eta=\bar\eta=0}
\end{equation}
where $\delta^L \bar\eta$ and $\delta^R \eta$ are the left and right derivative respectively, defined as
\begin{equation}
    \frac{\delta}{\delta^L \bar\eta} = \begin{pmatrix} \frac{\delta}{\delta{\bar\chi}} & \frac{\delta}{\delta\bar\phi} \\ \frac{\delta}{\delta\bar{\tilde\phi}} & \frac{\delta}{\delta\bar{\tilde \chi}} \end{pmatrix}\,, \qquad \frac{\delta}{\delta^R \eta} = \begin{pmatrix} \frac{\delta}{\delta{\chi}} & \frac{\delta}{\delta\tilde\phi} \\ \frac{\delta}{\delta \phi} & \frac{\delta}{\delta \tilde \chi} \end{pmatrix}\,,
\end{equation}
such that $\frac{\delta  (\theta_1 \theta_2)}{\delta^R \theta_2}=\theta_1$, where $\theta_1,\theta_2$ are Grassmann numbers.

Assuming that a consistent definition of integration over supermatrices exists, and that it leads to a well-defined and non-vanishing results for Gaussian integrals, we can solve the path integral in \eqref{eqn:ad2} with the standard technique of completing the square at the exponent. In particular, we find
\begin{equation}
\label{eqn:ad3}
    \cZ [\eta,\bar\eta] \propto \exp\int d\tau\,\bar\eta D_{\tau}^{-1}\eta\,,
\end{equation}
where the overall coefficient associated to the result of the supermatrix Gaussian integration is irrelevant, since the two-point function is defined as the derivative of the logarithm of $\cZ$.

Now, taking the double derivative of \eqref{eqn:ad3}, one can easily check that
\begin{equation}
\label{eqn:ad5}
    \frac{\delta^2 \log\cZ[\eta,\bar\eta]}{\delta^L \bar\eta \, \delta^R \eta}\Big|_{\eta=\bar\eta=0} = 2 D_{\tau}^{-1}\,.
\end{equation}
On the other hand, the following identity holds
\cite{Arefeva:1980zd,CRAIGIE1981204,Dorn:1986dt}
\begin{equation}
\label{eqn:ad6}
    D_{\tau}^{-1}=\left( \partial_{\tau} +i\cL \right)^{-1} = \theta(\tau)\cP e^{-i\int_0^{\tau} d\tau' \cL(\tau')}\,.
\end{equation}
In conclusion, combining \eqref{eqn:ad4} with \eqref{eqn:ad5} and inserting \eqref{eqn:ad6}, we find
\begin{equation}
\label{eqn:ad7}
    \frac{1}{2}\langle \Psi(\tau) \bar\Psi(0) \rangle_{1D} = \theta(\tau) \cP e^{-i\int_0^{\tau}d\tau'\cL(\tau')}\,.
\end{equation}
Finally, taking the trace and the ABJ(M) expectation value we reproduce \eqref{eqn:ad0}.


\section{Perturbative computations}
\label{app:renormalization}

\subsection{Gauge-fermion vertex corrections}
\label{sec:gamma}

Here we provide details on the evaluations of the gauge vertex corrections. 

We start from diagram in figure \ref{subfig:gauge1b} that corresponds to the following integral
\begin{equation}
    \Gamma^{\rm \subref{subfig:gauge1b}}_{\text{gauge}}=\frac{1}{18g^2}\int d\tau_1 \int d\tau_2 \int d^d x \, \langle\left(\bar z_1 A_{\mu}(\tau_1) \dot x_1^{\mu} z_1\right)\left(\bar z_2 A_{\nu}(\tau_2) \dot x_2^{\nu} z_2\right)\left(\epsilon^{\rho\sigma\tau} A_{\rho}A_{\sigma}A_{\tau} \right)(x)\rangle.
\end{equation}
It is easy to see that this contribution is vanishing due to the antisymmetry of the $\epsilon$ tensor. In fact, to begin with, we perform the contractions using the one-dimensional and gauge fields propagators. We obtain
\begin{equation}
\label{eq:apxC1}
    \Gamma^{\rm\subref{subfig:gauge1b}}_{\text{gauge}} \sim \int d\tau_1 \int^{\tau_1} d\tau_2 \int d^d x \bar z_1 \dot x^{\mu}_1 \dot x^{\nu}_2 z_2 A_{\rho} \, \epsilon^{\rho\sigma\tau}\epsilon_{\mu\sigma\omega}\epsilon_{\nu\tau\eta}\, \frac{(x_1-x)^{\omega}(x_2-x)^{\eta}}{|x_1-x|^d|x_2-x|^d}.
\end{equation}
Then, we take the $\tau_2\to\tau_1$ limit focusing only on the potentially divergent terms. The numerator of \eqref{eq:apxC1} turns out to be proportional to 
\begin{equation}
     \epsilon^{\rho\sigma\tau}\epsilon_{\mu\sigma\omega}\epsilon_{\nu\tau\eta} \, \dot x^{\mu}_1\dot x^{\nu}_1 (x_1-x)^{\omega} \left[ (x_1-x)^{\eta}+(\tau_2-\tau_1)\dot x^{\eta}_1  \right].
\end{equation}
By using the following relation
\begin{equation}
        \epsilon^{\rho\sigma\tau}\epsilon_{\mu\sigma\omega}\epsilon_{\nu\tau\eta}=(\delta^{\rho}_{\mu}\delta^{\tau}_{\omega}-\delta^{\rho}_{\omega}\delta^{\tau}_{\mu})\epsilon_{\nu\tau\eta}= \delta^{\rho}_{\mu}\epsilon_{\nu\omega\eta}-\delta^{\rho}_{\omega}\epsilon_{\nu\mu\eta}\,,
\end{equation}
it can be reduced to contractions between symmetric and antisymmetric tensors, which eventually lead to a vanishing result.

From the diagram in figure \ref{subfig:gauge1c} we have 
\begin{equation}
\begin{split}
    \Gamma^{\rm\subref{subfig:gauge1c}}_{\text{gauge}}&=\int d\tau_1\int d\tau_2\int d^d x \langle \left(\bar\psi \gamma^{\mu} \psi A_{\mu}\right)(x)\left( \bar z_1 \bar f_1 \tilde \varphi_1 \right)\left( \bar{\tilde\varphi}_2 f_2 z_2 \right) \rangle \\ &\,\,\begin{split}
        =-\int d\tau_1 \int^{\tau_1} d\tau_2 \int d^d x  &\left( \bar\alpha^i\alpha_i e^{  i\frac{\tau_{12}}{2}}\eta_1\xi_2 + \beta^j\bar\beta_j e^{-i\frac{\tau_{12}}{2}}\xi_1\eta_2  \right)\\&\times\langle \psi(\tau_2)\bar\psi(x) \rangle \gamma^{\mu} \langle \psi(x)\bar\psi(\tau_1) \rangle z_2 \bar z_1 A_{\mu}(x)\,.
    \end{split}
\end{split}
\end{equation}
Inserting the propagators and exploiting the properties of $\eta$ and $\xi$, we find
\begin{multline}
    \Gamma^{\rm\subref{subfig:gauge1c}}_{\text{gauge}}=N_2(\bar\alpha^i\alpha_i+\beta^j\bar\beta_j)\frac{\Gamma^2(\frac{3}{2}-\epsilon)}{4\pi^{3-2\epsilon}}\\ \times \int d\tau_1\int^{\tau_1}d\tau_2 \int d^d x \xi_2\gamma^{\nu}\gamma^{\mu}\gamma^{\rho}\eta_1 e^{i\frac{\tau_{12}}{2}}\frac{(x_2-x)_{\nu}(x-x_1)_{\rho}}{|x_2-x|^{d}|x-x_1|^d} \bar z_1 z_2  A_{\mu}(x).
\end{multline}
We then use the spinorial relation $\gamma^{\nu}\gamma^{\mu}\gamma^{\rho}=\delta^{\nu\mu}\gamma^{\rho}+\delta^{\mu\rho}\gamma^{\nu}-\delta^{\nu\rho}\gamma^{\mu}+i\epsilon^{\rho\mu\nu}$ in order to write the integrand as
\begin{equation}
\begin{split}
    \frac{(x_2-x)^{\mu}\gamma^{\rho}(x-x_1)_{\rho}+(x_2-x)_{\nu}\gamma^{\nu}(x-x_1)^{\mu}-(x_2-x)_{\rho}(x-x_1)^{\rho}\gamma^{\mu}}{|x_2-x|^d|x-x_1|^d}A_{\mu}(x)\bar z_1 z_2,
\end{split}
\end{equation}
where we dropped the $\epsilon^{\rho\mu\nu}$ term since it will not contribute in the $\tau_1\to\tau_2$ limit. By using the following integral \cite{Dorn:1986dt}
\begin{multline}
\label{eqn:integral}
    \int \! \! d^d y  \,\frac{f(y)(y-x_1)_{\mu}(y-x_2)_{\nu}}{|x_1-y|^d |x_2-y|^d}=\frac{2\pi^{d/2}}{(d-2)^2\Gamma(\frac{d}{2}-1)}(f(x_1)+O(|x_1-x_2|))\\ \times\left[ \delta_{\mu\nu}|x_1-x_2|^{2-d} + (x_1-x_2)_{\mu}(x_1-x_2)_{\nu}(2-d)|x_1-x_2|^{-d} \right],
\end{multline}
we obtain
\begin{multline}
     \Gamma^{\rm\subref{subfig:gauge1c}}_{\text{gauge}}=N_2(\bar\alpha^i\alpha_i+\beta^j\bar\beta_j)\frac{2\pi^{\frac{3}{2}-\epsilon}}{(1-2\epsilon)\Gamma(\frac{1}{2}-\epsilon)}\frac{\Gamma^2(\frac{3}{2}-\epsilon)}{2\pi^{3-2\epsilon}}\\ \times \int d\tau_1\int^{\tau_1}d\tau_2 e^{i\frac{\tau_{12}}{2}}\Big[ (x_1-x_2)^{\mu}(x_1-x_2)_{\nu}\xi_2\gamma^{\nu}\eta_1|\tau_{12}|^{-3+2\epsilon} \Big]A_{\mu}(x_1)\bar z_1 z_2\,.
\end{multline}
If we focus on the $\tau_2\to\tau_1$ limit we find
\begin{align}
    \Gamma^{\rm\subref{subfig:gauge1c}}_{\text{gauge}}&=-ig^2N_2(\bar\alpha^i\alpha_i+\beta^j\bar\beta_j)\frac{2\Gamma^2(\frac{3}{2}-\epsilon)}{(1-2\epsilon)\pi^{\frac{3}{2}-\epsilon}\Gamma(\frac{1}{2}-\epsilon)}\int \! \! d\tau_1\int^{\tau_1} \! \! \! \! d\tau_2 (\tau_{12})^{-1+2\epsilon}\dot x^{\mu}_1A_{\mu}(x_1)\bar z_1 z_1\nonumber\\ &= -\frac{g^2N_2}{4\pi \epsilon} (\bar\alpha^i\alpha_i+\beta^j\bar\beta_j)\int d\tau \ i\bar z A_{\mu} \dot x^{\mu} z.
\end{align}

Finally, from diagram \ref{subfig:gauge1d} we obtain
\begin{equation}
\begin{split}
    \Gamma^{\rm\subref{subfig:gauge1d}}_{\text{gauge}} &\sim g^2 \int d\tau_1 \int d^d x \langle(\bar z_1 C_J(x_1) \bar C^J(x_1) z_1)(\partial^{\mu} C_I \, \bar C^I A_{\mu})(x)\rangle \\ &\sim g^2\int d\tau_1 \bar z_1 z_1 \int d^d x \, A_{\mu}(x)  \, \partial^{\mu}_x\left(\langle C(x) \bar C(x_1) \rangle\right)^2,
\end{split}
\end{equation}
where in the second line we have exploited the identity $\langle C(x) \bar{C}(x_1)\rangle = \langle \bar{C}(x) C(x_1)\rangle$. Since in Lorentz gauge the integrand is a total $x$-derivative, the final result is zero. 


\subsection{Fermion vertex corrections}
\label{sec:gamma_af}

In this section we perform the explicit calculation of the one-loop corrections to the fermion vertices \eqref{eq:fermionvertices}, shown in figure \ref{fig:fermion1}. 

Focusing on the $\alpha_i$ corrections, the algebraic expression corresponding to diagram \ref{subfig:fermion1a} reads
\begin{equation}
\label{eqn:s1}
\begin{split}
    \Gamma^{\rm\subref{subfig:fermion1a}}_{\text{fermion}} &= \int d\tau_1 \int d\tau_2 \int d^d x \langle (\bar{\tilde z}_1 f_1 \varphi_1)(\bar\varphi_2 A_{\mu}(x_2)\dot x^{\mu}_2 \varphi_2)(\bar\psi \gamma^{\nu} \psi A_{\nu})(x) \rangle \\ & \to \, (\alpha_1-\alpha_2)\int d\tau_1 \int^{\tau_1} d\tau_2 \int d^d x \bar{\tilde z}_1 \varphi_2 e^{-i\frac{\tau_1}{2}} \xi_1 \langle \psi(x_1)\bar\psi(x) \rangle \langle A_{\mu}(x_2)A_{\nu}(x) \rangle \dot x^{\mu}_2 \gamma^{\nu} \psi(x).
\end{split}
\end{equation}
Here $\psi$ stands for any fermion component and we have taken into account that the contributions to $\alpha_1$ and $\alpha_2$ are the same, apart from a different overall sign. 

Reading the propagators from equation \eqref{eqn:propagator}, we obtain
\begin{align}
    \Gamma^{\rm\subref{subfig:fermion1a}}_{\text{fermion}}&= g^2N_1(\alpha_1-\alpha_2)\frac{\Gamma^2(\frac{3}{2}-\epsilon)}{4\pi^{3-2\epsilon}}\int d\tau_1 \int^{\tau_1} d\tau_2 \int d^d x \, \bigg[\bar{\tilde z}_1 \varphi_2 e^{-i\frac{\tau_1}{2}} \xi_1 \gamma^{\sigma}  \epsilon_{\mu\nu\rho}\frac{(x_1-x)_{\sigma}}{|x_1-x|^d} \nonumber\\
    &\times \frac{(x_2-x)^{\rho}}{|x_2-x|^d} \dot x^{\mu}_2 \gamma^{\nu} \psi(x)\bigg].
\end{align}
The $d$-dimensional integral can be evaluated using \eqref{eqn:integral}. This leads to
\begin{align}
    \Gamma^{\rm\subref{subfig:fermion1a}}_{\text{fermion}} &= g^2N_1(\alpha_1-\alpha_2) \frac{\Gamma^2(\frac{3}{2}-\epsilon)}{2(1-2\epsilon)^2\pi^{\frac{3}{2}-\epsilon}\Gamma(\frac{1}{2}-\epsilon)}\int d\tau_1 \int^{\tau_1} d\tau_2 \,\bigg[\bar{\tilde z}_1 \varphi_2 e^{-i\frac{\tau_1}{2}} \xi_1 \epsilon_{\mu\nu\rho}\gamma^{\sigma} \gamma^{\nu}\dot x^{\mu}_2 \nonumber\\ &\times \psi(x_1) \bigg( (\tau_1-\tau_2)^{-1+2\epsilon} \delta^{\rho}_{\sigma} -(1-2\epsilon)(x_1-x_2)_{\sigma}(x_1-x_2)^{\rho}(\tau_1-\tau_2)^{-3+2\epsilon} \bigg)\bigg].
\end{align}
Using $(\gamma^{\sigma})_{\alpha}^{\ \beta}(\gamma^{\nu})_{\beta}^{ \ \delta}=\delta^{\sigma\nu}\delta_{\alpha}^{\delta} + i \epsilon^{\sigma\nu\tau}(\gamma_{\tau})_{\alpha}^{\ \delta}$ we see that some terms drop out due to anti-symmetry. Eventually, in the $\tau_2\to\tau_1$ limit we obtain
\begin{align}
    \Gamma^{\rm\subref{subfig:fermion1a}}_{\text{fermion}} &= ig^2N_1(\alpha_1-\alpha_2) \frac{\Gamma^2(\frac{3}{2}-\epsilon)}{(1-2\epsilon)\pi^{\frac{3}{2}-\epsilon}\Gamma(\frac{1}{2}-\epsilon)}\int d\tau_1 \int^{\tau_1} d\tau_2 \bar{\tilde z}_1 \varphi_1 \frac{e^{-i\frac{\tau_1}{2}}}{(\tau_1-\tau_2)^{1-2\epsilon}} \xi_1 \psi(x_1) \nonumber \\ &= (\alpha_1-\alpha_2)\frac{g^2 N_1}{8\pi\epsilon}\int d\tau \ i \bar{\tilde z} e^{-i\frac{\tau}{2}}\xi \psi \varphi.
\end{align}
The evaluation of $\Gamma^{\rm \subref{subfig:fermion1b}}_{\text{fermion}}$ proceeds exactly in the same way, the only change being the replacement of $A_\mu$ with $\hat{A}_\mu$.


\subsection{Scalar vertex corrections}
\label{sec:gamma_f}

Here we compute in details the scalar vertex corrections of figure \ref{fig:scalarvertex}. 

Starting from diagram  \ref{subfig:scalarvertexa} we have
\begin{align}
     \Gamma^{\rm\subref{subfig:scalarvertexa}}_{\text{scalar}} &= g^4 M_{I}^{ \ K} M_{L}^{\ J} \int d\tau_1 \int d\tau_2 \big( \bar\varphi C_K \bar C^I \varphi \big)(x_1) \big( \bar\varphi C_J \bar C^L \varphi \big)(x_2)\nonumber
     \\&= -g^4 N_1 M_{I}^{ \ K} M_{K}^{\ J} \frac{\Gamma(\frac{1}{2}-\epsilon)}{4\pi^{\frac{3}{2}-\epsilon}}\int d\tau_1 \int^{\tau_1}d\tau_2\, \bar\varphi_2 C_J(x_2) \bar C^I(x_1) \varphi_1 (\tau_{12})^{-1+2\epsilon} \,,
\end{align}
which in the $\tau_2\to\tau_1$ limit gives
\begin{equation}
    \Gamma^{\rm\subref{subfig:scalarvertexa}}_{\text{scalar}} =- g^4 \frac{N_1}{8\pi\epsilon} M_{I}^{ \ K} M_{L}^{\ J} \int_{-L}^L d\tau\,  \bar\varphi C_J \bar C^I \varphi\,.
\end{equation}

Diagram \ref{subfig:scalarvertexb} contributes with
\begin{align}
    \Gamma^{\rm\subref{subfig:scalarvertexb}}_{\text{scalar}} &= \int d\tau_1 \int \tau_2 \int d^d x \big(i\bar\varphi A_{\mu}\dot x^{\mu}\varphi\big)(x_1)\big(i\bar\varphi A_{\nu}\dot x^{\nu}\varphi\big)(x_2)\big( A_{\rho} C_I \bar C^I A^{\rho} \big)(x)\nonumber\\
    &= g^2 N_1\frac{\Gamma^2(\frac{3}{2}-\epsilon)}{4\pi^{3-2\epsilon}} \int d\tau_1 \int^{\tau_1}  d\tau_2 \int d^d x\,\bigg[ \epsilon_{\rho\mu\sigma}\epsilon^{\rho\nu\lambda} \dot x_1^{\mu}\dot x_{2\nu} \frac{(x_1-x)^{\sigma}(x_2-x)_{\lambda}}{|x_1-x|^{3-2\epsilon}|x_2-x|^{3-2\epsilon}} \nonumber\\
    &\times\varphi_1\bar\varphi_2 C_I(x) C^I(x)\bigg]\,.
\end{align}
The $d^dx$ integral can be solved by using \eqref{eqn:integral}. In the $\tau_2\to\tau_1$ limit, we obtain
\begin{align}
    \Gamma^{\rm\subref{subfig:scalarvertexb}}_{\text{scalar}} &= g^2 N_1\frac{\Gamma^{2}(\frac{3}{2}-\epsilon)}{4\pi^{3-2\epsilon}} \frac{2\pi^{\frac{3}{2}-\epsilon}(2-2\epsilon)}{(1-2\epsilon)^2\Gamma(\frac{1}{2}-\epsilon)} \int d\tau_1 \int^{\tau_1}d\tau_2 (\tau_{12})^{-1+2\epsilon}\varphi_1\bar\varphi_1 C_I(x_1)\bar C^I(x_1)\nonumber\\ 
    &= g^2 \frac{N_1}{8\pi\epsilon} \int d\tau\, \bar\varphi C_I \bar C^I \varphi\,.
\end{align}

For diagram in figure \ref{subfig:scalarvertexc}, we first consider the contribution $\Gamma^{\rm\subref{subfig:scalarvertexc},1}_{\text{scalar}}$ in \eqref{eqn:Gammascalarc1}, that is the one obtained by using the ABJ(M) Yukawa vertex
$2g^2C_I\bar C^J\bar\psi^I\psi_J$. To begin with, we write 
\begin{align}
    \Gamma^{\rm\subref{subfig:scalarvertexc},1}_\text{scalar} &= -2ig^2\int d\tau_1 \int d\tau_2 \int d^dx \big(\bar\varphi \bar f \bar{\tilde z}\big)(x_1) \big(\tilde z f \varphi \big)(x_2)(C_I \bar C^J \bar\psi^I \psi_J) \nonumber\\ 
    &= 4i g^4 N_2 \bar\alpha^2\alpha_2 \int d\tau_1 \int^{\tau_1} d\tau_2 \int d^dx\,\bigg[ \bar\varphi_1 C_1(x) \bar C^1(x) \varphi_2 e^{i\frac{\tau_{12}}{2}}e^{i\tau_2} u^{\alpha}(\tau_1)v_{\beta}(\tau_2)\nonumber\\
    &\times\langle \psi^{\beta}(x_2)\bar\psi_{\delta}(x) \rangle \langle \psi^{\delta}(x)\bar\psi_{\alpha}(x_1) \rangle\bigg]\,.
\end{align}
We then proceed as we have done in section \ref{sec:gamma_af} for the fermion vertex corrections, so obtaining
\begin{align}
    \Gamma^{\rm\subref{subfig:scalarvertexc},1}_\text{scalar}&=-4g^4 N_2 \bar\alpha^2\alpha_2 \frac{\Gamma^2(\frac{3}{2}-\epsilon)}{4\pi^{3-2\epsilon}}\frac{2\pi^{\frac{3}{2}-\epsilon}}{(1-2\epsilon)^2\Gamma(\frac{1}{2}-\epsilon)}\int d\tau_1 \int^{\tau_1}d\tau_2\bigg[\bar\varphi_1 C_1(x_1) \bar C^1(x_1) \varphi_2 \nonumber\\
    &\times \cos\frac{\tau_{12}}{2}\bigg( (3-2\epsilon)|x_1-x_2|^{-1+2\epsilon} + (x_1-x_2)^2(-1+2\epsilon)|x_1-x_2|^{-3+2\epsilon} \bigg)\bigg]\,.
\end{align}
By computing the $\tau_2$-integral in the $\tau_2\to\tau_1$ limit we eventually find
\begin{equation}
     \Gamma^{\rm\subref{subfig:scalarvertexc},1}_\text{scalar}= -g^4\frac{N_2}{2\pi\epsilon}\bar\alpha^2\alpha_2\int d\tau\, \bar\varphi C_1 \bar C^1 \varphi\,.
\end{equation}
The contribution $\Gamma^{\rm \subref{subfig:scalarvertexc},2}_{\text{scalar}}$ in \eqref{eqn:Gammascalarc1}, coming from the ABJ(M) vertex $-g^2C_I\bar C^I\bar\psi^J\psi_J$, corresponds to performing contractions in the following string
\begin{equation}
    \Gamma^{\rm\subref{subfig:scalarvertexc},2}_{\text{scalar}} = ig^2\int d\tau_1 \int d\tau_2 \int d^dx \, \langle \big(\bar\varphi \bar f \bar{\tilde z}\big)(x_1) \big(\tilde z f \varphi \big)(x_2)(C_I \bar C^I \bar\psi^J \psi_J) \rangle \,.
\end{equation}
The computation is analogous to the previous one, so we do not replicate it. The final result reads
\begin{equation}
    \Gamma^{\rm\subref{subfig:scalarvertexc},2}_{\text{scalar}} =g^4\frac{N_2}{4\pi\epsilon}\bar\alpha^2\alpha_2\int d\tau\, \bar\varphi C_I \bar C^I \varphi\,.
\end{equation}


\newpage


\bibliographystyle{JHEP}
\bibliography{refs}
\newpage 

\end{document}